\documentclass[a4paper,fleqn, usenatbib]{mnras}

\usepackage[T1]{fontenc}
\usepackage{ae,aecompl}
\usepackage{longtable}

\usepackage{graphicx}	
\usepackage{amsmath}	
\usepackage{amssymb}	
\usepackage{subfigure}

\title[BLR Modeling]{Estimation of the size and structure of the broad line region using Bayesian approach}

\author[Mandal et al.]{Amit Kumar Mandal$^{1,2}$\thanks{E-mail: amitkumar@iiap.res.in},
Suvendu Rakshit$^{3,4}$,
C. S. Stalin$^{2}$,
\newauthor
R. G. Petrov$^{5}$,
Blesson Mathew$^{1}$,
Ram Sagar$^{2}$
\\\\
$^{1}$Department of Physics, CHRIST (Deemed to be University), Hosur Road, Bangalore 560 029, India\\
$^{2}$Indian Institute of Astrophysics, Block II, Koramangala, Bangalore, 560 034, India\\
$^{3}$Aryabhatta Research Institute of Observational Sciences, Manora Peak, Nainital 263002, India\\
$^{4}$Finnish Centre for Astronomy with ESO (FINCA), University of Turku, Quantum, Vesilinnantie 5, 20014 University of Turku, Finland\\
$^{5}$Observatoire de la C$\hat{o}$te d'Azur, CNRS, Laboratoire Lagrange, Universit$\acute{e}$ C$\hat{o}$te d'Azur, Parc Valrose, B$\hat{a}$t. H. Fizeau, F-06108 Nice, France \\
\\
}

\date{Accepted 2020 December 30. Received 2020 December 28; in original form 2019 November 20}

\pubyear{2020}

\begin{document}
\label{firstpage}
\pagerange{\pageref{firstpage}--\pageref{lastpage}}
\maketitle

\begin{abstract}
Understanding the geometry and kinematics of the broad line region (BLR) of 
active galactic nuclei (AGN) is important to estimate black hole masses in
AGN and study the accretion process. The technique of reverberation 
mapping  (RM)
has provided estimates of BLR size for more than 100 AGN now, however, the structure 
of the BLR has been studied for only a handful number of objects. Towards 
this, we investigated the geometry of the BLR for a large sample of 57 
AGN using archival RM data. We performed systematic modeling of the continuum 
and emission line light curves using a Markov Chain Monte Carlo method based on 
Bayesian statistics implemented in PBMAP (Parallel Bayesian code for reverberation$-$
MAPping data) code to constrain BLR geometrical 
parameters and recover velocity integrated transfer function. We found that 
the recovered transfer functions have various shapes such as single-peaked, 
double-peaked and top-hat suggesting that AGN have very different BLR 
geometries.  Our model lags are in general consistent with that estimated 
using the conventional cross-correlation methods. The BLR sizes obtained from 
our modeling approach is related to the luminosity with a slope of 
$0.583\pm0.026$ and $0.471\pm0.084$ based on H$\beta$ and H$\alpha$ lines, 
respectively. We found a non-linear response of emission line fluxes to the ionizing optical continuum for 93$\%$ objects. The estimated virial factors for the AGN studied in this work range 
from 0.79 to 4.94 having a mean at $1.78\pm 1.77$ consistent with the values found in the literature.       

\end{abstract}

\begin{keywords}
galaxies: active $-$ galaxies: Seyfert $-$ (galaxies:) model: Bayesian  
\end{keywords}



\section{Introduction}
Active Galactic Nuclei (AGN) are believed to be powered by the accretion of 
matter on to super massive black hole (SMBH; 10$^6$ $-$ 10$^{10}$  M$_{\odot}$) located at 
the center of galaxies \citep{1969Natur.223..690L,1964ApJ...140..796S}. 
This extreme physical process is responsible for the radiation we receive 
from AGN over a wide range of energies predominantly emitted in 
X-ray, UV and optical wavelengths. The SMBH is surrounded by
the accretion disk and the optical/UV emission seen in the 
spectral energy distribution of an AGN is attributed to the 
thermal emission from the accretion disk. Farther from the accretion
disk on scales of about $\sim$ 0.01 parsec (typical for the Seyfert galaxies category of 
AGN) lies the broad line region (BLR) that produces the line emission. 
The UV/optical continuum from the accretion disk photoionizes the 
gas clouds in the BLR, and as these gas clouds are situated deep
within the potential well of the black hole, the emission lines
from them are broadened by few thousands of km/sec. The BLR is surrounded
by the dusty torus which is responsible for the different manifestations of the 
AGN when viewed at different orientations \citep{1993ARA&A..31..473A,1995PASP..107..803U}. 
According to \citet{1987ApJ...320..537B}, the bump or excess in the near 
infrared (NIR) continuum is thermally produced by the hot dust grains heated by 
the UV/optical radiation from the central region. For a typical UV luminosity of 
$10^{42} - 10^{44}$ $\mathrm{erg ~s^{-1}}$, the inner extent of the torus is around 
$0.01 - 0.1$ pc or $10 - 100$ light-days \citep{2006ApJ...639...46S}. Thus, the 
central region of AGN is highly compact, and difficult to probe using any 
current imaging techniques. However, recently, from interferometric 
observations carried out in IR using  the GRAVITY instrument on the European Very Large Telescope (VLT), 
\citet{2018Natur.563..657G} were able to resolve the central region in 3C 273 
on scales of about 0.12 parsec.  On 
much larger scales in AGN  are the clouds
in the narrow line region that are responsible for the narrow emission lines we 
see in the spectrum of AGN with widths of few hundreds of km/sec.

One of the defining characteristics of AGN is that they show flux variations \citep{1995ARA&A..33..163W,1997ARA&A..35..445U}. 
This is known since their discovery, though, the causes of flux variations 
 are not yet understood. In spite of that, the flux variability 
characteristic of AGN provides a very useful way to probe the spatially 
unresolved inner regions in them.
A technique that uses the variability of AGN  to probe their central
regions is known as reverberation mapping \citep[RM;][]{1982ApJ...255..419B,
1993PASP..105..247P}. This method is based on the variation 
of the line-fluxes from the BLR in response to the the continuum UV-optical flux 
variations from the accretion disk. The time delay ($\tau$) measured using the traditional
cross-correlation techniques (\citealt{1988ApJ...333..646E, 1993PASP..105..247P}) 
between the continuum flux and line flux variations is the average light travel 
time from the accretion disk to the BLR, which in principle gives the average 
radius of the BLR. Having the BLR size ($R_{\mathrm{BLR}}=\tau/c$, where c is 
the velocity of light) and the width of the broad emission line ($\Delta V$), 
measurable from the spectrum, the mass of the black hole (M$_{BH}$) in an AGN can 
be estimated using  the virial relation,
\begin{gather}
M_{\mathrm{BH}} = f_{\mathrm{BLR}} \left(\frac{\Delta V^{2}R_{\mathrm{BLR}}}{G}\right)
\label{eq:blc_v}
\end{gather}
where $G$ represents the gravitational constant and $f_{\mathrm{BLR}}$ is the 
virial factor that depends on the geometry and kinematics of BLR. 
The method of RM has been used to measure M$_{BH}$ in more than 100 AGN with most
of the measurements coming from the compilation\footnote{http://www.astro.gsu.edu/AGNmass/} 
of  \citet{2015PASP..127...67B}, the Sloan Digital Sky Survey Reveberation Measurement project; SDSS$-$RM
\citep{2017ApJ...851...21G} and the Super-Eddington Accretion in Massive Black Holes
project \citep{2014ApJ...782...45D,2016ApJ...825..126D,2018ApJ...856....6D,2014ApJ...793..108W}.
The $f_{\mathrm{BLR}}$ in Equation \ref{eq:blc_v} is calibrated considering both AGN and 
local quiescent galaxies follow the same 
$M-\sigma_*$ relation \citep[e.g.,][]{2004ApJ...615..645O,2015ApJ...801...38W}, where $\sigma_*$ is the stellar velocity dispersion. 
However, it is not clear whether a constant $f_{\mathrm{BLR}}$ can be used to 
estimate M$_{BH}$ for all AGN considering the complex geometry and 
kinematics of individual AGN. \citet{2014ApJ...789...17H} performed a 
recalibration of the virial factor and found that it depends on the bulge type 
of the host galaxy, the $f_{\mathrm{BLR}}$ for classical bulges and elliptical 
bulge objects are twice larger than pseudo bulge objects. \citet{2014MNRAS.445.3073P} 
performed dynamical modeling of BLR and found that $f_{\mathrm{BLR}}$ 
correlates with the inclination of the objects and has different values for 
different objects. 
In addition to getting M$_{BH}$ values, RM observations involving optical and IR photometric observatios as well as IR spectroscopic monitoring observations  can be used to constrain the inner edge of the dusty torus in AGN\citep{2006ApJ...639...46S,2014ApJ...788..159K,2014A&A...561L...8P,2018MNRAS.475.5330M, 2019arXiv190801627L, 2020MNRAS.tmp.3605M}. Recently, from high resolution radio observations using the Very Large Array, \cite{2019ApJ...874L..32C}  have imaged the torus in Cygnus A. \citet{2019arXiv191000593G} partially resolved the size and structure of hot dust using VLTI/GRAVITY and reported the increase of the physical radius with bolometric luminosity in 8 Type 1 AGN.

Most of the RM studies available in literature are mainly 
focused on estimating BLR sizes and M$_{BH}$ which are solely driven
by the quality of the available RM data both in terms of time 
resolution and signal-to-noise ratio (SNR). This approach makes the geometry 
and 
kinematics of the central engine of BLR remain unknown. However, for sources
with densely sampled and good SNR spectra,  it is in principle possible to 
calculate time delays as a function of velocity across the emission line profile 
and better constrain the geometry and kinematics of the gas in the BLR 
\citep[e.g.,][]{1996MNRAS.283..748U,2010ApJ...720L..46B,2013ApJ...764...47G,
2018ApJ...864..109X}. Also, modeling of RM  data using 
Bayesian approach has led to constrain the geometry and dynamics of the 
BLR, estimate AGN parameters, as well as,  $f_{\mathrm{BLR}}$ for individual objects
\citep{2011ApJ...730..139P,2011ApJ...733L..33B,2012ApJ...754...49P,2014MNRAS.445.3073P,2018ApJ...869..137L}.  
A geometrical modeling code, PBMAP (Parallel Bayesian code for 
reverberation$-$MAPping data) developed by \cite{2013ApJ...779..110L}  
in addition to providing several  parameters, 
e.g., BLR size, inclination angle ($\theta_{\mathrm{inc}}$), opening angle 
($\theta_{\mathrm{opn}}$), also includes non-linear response of the line emission 
to the continuum and allow for detrending the light curves \citep{1999PASP..111.1347W,2010ApJ...721..715D,2013ApJ...779..110L,2019arXiv190904511L}. This technique was
applied to a sample of 40 objects using archival H$\beta$ line 
and continuum light curves
by \citet{2013ApJ...779..110L} and they were able to recover velocity integrated 
transfer function. They found that the BLR structure for H$\beta$ line is 
mainly disk-like. Such a flattened disk like BLR is also seen in 
the high resolution observations of 3C 273 with the VLT \citep{2018Natur.563..657G}. Also, 
in \cite{2013ApJ...779..110L} the observed line fluxes were better reproduced using the 
non-linear response of the line to the continuum that has not been considered 
in previous RM studies. It is therefore important to extend this approach to 
a large number of AGN to examine differences if any in the geometry of the BLR  between
different AGN. Towards this, we carried out a systematic modeling of the 
RM data available in the literature to investigate the 
geometry of the BLR. For this we used 
the  PBMAP code to model the emission line and continuum light curves by constraining several BLR parameters such as (a) the BLR size, (b) $\theta_{\mathrm{inc}}$,
(c) $\theta_{\mathrm{opn}}$ and the non-linear response index. We also 
calculated  $f_{\mathrm{BLR}}$ for a number of sources by constraining the geometry of the BLR. The 
paper is structured as follows. In Section \ref{sec:sample}, we describe the 
sample selection. We present our analysis in 
Section \ref{sec:model}. Our results are given in Section \ref{sec:results} 
followed by the summary  in Section \ref{sec:summary}.  For the cosmological 
parameters we assumed $H_0 = 73~\mathrm{km~s^{-1}~Mpc^{-1}}$ , $\Omega_{\mathrm{m}} 
= 0.27$ and $\Omega_{\Lambda} = 0.73$. 

\section{The Sample}\label{sec:sample}
The objective of this work is to constrain the geometry of the BLR as well as 
other characteristic properties of AGN by modeling the continuum and line 
light curves. This requires spectrophotometric monitoring observations of AGN. 
We therefore, collected data for all objects that are in the black hole mass database 
by \citet{2015PASP..127...67B} and the SDSS-RM program 
\citep{2016ApJ...818...30S,2017ApJ...851...21G} from the literature with the following two 
additional constraints (i)  the objects must have continuum and line light 
curves. The line light curves can be either of Mg II, $\mathrm{H\beta}$ or $\mathrm{H\alpha}$ 
and (ii) BLR structure of the objects was not investigated previously. 
With the above constraints we arrived at a sample of 57 objects. Of 
the 57 objects, 22 have data for both  $\mathrm{H\beta}$ and H$\alpha$ lines, 3 have data for both $\mathrm{H\beta}$ and MgII lines, 25 have only $\mathrm{H\beta}$, 4 have only $\mathrm{H\alpha}$ and 3 have data for only MgII line with a total of 82 independent measurements.  The data for these objects were 
from \citet{2016ApJ...818...30S}, \citet{2017ApJ...851...21G}, \citet{2014ApJ...793..108W}, 
\citet{2012ApJ...755...60G}, \citet{2014ApJ...796....8B}, 
\citet{2014ApJ...795..149P}, \citet{2013ApJ...773...24R}, \citet{2014ApJ...782...45D} and \citet{2013A&A...559A..10S}. For the objects 
taken from \citet{2016ApJ...818...30S}, the continuum light curves pertain to 
the measurements at 5100 \AA, while for the objects taken from 
\citet{2017ApJ...851...21G}, the continuum light curves belong to the 
photometric monitoring observations done in $g$ and $i$ bands. The details of the objects selected for this study are given in Table \ref{tab:table-1}.

For this work, we directly downloaded the continuum and line light curves 
from the literature in their original form i.e. magnitude or flux versus time, 
without any further special treatment. Note that different authors followed 
different methods for their data analysis. In some cases the presence of  
constant flux component in the continuum light curves from the host-galaxy 
were removed by the respective authors 
\citep{2014ApJ...796....8B, 2014ApJ...793..108W,2016ApJ...818...30S} by  
difference imaging analysis of the images. However, in some cases the 
removal of the host galaxy  was not done e.g., 
the $\mathrm{f_{5100}}$ light curves from 
\citet{2012ApJ...755...60G, 2013ApJ...773...24R, 2014ApJ...782...45D, 2017ApJ...851...21G} 
are not host galaxy subtracted. We note that host galaxies contribute a constant flux to the continuum light curves, however, this can vary depending on the seeing variations between different epochs of observations \citep{2001sac..conf....3P}. Though this can have some effect on the deduced amplitude of variations \citep{2001sac..conf....3P}, it does not affect the estimation of the model parameters and the main conclusions of the paper. Similarly, as our sample  comes from different sources, not all broad emission line light curves have their 
narrow component subtracted. For example,  \citet{2016ApJ...818...30S, 2017ApJ...851...21G} subtracted the narrow component 
while \citet{2014ApJ...782...45D} used the total H$\beta$ profile to generate  the H$\beta$ light curves.

\begin{table*}
\caption{Details of the objects used in this study. Here $\tau_{\mathrm{cent}}$ is in days, FWHM and $\sigma$ are in km$s^{-1}$ and log$L_{5100 \AA}$ is in erg$s^{-1}$.}
\label{tab:table-1}
\begin{tabular}{ccccccccccc} \hline
No.  &  $\alpha_{2000}$ & $\delta_{2000}$ & Type  & z  & line  & $\tau_{\mathrm{cent}}$ & FWHM  & $\sigma$ & log$L_{5100 \AA}$ & Reference  \\\hline
1 &  00:10:31.01 & +10:58:29.5 & Sy1 & 0.090 & H${\beta}$ & $12.6^{+3.9}_{-3.9}$ & $5054 \pm 145$ & $3321 \pm 107$ & $44.320 \pm 0.050$ & D  \\
2 &  02:30:05.52 & -08:59:53.2 & Sy1 & 0.016 & H${\beta}$ & $4.8^{+7.4}_{-3.7}$ & - & - & $42.950 \pm 0.050$ & C \\
3 &  06:52:12.32 & +74:25:37.2 & Sy1 & 0.019 & H${\beta}$ & $10.1^{+1.1}_{-1.1}$ & $9744 \pm 3700$ & $3714 \pm 68$ & $43.75 \pm 0.060$ & D \\
4 &  11:39:13.92 & +33:55:51.1 & Sy1 & 0.033 & H${\beta}$ & $12.5^{+0.5}_{-11.5}$, $23.3^{+7.5}_{-5.8}$ & - & - & $42.670 \pm 0.090$ & G, C \\
5 &  12:42:10.61 & +33:17:02.7 & Sy1 & 0.044 & H${\beta}$ & $11.4^{+2.9}_{-1.9}$ & - & - & $43.590 \pm 0.040$ & H   \\
6 &  13:42:08.39 & +35:39:15.3 & Sy1 & 0.003 & H${\beta}$ & $2.22^{+1.19}_{-1.61}$ & $4615 \pm 330$ & $1544 \pm 98$ & $41.534 \pm 0.144$ & E  \\ 
  &              &             &     &       & H${\alpha}$ & $2.06^{+1.42}_{-1.31}$ & - & - & $41.534 \pm 0.144$ & E \\ 		
7 &  14:05:18.02 & +53:15:30.0 & QSO  & 0.467 & H${\beta}$ & $41.6^{+14.8}_{-8.3}$ & $3131 \pm 44$ & $1232 \pm 16$ & $44.300 \pm 0.001$ & B \\	
   &             &             &      &        & H${\alpha}$ & - & - & - & $44.300 \pm 0.001$ & B \\
8 &  14:05:51.99 & +53:38:52.1 & QSO & 0.455 & H${\alpha}$ & $53.0^{+8.7}_{-5.7}$ & $3489 \pm 84$ & $1590 \pm 24$ & $43.900 \pm 0.002$ & B \\
9 &  14:07:59.07 & +53:47:59.8 & Sy1 & 0.173 & H${\beta}$ & $19.2^{+4.3}_{-12.8}$& $5115 \pm 59$ & $1790 \pm 10$ & $43.541 \pm 0.001$ & A    \\
10 &  14:08:12.09 & +53:53:03.3 & Sy1 & 0.116 & H${\beta}$ & $10.5^{+1.0}_{-2.2}$ & $3111 \pm 36$ & $1409 \pm 11$ & $43.120 \pm 0.001$ & B \\ 
   &              &             &     &       & H${\alpha}$ & $8.3^{+4.9}_{-6.3}$ & $2794 \pm 15$ & $1185 \pm 7$ & $43.120 \pm 0.001$ & B \\
11 &  14:09:04.43 & +54:03:44.2 & QSO & 0.658 & H${\beta}$ & $11.6^{+8.6}_{-4.6}$ & $12673 \pm 455$ & $5284 \pm 54$ & $44.120 \pm 0.003$ & B \\
12 &  14:09:15.70 & +53:27:21.8 & Sy1 & 0.258 & H${\alpha}$ & $42.1^{+2.7}_{-2.1}$ & $6279 \pm 20$ & $3232 \pm 40$ & $43.300 \pm 0.002$ & B \\
13 &  14:10:04.27 & +52:31:41.0 & QSO & 0.527 & H${\beta}$ & $53.5^{+4.2}_{-4.0}$ & $3172 \pm 85$ & $2126 \pm 35$ & $44.190 \pm 0.001$ & B \\
   &              &             &     &       & H${\alpha}$ & - & - & - & $44.190 \pm 0.001$ & B \\
14 &  14:10:18.04 & +53:29:37.5 & QSO & 0.470  & Mg II & $32.3^{+12.9}_{-5.3}$ & - & - &  & A \\
   &              &             &     &        & H${\beta}$ & $16.2^{+2.9}_{-4.5}$ & $2377 \pm 288$ & $1781 \pm 38$ & $43.550 \pm 0.003$ & B\\
   &              &             &     &        & H${\alpha}$ & $22.1^{+7.7}_{-7.3}$ & $2103 \pm 365$ & $1738 \pm 31$ & $43.550 \pm 0.003$ & B \\
15 &  14:10:31.33 & +52:15:33.8 & Sy2 & 0.608 & H${\beta}$ & $35.8^{+1.1}_{-10.3}$ & $2578 \pm 112$ & $1619 \pm 38$ & $43.990 \pm 0.002$ & B \\
16 &  14:10:41.25 & +53:18:49.0 & QSO & 0.359 & H${\beta}$ & $21.9^{+4.2}_{-2.4}$ & $4183 \pm 51$ & $1909 \pm 12$ & $43.790 \pm 0.001$ & B  \\
   &              &             &     &       & H${\alpha}$ & $21.0^{+1.4}_{-2.8}$ & $3642 \pm 26$ & $1318 \pm 11$ & $43.790 \pm 0.001$ & B  \\
17 &  14:11:12.72 & +53:45:07.1 & QSO & 0.587 & H${\beta}$ & $18.6^{+7.1}_{-3.8}$ & $2089 \pm 77$ & $1221 \pm 36$& $44.092 \pm 0.002$ & A  \\
18 &  14:11:15.19 & +51:52:09.0 & QSO & 0.572 & H${\beta}$ & $49.1^{+11.1}_{-2.0}$ & $3234 \pm 164$ & $1423 \pm 32$ & $44.280 \pm 0.001$ & B \\
   &              &             &     &       & H${\alpha}$ & - & - & - & $44.280 \pm 0.001$ & B \\
19 &  14:11:23.42 & +52:13:31.7 & Sy1 & 0.472 & H${\beta}$ & $13.0^{+1.4}_{-0.8}$ & $4123 \pm 40$ & $1443 \pm 11$ & $44.100 \pm 0.001$ & B \\
   &              &             &     &       & H${\alpha}$ & $22.6^{+0.6}_{-1.5}$ & $3483 \pm 44$ & $1346 \pm 13$ & $44.100 \pm 0.001$ & B \\
20 &  14:11:35.89 & +51:50:04.5 & QSO & 0.650 & H${\beta}$ & $17.6^{+8.6}_{-7.4}$ & $3422 \pm 491$ & $1527 \pm 22$ & $44.010 \pm 0.003 $ & B \\
21 &  14:12:14.20 & +53:25:46.7 & QSO & 0.458  & Mg II & $36.7^{+10.4}_{-4.8}$ & - & - & & A  \\
22 &  14:12:53.92 & +54:00:14.4 & Sy1 & 0.187 & H${\beta}$ & $21.5^{+5.8}_{-7.7}$& $5120 \pm 130$ & $1758 \pm 22$ & $42.972 \pm 0.003$ & A    \\
23 &  14:13:14.97 & +53:01:39.4 & QSO & 1.026 & H${\beta}$ & $43.9^{+4.9}_{-4.3}$ & $11002 \pm 1743$ & $6543 \pm 34$ & $44.500 \pm 0.038$ & B  \\
24 &  14:13:18.96 & +54:32:02.4 & QSO & 0.362 & H${\beta}$ & $20.0^{+1.1}_{-3.0}$ & $2730 \pm 137$ & $1353 \pm 23$ & $43.910 \pm 0.001$ & B \\
25 &  14:13:24.28 & +53:05:27.0 & QSO & 0.456 & H${\beta}$ & $25.5^{+10.9}_{-5.8}$ & $7758 \pm 77$ & $6101 \pm 48$ & $43.910 \pm 0.002$ & B \\
   &              &             &     &       & H${\alpha}$ & $56.6^{+7.3}_{-15.1}$ & $5604 \pm 31$ & $4569 \pm 51$ & $43.910 \pm 0.002$ & B \\
26 &  14:14:17.13 & +51:57:22.6 & QSO & 0.604 & Mg II & $29.1^{+3.6}_{-8.8}$ & - & - &  & A  \\
   &              &             &     &       & H${\beta}$ & $15.6^{+3.2}_{-5.1}$ & $7451 \pm 221$ & $2788 \pm 48$ & $43.370 \pm 0.012$ & B \\
27 &  14:15:32.36 & +52:49:05.9 & Sy1 & 0.715 & H${\beta}$ & $26.5^{+9.9}_{-8.8}$ & $1626 \pm 243$ & $857 \pm 32$ & $44.110 \pm 0.003$ & B \\
28 &  14:16:25.71 & +53:54:38.5 & Sy1 & 0.263 & H${\beta}$ & $21.9^{+7.9}_{-10.4}$ & $3752 \pm 93$ & $1636 \pm 11$ & $43.929 \pm 0.018$ & A \\
   &              &             &     &       & H${\alpha}$ & $32.2^{+15.6}_{-12.6}$ & $2632 \pm 28$ & $1298 \pm 8$ & $43.929 \pm 0.018$ & " \\
29 &  14:16:44.17 & +53:25:56.1 & QSO & 0.425 & Mg II & $17.2^{+2.7}_{-2.7}$ & - & - & & A  \\
30 &  14:16:45.15 & +54:25:40.8 & QSO & 0.244 & H${\beta}$ & $10.9^{+20.9}_{-6.6}$ & $4981 \pm 97$ & $1902 \pm 20$& $43.178 \pm 0.002$ & A    \\ 
   &              &             &     &       & H${\alpha}$ & $10.6^{+2.3}_{-2.4}$ & $6027 \pm 19$ & $3927 \pm 30$ & $43.178 \pm 0.002$ & B    \\ 
31 &  14:16:45.58 & +53:44:46.8 & Sy1 & 0.442 & H${\beta}$ & $23.3^{+2.7}_{-11.2}$ & $1854 \pm 70$ & $990 \pm 19$ & $43.646 \pm 0.008$ & A \\
   &              &             &     &       & H${\alpha}$ & $16.7^{+4.1}_{-5.5}$ & $1575 \pm 60$ & $796 \pm 23$ & $43.646 \pm 0.008$ & B \\
32 &  14:16:50.93 & +53:51:57.0 & QSO & 0.527 & Mg II & $25.1^{+2.0}_{-2.6}$ & - & - & & A \\
33 &  14:17:06.68 & +51:43:40.1 & Sy1 & 0.532  & H${\beta}$ & $14.1^{+12.9}_{-9.5}$ & $1661 \pm 104$ & $743 \pm 24$ & $44.155 \pm 0.001$ & A   \\
34 &  14:17:12.30 & +51:56:45.5 & Sy1 & 0.554 & H${\beta}$ & $12.5^{+1.8}_{-2.6}$ & $17614 \pm 153$ & $9475 \pm 33$ & $43.180 \pm 0.012$ & B \\
   &              &             &     &       & H${\alpha}$ & - & - & - & $43.180 \pm 0.012$ & B \\ 
35 &  14:17:24.59 & +52:30:24.9 & Sy1 & 0.482 & H${\beta}$ & $10.1^{+12.5}_{-2.7}$ & $4930 \pm 163$ & $2036 \pm 39$ & $43.960 \pm 0.002$ & B \\
   &              &             &     &       & H${\alpha}$ & - & - & - & $43.960 \pm 0.002$ & B \\
36 &  14:17:29.27 & +53:18:26.5 & QSO & 0.237 & H${\beta}$ & $5.5^{+5.7}_{-2.1}$ & $9448 \pm 367$ & $6318 \pm 38$ & $43.260 \pm 0.002$ & B \\
   &              &             &     &       & H${\alpha}$ & $45.0^{+23.7}_{-3.9}$ & $8898 \pm 66$ & $5157 \pm 40$ & $43.260 \pm 0.002$ & B \\
37 &  14:17:51.14 & +52:23:11.1 & QSO & 0.281 & H${\alpha}$ & $10.1^{+2.4}_{-1.9}$ & $7868 \pm 66$ & $3384 \pm 71$ & $42.700 \pm 0.007 $ & B \\
38 &  14:18:56.19 & +53:58:45.0 & QSO & 0.976 & H${\beta}$ & $15.8^{+6.0}_{-1.9}$ & $7156 \pm 61$ & $7568 \pm 70$ & $45.350 \pm 0.002$ & B \\
39 &  14:19:23.37 & +54:22:01.7 & Sy1 & 0.152 & H${\beta}$ & $11.8^{+0.7}_{-1.5}$ & $2709 \pm 55$ & $1205 \pm 9$ & $43.090 \pm 0.001$ & B \\
   &              &             &     &       & H${\alpha}$ & $80.2^{+4.9}_{-6.3}$ & $2643 \pm 23$ & $1018 \pm 7$ & $43.090 \pm 0.001$ & B \\
40 &  14:19:41.11 & +53:36:49.6 & QSO & 0.646 & H${\beta}$ & $30.4^{+3.9}_{-8.3}$ & $2553 \pm 136$ & $1232 \pm 30$ & $44.490 \pm 0.017$ & B \\
41 &  14:19:52.23 & +53:13:40.9 & QSO & 0.884 & H${\beta}$ & $32.9^{+5.6}_{-5.1}$ & $21468 \pm 2120$ & $7681 \pm 64$ & $44.220 \pm 0.006$ & B  \\

\hline
\end{tabular}

\end{table*}

\begin{table*}
\contcaption{}
\begin{tabular}{ccccccccccc} \hline
No. & $\alpha_{2000}$ & $\delta_{2000}$ & Type & z & line &  $\tau_{\mathrm{cent}}$ & FWHM & $\sigma$ & $L_{5100 \AA}$ & Reference  \\ \hline

42 &  14:19:55.62 & +53:40:07.2 & QSO & 0.418 &  H${\beta}$ & $10.7^{+5.6}_{-4.4}$ & $5136 \pm 226$ & $2291 \pm 33$ & $43.360 \pm 0.003$ & B \\ 
  &  &  &  &  &  H${\alpha}$ & - & - & - & $43.360 \pm 0.003$ & B \\
43 & 14:20:10.25 & +52:40:29.6 & QSO & 0.548 & H${\beta}$ & $12.8^{+5.7}_{-4.5}$ & $10477 \pm 114$ & $6259 \pm 23$ & $44.060 \pm 0.001$ & B \\
   &             &             &     &       & H${\alpha}$ & $32.22^{+7.75}_{-11.55}${*} & - & - & $44.060 \pm 0.001$ & B \\
44 & 14:20:23.88 & +53:16:05.1 & QSO & 0.734 & H${\beta}$ & $8.5^{+3.2}_{-3.9}$ & $11017 \pm 109$ & $7165 \pm 36$ & $44.190 \pm 0.005$ & B \\
45 & 14:20:38.52 & +53:24:16.5 & QSO & 0.265 & H${\beta}$ & $29.6^{+2.5}_{-15.7}$ & $2975 \pm 64$ & $1362 \pm 33$ & $43.424 \pm 0.001$ & A \\
   &             &             &     &       & H${\alpha}$ & $20.2^{+10.5}_{-9.3}$ & $2808 \pm 41$ & $1320 \pm 17$ & $43.424 \pm 0.001$ & B \\
46 & 14:20:39.80 & +52:03:59.7 & QSO & 0.474 & H${\beta}$ &  $14.2^{+6.5}_{-8.1}$,  $20.7^{+0.9}_{-3.0}$ & $3696 \pm 55$ & $1360 \pm 20$ & $44.109 \pm 0.001$ & A, B\\
   &             &             &     &       & H${\alpha}$ & $24.2^{+10.2}_{-5.3}$ & $3118 \pm 80$ & $1352 \pm 24$ & $44.109 \pm 0.001$ & B \\
47 & 14:20:43.53 & +52:36:11.4 & QSO & 0.337 & H${\alpha}$ & $5.7^{+0.5}_{-0.5}$ & $2971 \pm 114$ & $1372 \pm 40$ & $43.370 \pm 0.002$ & B \\
48 & 14:20:49.28 & +52:10:53.3 & QSO & 0.751 & Mg II & $34.0^{+6.7}_{-12.0}$ & - & - &  & A \\
   &             &             &     &       & H${\beta}$ & $46.0^{+9.5}_{-9.5}$ & $7625 \pm 136$ & $5013 \pm 49$ & $44.420 \pm 0.002$ & B \\
49 & 14:20:52.44 & +52:56:22.4 & QSO & 0.676 & H${\beta}$ & $11.9^{+1.3}_{-1.0}$ & $13483 \pm 141$ & $7195 \pm 40$ & $45.030 \pm 0.001$ & B \\
50 & 14:21:03.53 & +51:58:19.5 & Sy1 & 0.263 & H${\beta}$ & $75.2^{+3.2}_{-3.3}$ & $3340 \pm 82$ & $1089 \pm 22$ & $43.600 \pm 0.018$ & B \\
   &             &             &     &       & H${\alpha}$ & - & - & - & $43.600 \pm 0.018$ & B \\
51 & 14:21:12.29 & +52:41:47.3 & QSO & 0.843 & H${\beta}$ & $14.2^{+3.7}_{-3.0}$ & $10839 \pm 153$ & $3658 \pm 56$ & $44.290 \pm 0.008$ & B \\
52 & 14:21:35.90 & +52:31:38.9 & Sy1 & 0.249 & H${\beta}$ & $3.9^{+0.9}_{-0.9}$ & $2078 \pm 35$ & $1026 \pm 14$ & $43.440 \pm 0.001$ & B \\
   &             &             &     &       & H${\alpha}$ & $5.9^{+1.6}_{-1.0}$ & $2142 \pm 11$ & $907 \pm 6$ & $43.440 \pm 0.001$ & B \\
53 & 14:24:17.22 & +53:02:08.9 & QSO & 0.890 & H${\beta}$ & $36.3^{+4.5}_{-5.5}$ & $2752 \pm 90$ & $1252 \pm 11$ & $44.060 \pm 0.060$ & B \\
54 & 15:36:38.40 & +54:33:33.2 & Sy1 & 0.039 & H${\beta}$ & $20.0^{+8.7}_{-3.2}$ & - & - & $43.59 \pm 0.030$ & C  \\
55 & 15:59:09.62 & +35:01:47.6 & Sy1 & 0.031 &  H${\beta}$ & $12.2^{+3.5}_{-16.7}$ & - & - & $43.000 \pm 0.060$ & C \\
56 & 17:19:14.49 & +48:58:49.4 & Sy1 & 0.024 & H${\beta}$ & $20.61^{+54.33}_{-18.71}$ & - & - & $42.460 \pm 0.140$ &  I \\ 
57 & 23:03:15.67 & +08:52:25.3 & Sy1 & 0.016 & H${\beta}$ & $10.8^{+3.4}_{-1.3}$ & - & - & $43.330 \pm 0.030$ & F   \\ 
\hline
\end{tabular}

\raggedright A:\citet{2016ApJ...818...30S}, B:\citet{2017ApJ...851...21G}, C:\citet{2014ApJ...793..108W}, D:\citet{2012ApJ...755...60G}, E:\citet{2014ApJ...796....8B}, F:\citet{2014ApJ...795..149P}, G:\citet{2013ApJ...773...24R}, H:\citet{2014ApJ...782...45D}, I: \citet{2013A&A...559A..10S}. 
\raggedright *lag derived from ICCF analysis (see Appendix \ref{sec:ID301}). \\
\raggedright  Note: Col. (1): Number. Col. (2): RA. Col. (3): Dec. Col. (4): Type of the object. Col. (5): Redshift. Col. (6): Emission line. Col. (7): Centroid lag obtained from CCF analysis retrieved from literature. Col. (8): FWHM. Col. (9): Line dispersion. Col. (10): Optical host-galaxy corrected luminosity at 5100 \AA. Col. (11): Reference.
\end{table*}

\section{Analysis}\label{sec:model}
\subsection{Variability}
We characterized the line and continuum variability of our sample of sources
using the $F_{var}$ parameter \citep{2003MNRAS.345.1271V} 
and it is defined as
\begin{equation}
F_{var} = \sqrt\frac{(\sigma^2 - \bar{\epsilon_{err}^2)}} {\bar{x}^2}
\end{equation}
where $\bar{x}$ is the average flux in the light curve. The 
sample variance $\sigma^2$ and the mean error $\bar{\epsilon^{2}_{err}}$  
are given as 
\begin{equation}
\sigma^{2} = \frac{1}{N-1}\sum_{i=1}^{N}(x_{i} - \bar{x})^{2}
\end{equation}
\begin{equation}
\bar{\epsilon^{2}_{err}} = \frac{1}{N}\sum_{i=1}^{N}{\epsilon^{2}_{i}}
\end{equation}
where $\epsilon_i$ is the uncertainty in each flux measurement. 
The uncertainty in $F_{var}$ is calculated as \citep{2017MNRAS.466.3309R}
\begin{equation}
err(F_{var}) = \sqrt{\Bigg(\sqrt{\frac{1}{2N}}\frac{\bar{\epsilon_{err}^{2}}}{\bar{x}^{2}F_{var}}\Bigg)^{2} + \Bigg(\sqrt{\frac{\bar{\epsilon_{err}^{2}}}{N}}\frac{1}{\bar{x}}\Bigg)^{2}}
\end{equation}
The results of our variability analysis are given in Table \ref{tab:table-2} where we also mention the values of $R$, which is the ratio between the maximum and minimum fluxes in the light curves.

\subsection{Reconstruction of Light Curve}

We used the PBMAP code developed by \citet{2013ApJ...779..110L} to perform the 
light curve modeling. A detailed description of this code can be found in \citet{2013ApJ...779..110L}. 
However, we describe briefly the methodology here. The data that are needed 
for the code are the observed continuum and line light curves. Using the 
irregularly sampled continuum light curve, the code reconstructs the continuum 
light curve using the damped random walk model (DRW; \citealt{2009ApJ...698..895K}) 
following a Bayesian approach. Many investigations in the literature suggest that the optical flux variations in AGN can be well explained by DRW \cite[e.g.,][]{2010ApJ...708..927K,2010ApJ...721.1014M,2011ApJ...735...80Z}.

The AGN continuum variability is modeled as a random process in which the co-variance matrix S of the signal can be expressed as
\begin{equation}
S(t_i - t_j) = \sigma_d^2 exp\left[-\left(\frac{|t_i - t_j|}{\tau_d}\right)^{\alpha}\right]
\end{equation}
Here, $t_i$ and $t_j$ are the two epochs and the co-variance function depends on the time difference $t_i - t_j$, $\tau_d$ is the damping timescale, $\sigma_d$ is the standard deviation of variation and $\alpha$ is the smoothening parameter which is 
fixed to unity in the model calculation as it is shown to be sufficient for variability \citep{2009ApJ...698..895K}.
Then using a model of BLR and the reconstructed continuum light curve, the code reconstructs the line light curve. In the code (a) the BLR is modeled as an  axisymmetric disk consisting of a large 
number of point-like, discrete clouds of equal density, which re-radiate the 
UV/optical continuum as emission lines, (b) the BLR clouds subtend a solid 
angle denoted by the opening angle ($\theta_{opn}$), and the BLR is viewed at 
an inclination angle ($\theta_{inc}$) and (c) the central source that ionizes the 
BLR is point like, thereby emitting isotropically. The model regenerates
the velocity integrated line light curve represented as below

\begin{gather}
f_l(t)= A\int \Psi(\tau)f_c^{(1+\gamma)}(t-\tau)d\tau
\label{eq:nir_eq}
\end{gather}

with the transfer function 

\begin{gather}
\psi(\tau)=\sum_{i}\delta(\tau-\tau_i)w_i\left(\frac{I_i}{R_i^2}\right)^{1+\gamma}
\label{eq:nir_eq}
\end{gather}

where $\tau_i$ represents the time lag from the $i^{th}$ cloud at 
a distance $R_i$ to the central source, $A$ denotes the response coefficient, 
$w_i$ is the weight of the cloud to the response of the continuum, 
$I_i$ depicts any possible anisotropic effects and deviations from the continuum, and $ \gamma$ presents the non-linearity 
of the response. The weight $w_i$ is fixed to unity and $I_i$ is neglected in all calculations. A value of $\gamma$=0 points to the linear response of 
BLR to the continuum variations. The reduced $ \chi^2 $ value ($ \chi^2 / dof $) which is defined in \citet{2013ApJ...779..110L} is used to determine the quality of the generated light curves. 
When the value of $\chi^2 / dof$ was large ($>$ 1.5), we applied
detrending to the light curves before subjecting them to model fits. The detrending was done by removing a first-order polynomial fit to the light curve as mentioned in \citet{2013ApJ...779..110L}. This led to improvement in the results which has also been noticed by  \citet{2013ApJ...779..110L}. We 
found that for 6 objects, namely J0652+744, J1242+332, J1416+539,J1418+539, J1421+525 and J2303+088 though the $\chi^2 / dof$ is slightly larger after detrending, we obtained better estimates of BLR size
with smaller uncertainty. The final fitting results are given in Table \ref{tab:table-3} for H$\beta$  and in Table \ref{tab:table-4} for H$\alpha$ and MgII.

\begin{table*}
\caption{Result of the analysis of variability. The median SNR of the light curves are mentioned.}

\small
\setlength{\tabcolsep}{5pt}

\begin{tabular}{cccccccccccrr} \hline
\label{tab:table-2} 
    & $\alpha_{2000}$   & $\delta_{2000}$ &  \multicolumn{2}{c}{$F_{var}$} & \multicolumn{2}{c}{R} & \multicolumn{2}{c}{span ($\Delta t$)(days)} & \multicolumn{2}{c}{$\delta t_{mean}$(days)} & SNR & SNR \\
No  &         &     & continuum & line & continuum & line & continuum & line &  continuum & line &  continuum & line\\ \hline
1  & 00:10:31.01 & +10:58:29.5  & 0.109 & 0.102 & 1.482 & 1.442 & 137.79 & 127.28 & 0.67 & 1.61 & 64 & 36 \\
2  & 02:30:05.52 & -08:59:53.2  & 0.062 & 0.042 & 1.438 & 1.283 & 116.75 & 116.75 & 1.54 & 1.54 & 169 & 266 \\
3  & 06:52:12.32 & +74:25:37.2  & 0.130 & 0.097 & 1.667 & 1.399 & 138.85 & 117.86 & 0.68 & 1.25 & 128 & 65 \\
4  & 11:39:13.92 & +33:55:51.1  & 0.085 & 0.093 & 1.510 & 1.500 & 146.79 & 146.79 & 4.45 & 4.45 & 68 & 72 \\
5  & 12:42:10.61 & +33:17:02.7  & 0.058 & 0.048 & 1.260 & 1.284 & 146.71 & 146.71 & 2.93 & 2.93 & 131 & 98 \\
6  & 13:42:08.39 & +35:39:15.3  & 0.071 & 0.171 & 1.337 & 1.841 & 47.01 & 47.01 & 1.62 & 1.62 & 164 & 290 \\
7   &             &              &       & 0.164 &       & 1.909 & 47.01 & 47.01 & 1.62 & 1.62 & 164 & 983 \\
8  & 14:05:18.02 & +53:15:30.0  & 0.021 & 0.059 & 1.171 & 1.586 & 206.74 & 176.98 & 0.64 & 5.71 & 139 & 17 \\
9   &             &              &       & 0.049 &       & 1.490 & 206.74 & 173.00 & 0.64 & 5.77 & 139 & 18 \\
10 & 14:05:51.99 & +53:38:52.1  & 0.024 & 0.046 & 1.421 & 1.449 & 206.75 & 173.00 & 0.66 & 5.77 & 53 & 21 \\
11 & 14:07:59.07 & +53:47:59.8  & 0.049 & 0.126 & 1.217 & 1.713 & 176.98 & 176.98 & 5.71 & 5.71 & 987 & 70 \\
12 & 14:08:12.09 & +53:53:03.3  & 0.026 & 0.103 & 1.359 & 1.639 & 195.03 & 176.98 & 0.94 & 5.71 & 158 & 28 \\
13   &             &              & 0.022 & 0.056 & 1.144 & 1.363 & 206.75 & 173.00 & 0.79 & 5.77 & 90 & 22 \\
14 & 14:09:04.43 & +54:03:44.2  & 0.114 & 0.130 & 1.696 & 2.135 & 206.74 & 176.98 & 0.92 & 5.71 & 37 & 15 \\
15 & 14:09:15.70 & +53:27:21.8  & 0.030 & 0.059 & 1.180 & 1.337 & 206.75 & 173.00 & 0.50 & 5.77 & 148 & 26 \\
16 & 14:10:04.27 & +52:31:41.0  & 0.012 & 0.032 & 1.142 & 1.197 & 206.74 & 176.98 & 0.63 & 5.71 & 143 & 35\\
17  &             &              & 0.012 & 0.019 & 1.142 & 1.313 & 206.74 & 173.00 & 0.63 & 5.77 & 143 & 20 \\
18 & 14:10:18.04 & +53:29:37.5  & 0.064 & 0.170 & 1.482 & 2.097 & 176.98 & 176.98 & 5.71 & 5.71 & 211 & 21 \\
19   &             &              & 0.023 & 0.039 & 1.222 & 1.384 & 206.75 & 176.98 & 0.42 & 5.71 & 63 & 15 \\
20   &             &              & 0.023 & 0.037 & 1.222 & 1.381 & 206.75 & 173.00 & 0.42 & 5.77 & 63 & 19  \\
21 & 14:10:31.33 & +52:15:33.8  & 0.037 & 0.164 & 1.391 & 2.449 & 206.74 & 176.98 & 0.90 & 5.71 & 57 & 16 \\
22 & 14:10:41.25 & +53:18:49.0  & 0.070 & 0.121 & 2.014 & 1.645 & 195.02 & 176.98 & 0.77 & 5.71 & 23 & 51 \\
23   &             &              & 0.070 & 0.050 & 2.014 & 1.241 & 195.02 & 173.00 & 0.77 & 5.77 & 23 & 75 \\
24 & 14:11:12.72 & +53:45:07.1  & 0.030 & 0.044 & 1.128 & 1.261 & 176.98 & 176.98 & 6.32 & 6.32 & 386 & 60 \\
25 & 14:11:15.19 & +51:52:09.0  & 0.035 & 0.046 & 1.434 & 1.226 & 195.01 & 176.98 & 0.77 & 5.71 & 31 & 43 \\
26   &             &              & 0.030 & 0.079 & 1.253 & 4.246 & 206.74 & 173.00 & 0.63 & 5.77 & 49 & 7 \\
27 & 14:11:23.42 & +52:13:31.7  & 0.072 & 0.125 & 1.356 & 1.561 & 195.01 & 176.98 & 0.95 & 5.71 & 40 & 33 \\
28   &             &              & 0.071 & 0.065 & 1.515 & 1.286 & 206.79 & 173.00 & 0.66 & 5.77 & 41 & 49 \\
29 & 14:11:35.89 & +51:50:04.5  & 0.055 & 0.067 & 2.003 & 3.174 & 195.07 & 176.98 & 0.51 & 5.71 & 14 & 6\\
30 & 14:12:14.20 & +53:25:46.7  & 0.028 & 0.112 & 1.130 & 1.753 & 176.98 & 176.98 & 5.71 & 5.71 & 852 & 69 \\
31 & 14:12:53.92 & +54:00:14.4  & 0.055 & 0.136 & 1.232 & 1.912 & 176.98 & 176.98 & 5.90 & 5.90 & 446 & 22 \\
32 & 14:13:14.97 & +53:01:39.4  & 0.045 & 0.149 & 4.060 & 2.769 & 195.06 & 176.98 & 0.59 & 5.71 & 6 & 7\\
33 & 14:13:18.96 & +54:32:02.4  & 0.048 & 0.041 & 1.408 & 1.294 & 191.86 & 176.98 & 0.66 & 5.71 & 64 & 29 \\
34 & 14:13:24.28 & +53:05:27.0  & 0.029 & 0.055 & 1.250 & 1.298 & 195.06 & 176.98 & 0.79 & 5.71 & 60 & 29 \\
35   &             &              & 0.029 & 0.098 & 1.203 & 1.440 & 206.75 & 173.00 & 0.69 & 5.77 & 134 & 38 \\
36 & 14:14:17.13 & +51:57:22.6  & 0.094 & 0.120 & 1.414 & 1.553 & 176.98 & 176.98 & 6.10 & 6.10 & 79 & 25 \\
37   &             &              & 0.032 & 0.406 & 3.880 & 8.170 & 195.07 & 176.98 & 0.82 & 5.71 & 8 & 4 \\
38 & 14:15:32.36 & +52:49:05.9  & 0.114 & 0.091 & 3.890 & 1.769 & 195.06 & 176.98 & 0.71 & 5.71 & 13 & 20 \\
39 & 14:16:25.71 & +53:54:38.5  & 0.068 & 0.037 & 1.270 & 1.173 & 176.98 & 176.98 & 5.90 & 5.90 & 792 & 152 \\
40   &             &              & 0.053 & 0.010 & 1.290 & 1.104 & 206.75 & 173.00 & 1.02 & 5.77 & 84 & 54 \\
41 & 14:16:44.17 & +53:25:56.1  & 0.044 & 0.095 & 1.165 & 1.550 & 176.98 & 176.98 & 5.90 & 5.90 & 301 & 30 \\
42 & 14:16:45.15 & +54:25:40.8  & 0.072 & 0.282 & 1.308 & 3.842 & 176.98 & 176.98 & 5.71 & 5.71 & 440 & 21 \\
43   &             &              & 0.041 & 0.138 & 1.281 & 1.751 & 206.82 & 173.00 & 0.54 & 5.77 & 93 & 16 \\
44 & 14:16:45.58 & +53:44:46.8  & 0.076 & 0.141 & 1.628 & 1.944 & 176.98 & 176.98 & 5.71 & 5.71 & 270 & 21 \\
45   &             &              & 0.025 & 0.031 & 1.198 & 1.434 & 206.82 & 173.00 & 0.53 & 5.77 & 76 & 15 \\
46 & 14:16:50.93 & +53:51:57.0  & 0.037 & 0.106 & 1.160 & 1.566 & 176.98 & 176.98 & 5.90 & 5.90 & 181 & 37 \\
47 & 14:17:06.68 & +51:43:40.1  & 0.038 & 0.045 & 1.142 & 1.217 & 176.98 & 176.98 & 5.71 & 5.71 & 473 & 57 \\
48 & 14:17:12.30 & +51:56:45.5  & 0.115 & 0.348 & 12.431 & 5.887 & 206.78 & 176.98 & 0.42 & 5.90 & 9 & 4 \\
49   &             &              & 0.115 & 0.602 & 12.431 & 9.087 & 206.78 & 173.00 & 0.42 & 5.97 & 9 & 3 \\
50 & 14:17:24.59 & +52:30:24.9  & 0.064 & 0.044 & 1.359 & 1.305 & 206.78 & 176.98 & 0.48 & 5.71 & 65 & 40 \\
51   &             &              & 0.064 & 0.018 & 1.359 & 1.287 & 206.78 & 173.00 & 0.48 & 5.77 & 65 & 29 \\
52 & 14:17:29.27 & +53:18:26.5  & 0.014 & 0.050 & 1.147 & 1.426 & 195.06 & 176.98 & 0.67 & 5.71 & 118 & 33 \\
53   &             &              & 0.015 & 0.022 & 1.077 & 1.197 & 206.75 & 173.00 & 0.63 & 5.77 & 294 & 29 \\
54 & 14:17:51.14 & +52:23:11.1  & 0.029 & 0.043 & 1.399 & 1.188 & 195.09 & 173.00 & 0.44 & 5.77 & 33 & 36 \\
55 & 14:18:56.19 & +53:58:45.0  & 0.014 & 0.058 & 1.644 & 1.435 & 206.78 & 176.98 & 0.69 & 5.71 & 112 & 35 \\
56 & 14:19:23.37 & +54:22:01.7  & 0.062 & 0.109 & 1.308 & 1.479 & 206.78 & 176.98 & 0.69 & 5.71 & 87 & 26 \\
57   &             &              & 0.062 & 0.061 & 1.308 & 1.305 & 206.78 & 173.00 & 0.69 & 5.77 & 87 & 24 \\
58 & 14:19:41.11 & +53:36:49.6  & 0.056 & 0.041 & 3.212 & 1.443 & 195.11 & 176.98 & 0.56 & 5.71 & 32 & 20 \\
59 & 14:19:52.23 & +53:13:40.9  & 0.088 & 0.035 & 1.859 & 1.516 & 206.78 & 176.98 & 0.67 & 5.71 & 24 & 12 \\
60 & 14:19:55.62 & +53:40:07.2  & 0.047 & 0.131 & 1.253 & 2.014 & 206.78 & 153.02 & 0.84 & 5.67 & 50 & 11 \\
\hline
\end{tabular}

\end{table*}

\begin{table*}
\contcaption{}

\small
\setlength{\tabcolsep}{5pt}

\begin{tabular}{cccccccccccrr} \hline
    & $\alpha_{2000}$   & $\delta_{2000}$ &  \multicolumn{2}{c}{$F_{var}$} & \multicolumn{2}{c}{R} & \multicolumn{2}{c}{span ($\Delta t$) (days)} & \multicolumn{2}{c}{$\delta t_{mean}$ (days)} & SNR & SNR \\
No. &         &     & continuum & line & continuum & line & continuum & line &  continuum & line &  continuum & line \\ \hline
61 & 14:19:55.62 & +53:40:07.2   & 0.047 & 0.038 & 1.253 & 1.617 & 206.78 & 148.05 & 0.84 & 5.92 & 50 & 14 \\
62 & 14:20:10.25 & +52:40:29.6  & 0.109 & 0.201 & 2.217 & 2.753 & 195.10 & 176.98 & 0.55 & 5.71 & 25 & 26 \\
63   &             &              & 0.106 & 0.107 & 1.758 & 1.738 & 206.78 & 143.98 & 0.53 & 5.76 & 40 & 14 \\
64 & 14:20:23.88 & +53:16:05.1  & 0.116 & 0.088 & 1.850 & 1.914 & 206.78 & 176.98 & 0.68 & 5.71 & 23 & 10 \\
65 & 14:20:38.52 & +53:24:16.5  & 0.057 & 0.085 & 1.302 & 1.426 & 176.98 & 176.98 & 5.71 & 5.71 & 516 & 49 \\
66   &             &              & 0.022 & 0.029 & 1.368 & 1.218 & 206.79 & 173.00 & 0.46 & 5.77 & 84 & 26 \\
67 & 14:20:39.80 & +52:03:59.7  & 0.070 & 0.105 & 1.245 & 1.601 & 176.98 & 176.98 & 5.90 & 5.90 & 415 & 57 \\
68   &             &              & 0.075 & 0.045 & 1.440 & 1.272 & 206.77 & 173.00 & 0.80 & 5.77  & 55 & 45 \\
69 & 14:20:43.53 & +52:36:11.4  & 0.013 & 0.049 & 1.161 & 1.268 & 195.10 & 173.00 & 0.41 & 5.77 & 68 & 32 \\
70 & 14:20:49.28 & +52:10:53.3  & 0.074 & 0.116 & 1.326 & 1.532 & 176.98 & 176.98 & 6.10 & 6.10 & 249 & 44 \\
71   &             &              & 0.113 & 0.148 & 1.684 & 2.295 & 195.09 & 176.98 & 0.70 & 5.71 & 24 & 14 \\
72 & 14:20:52.44 & +52:56:22.4  & 0.018 & 0.029 & 1.128 & 1.142 & 206.78 & 176.98 & 0.68 & 5.71 & 71 & 72 \\
73 & 14:21:03.53 & +51:58:19.5  & 0.010 & 0.069 & 1.160 & 1.699 & 206.77 & 176.98 & 0.80 & 5.71 & 149 & 34 \\
74   &             &              & 0.010 & 0.024 & 1.160 & 1.187 & 206.77 & 173.00 & 0.80 & 5.77 & 149 & 44 \\
75 & 14:21:12.29 & +52:41:47.3  & 0.037 & 0.110 & 1.375 & 2.033 & 195.10 & 176.98 & 0.54 & 5.90 & 33 & 8 \\
76 & 14:21:35.90 & +52:31:38.9  & 0.021 & 0.293 & 1.184 & 3.738 & 206.77 & 176.98 & 0.66 & 5.71 & 89 & 12 \\
77   &             &              & 0.021 & 0.123 & 1.184 & 1.604 & 206.77 & 173.00 & 0.66 & 5.77 & 89 & 33 \\
78 & 14:24:17.22 & +53:02:08.9  & 0.148 & 0.710 & 2.665 & 1.017 & 206.78 & 176.98 & 1.04 & 5.90 & 15 & 5 \\
79 & 15:36:38.40 & +54:33:33.2  & 0.045 & 0.047 & 1.184 & 1.249 & 109.65 & 109.65 & 2.49 & 2.49 & 185 & 284 \\
80 & 15:59:09.62 & +35:01:47.6  & 0.053 & 0.035 & 1.284 & 1.147 & 55.86 & 55.86 & 2.15 & 2.15 & 124 & 128 \\
81 & 17:19:14.49 & +48:58:49.4  & 0.309 & 0.099 & 3.964 & 1.731 & 8138.42 & 8392.42 & 71.39 & 71.73 & 11 & 33 \\
82 & 23:03:15.67 & +08:52:25.3  & 0.035 & 0.062 & 1.204 & 1.381 & 137.80 & 108.86 & 0.50 & 1.51 & 140 & 51 \\
\hline
\end{tabular}
\raggedright  Note: Col. (1): Number. Col. (2): RA. Col. (3): Dec. Col. (4): Excess variance for continuum. Col. (5): Excess variance for line. Col. (6): Maximum to minimum flux ratio for continuum. Col. (7): Maximum to minimum flux ratio for line. Col. (8): Total duration of observation for continuum. Col. (9): Total duration of observation for line. Col. (10): Mean cadence of observation for continuum. Col. (11): Mean cadence of observation for line. Col. (12): Median signal to noise ratio for continuum. Col. (13): Median signal to noise ratio for line.
\end{table*}

\begin{table*}
\caption{Results of BLR modeling. The line light curves are from H$\beta$. For objects with *, detrending was done, while for 
others detrending was not done. For the objects with $\bullet$ the BLR model fits are shown in Fig. \ref{fig:fig-2}  and Fig. \ref{fig:fig-rep}.}
\resizebox{18cm}{!}{
\begin{tabular}{|cccccccccc|}\hline
\label{tab:table-3}
$\alpha_{2000}$  & $\delta_{2000}$ & continuum & $\log(\tau_d)$(days)& $\tau_{\mathrm{model}}$ (days) & $\theta_{\mathrm{inc}}$(degree) & $\theta_{\mathrm{opn}}(degree)$ & $\gamma$ & $f_{\mathrm{BLR}}$ & $\chi^2 / dof$\\ \hline

00:10:31.01 & +10:58:29.5 & --- & $2.78 \pm 2.88 $ & $11.18 \pm 5.7$ & $46.10 \pm 25.43$ & $46.87 \pm 25.88$ & $-0.55 \pm 0.10$ & --- & 1.13 \\
02:30:05.52 & -08:59:53.2 & --- & $0.61 \pm 0.58 $ & $50.61 \pm 25.30$ & $67.61 \pm 21.52$ & $36.24 \pm 26.93$ & $0.07 \pm 0.08$ &$0.83 \pm 0.36$ & 1.20 \\
02:30:05.52 & -08:59:53.2* & --- & --- & $54.28 \pm 19.54$ & $75.22 \pm 17.54$ & $28.96 \pm 23.49$ & --- & --- & 1.20 \\
06:52:12.32 & +74:25:37.2 & --- & --- & $43.0 \pm 18.49$ & $44.72 \pm 24.61$ & $44.17 \pm 26.18$ & --- & --- & 1.47\\
06:52:12.32 & +74:25:37.2* & --- & $2.81 \pm 2.32$ & $11.96 \pm 4.54$ & $32.96 \pm 23.22$ & $38.73 \pm 24.87$ & $-0.31 \pm 0.09$ & $1.45 \pm 1.19$ & 1.62\\
11:39:13.92 & +33:55:51.1 & --- & $0.72 \pm 0.80$ & $10.37 \pm 5.7$ & $52.09 \pm 24.59$ & $47.74 \pm 25.90$ & $0.12 \pm 0.15$ & --- & 0.90\\
11:39:13.92 & +33:55:51.1 & --- & $0.09 \pm 0.20 $ & $5.63 \pm 3.66$ & $56.40 \pm 24.31$ & $47.40 \pm 26.27$ & $0.41 \pm 0.19$ & --- & 0.85 \\
11:39:13.92 & +33:55:51.1* & --- & --- & $5.35 \pm 3.96$ & $54.51 \pm 24.86$ & $47.32 \pm 26.31$ & --- & --- & 0.90 \\
12:42:10.61 & +33:17:02.7 & --- & --- & $52.3 \pm 16.74$ & $73.75 \pm 12.91$ & $35.61 \pm 22.71$ & --- & --- & 0.89 \\
12:42:10.61 & +33:17:02.7* & --- & $1.32 \pm 1.32 $ & $51.26 \pm 17.43$ & $77.33 \pm 12.85$ & $25.20 \pm 22.81$ & $-0.02 \pm 0.12$ & $0.88 \pm 0.25 $ & 1.60 \\
13:42:08.39 & +35:39:15.3 & --- & $0.10 \pm 0.10 $ & $2.20 \pm 1.76$ & $51.1 \pm 24.92$ & $48.5 \pm 25.82$ & $1.67 \pm 0.28$ & --- & 1.08 \\
14:05:18.02 & +53:15:30.0 & g &  $1.05 \pm 1.26$  & $23.73 \pm 19.46$ & $47.88 \pm  24.79$ & $48.19 \pm 25.6 $ & $1.70 \pm 0.82$  & ---              & 1.06\\
14:05:18.02 & +53:15:30.0 & i &  ---              & $24.70 \pm 28.90$ & $45.76 \pm  25.24$ & $47.05 \pm 25.98$ & ---                & ---              & 1.08\\
14:07:59.07 & +53:47:59.8$\bullet$ & --- & $0.99 \pm 1.08$ & $18.74 \pm 7.50$ & $65.02 \pm 22.13$ & $36.95 \pm 26.54$ & $1.76 \pm 0.29$ & $0.85 \pm 0.38$ & 1.24\\
14:08:12.09 & +53:53:03.3 & g &  ---              & $11.83 \pm  6.50$ & $39.30 \pm  25.34$ & $43.51 \pm 26.13$ & ---                & ---              & 1.51\\
14:08:12.09 & +53:53:03.3 & i &  $2.93 \pm 2.82$  & $11.25 \pm  6.86$ & $40.91 \pm  25.47$ & $44.23 \pm 25.76$ & $2.77 \pm 0.18$ & ---              & 1.17\\	
14:09:04.43 & +54:03:44.2 & g &  $2.68 \pm 2.82$  & $15.56 \pm  8.40$ & $44.16 \pm  25.21$ & $46.39 \pm 25.96$ & $0.20 \pm 0.14$ & ---              & 1.01\\
14:09:04.43 & +54:03:44.2 & i &  ---              & $19.77 \pm  9.29$ & $49.62 \pm  24.77$ & $48.59 \pm 25.84$ & ---                & ---              & 0.86\\
14:10:04.27 & +52:31:41.0 & g &  $0.91 \pm 1.36 $ & $44.11 \pm 11.03$ & $51.84 \pm  24.39$ & $50.25 \pm 25.49$ & $2.36 \pm 0.47$ & ---              & 1.13\\
14:10:18.04 & +53:29:37.5 & g &  $1.33 \pm 1.62 $ & $15.56 \pm  7.00$ & $48.48 \pm  24.93$ & $47.89 \pm 25.78$ & $2.43 \pm 0.41$ & ---              & 0.89\\
14:10:31.33 & +52:15:33.8* & g & $1.09 \pm 1.46 $ & $30.72 \pm 13.52$ & $29.99 \pm  22.58$ & $34.88 \pm 25.56$ & $2.46 \pm 0.53$ & $ 1.73 \pm 1.62$ & 2.98\\
14:10:41.25 & +53:18:49.0 & i &  $2.65 \pm 2.58 $ & $27.75 \pm 11.66$ & $32.66 \pm  22.51$ & $41.74 \pm 24.78$ & $0.96 \pm 0.19$ & ---              & 1.02\\
14:11:12.72 & +53:45:07.1$\bullet$ & --- & $0.62 \pm 0.84$ & $21.72 \pm 7.60$ & $26.37 \pm 22.59$ & $32.47 \pm 23.17$ & $0.90 \pm 0.29$ & $2.06 \pm 2.05 $ & 0.97\\
14:11:15.19 & +51:52:09.0 & i &  $1.78 \pm 2.20 $ & $42.00 \pm 13.86$ & $62.27 \pm  22.28$ & $45.13 \pm 26.42$ & $ 0.83 \pm 0.26$ & ---              & 0.98\\
14:11:23.42 & +52:13:31.7 & i &  $2.69 \pm 2.64 $ & $15.70 \pm  7.53$ & $38.03 \pm  24.10$ & $45.02 \pm 24.90$ & $1.20 \pm 0.15$ & ---              & 0.75\\
14:11:35.89 & +51:50:04.5 & g &  ---              & $17.45 \pm 19.72$ & $45.30 \pm  25.66$ & $45.74 \pm 25.90$ & ---                & ---              & 1.15\\
14:11:35.89 & +51:50:04.5 & i &  $1.60 \pm 2.26 $ & $17.10 \pm 15.39$ & $45.66 \pm  25.68$ & $45.53 \pm 26.04$ & $1.31 \pm 0.73$  & ---              & 1.06\\
14:12:53.92 & +54:00:14.4 & --- & $0.62 \pm 0.66$ & $24.74 \pm 10.14$ & $42.23 \pm 26.23$ & $44.35 \pm 26.0$ & $2.18 \pm 0.39$ & --- & 1.01\\
14:13:14.97 & +53:01:39.4 & g &  ---              & $33.58 \pm 17.13$ & $49.58 \pm  28.67$ & $35.84 \pm 27.01$ & ---                & ---              & 1.65\\
14:13:14.97 & +53:01:39.4 & i &  $1.31 \pm 1.90 $ & $33.25 \pm 14.30$ & $52.24 \pm  26.20$ & $42.56 \pm 26.84$ & $1.83 \pm 0.71$ & ---              & 1.42\\
14:13:18.96 & +54:32:02.4 & g &  $1.72 \pm 2.06 $ & $20.93 \pm  7.32$ & $30.25 \pm  26.58$ & $31.74 \pm 25.12$ & $1.14 \pm 0.48$ & $1.88 \pm 2.00$  & 1.32\\
14:13:18.96 & +54:32:02.4 & i &  ---              & $19.13 \pm  7.08$ & $47.07 \pm  24.97$ & $44.13 \pm 25.80$ & ---                & ---              & 1.53\\
14:13:24.28 & +53:05:27.0 & g &  ---              & $29.79 \pm 12.51$ & $53.74 \pm  25.36$ & $45.69 \pm 26.19$ & ---                & ---              & 1.02\\
14:13:24.28 & +53:05:27.0 & i &  $1.18 \pm 1.62 $ & $21.85 \pm  9.83$ & $53.83 \pm  24.75$ & $47.42 \pm 26.18$ & $0.48 \pm 0.24$  & ---              & 0.86\\
14:14:17.13 & +51:57:22.6 & i &  $2.01 \pm 1.84 $ & $11.68 \pm 12.14$ & $41.83 \pm  25.69$ & $43.98 \pm 26.08$ & $1.97 \pm 0.66$ & ---              & 2.00\\
14:14:17.13 & +51:57:22.6* & i &  ---             & $11.11 \pm 12.66$ & $43.80 \pm  25.71$ & $44.96 \pm 26.01$ & ---                &                  & 1.97\\
14:15:32.36 & +52:49:05.9 & i &  $-0.01 \pm 0.14$ & $7.32 \pm  8.64$  & $44.56 \pm  25.95$  & $44.63 \pm 26.17$ & $0.07 \pm 0.27$ & ---             & 1.18\\
14:16:25.71 & +53:54:38.5 & --- & --- & $37.21 \pm 12.28$ & $41.60 \pm 22.29$ & $52.74 \pm 24.13$ & --- & --- & 1.02\\
14:16:25.71 & +53:54:38.5* & --- & $1.80 \pm 2.12$ & $26.49 \pm 9.8$ & $68.47 \pm 18.79$ & $39.71 \pm 25.38$ & $0.03 \pm 0.11$ & $0.79 \pm 0.30 $ & 1.61\\
14:16:45.15 & +54:25:40.8 & --- & --- & $22.02 \pm 11.89$ & $58.06 \pm 22.50$ & $46.67 \pm 26.53$ & --- & --- & 1.75\\
14:16:45.15 & +54:25:40.8* & --- & $0.77 \pm 0.90$ & $19.14 \pm 10.34$ & $53.12 \pm 23.09$ & $49.88 \pm 25.75$ & $2.71 \pm 0.21$ & --- & 1.65\\
14:16:45.58 & +53:44:46.8 & --- & $0.62 \pm 0.74$ & $17.53 \pm 10.87$ & $41.63 \pm 23.84$ & $48.32 \pm 24.78$& $2.26 \pm 0.38$ & --- & 1.10\\
14:17:06.68 & +51:43:40.1 & --- & --- & $25.11 \pm 9.29$ & $50.71 \pm 22.7$ & $54.20 \pm 25.50$ & --- & --- & 2.36\\
14:17:06.68 & +51:43:40.1*$\bullet$ & --- & $1.20 \pm 1.32$ & $22.95 \pm 8.95$ & $57.91 \pm 21.62$ & $51.79 \pm 25.78$ & $1.44 \pm 0.41$ & --- & 1.50\\
14:17:12.30 & +51:56:45.5 & g &  $2.67 \pm 2.56 $ & $17.10 \pm 8.55$  & $42.17 \pm  26.37$  & $44.16 \pm 25.0$  & $1.89 \pm 0.40$ & ---             & 1.24\\
14:17:24.59 & +52:30:24.9 & g &  $2.69 \pm 2.42 $ & $19.04 \pm 10.28$ & $39.66 \pm  23.99$  & $43.85 \pm 26.19$ & $0.37 \pm 0.46$ & ---             & 1.39\\
14:17:29.27 & +53:18:26.5 & g &  ---              & $13.29 \pm 13.96$ & $48.16 \pm  25.21$  & $46.32 \pm 26.17$ & ---               & ---             & 1.43\\
14:17:29.27 & +53:18:26.5 & i &  $1.09 \pm 1.30 $ & $13.03 \pm 15.38$ & $45.75 \pm  26.07$  & $45.24 \pm 26.06$ & $1.90 \pm 0.53$  & ---             & 1.34 \\
14:18:56.19 & +53:58:45.0 & g &  ---              & $34.78 \pm 13.56$ & $39.74 \pm  23.28$  & $44.89 \pm 25.30$ & ---                & ---             & 2.29\\
14:18:56.19 & +53:58:45.0* & g & $0.60 \pm 0.74 $ & $17.62 \pm 9.89$  & $40.6 \pm   24.97$  & $43.99 \pm 25.66$ & $2.33 \pm 0.53$ & ---             & 2.32\\
14:19:23.37 & +54:22:01.7 & g &  $2.88 \pm 2.58 $ & $15.62 \pm 10.93$ & $48.03 \pm  25.69$  & $46.87 \pm 25.82$ & $0.93 \pm 0.20$  & ---             & 1.35\\
14:19:41.11 & +53:36:49.6 & g &  ---              & $26.35 \pm 20.56$ & $45.01 \pm  25.78$  & $45.11 \pm 26.06$ & ---                & ---             & 1.51\\
14:19:41.11 & +53:36:49.6 & i &  $2.36 \pm 2.50 $ & $33.84 \pm 21.66$ & $40.19 \pm  25.57$  & $43.81 \pm 25.54$ & $0.30 \pm 0.54$ & ---             & 1.33\\
14:19:52.23 & +53:13:40.9 & g &  $2.63 \pm 2.58 $ & $21.47 \pm 30.70$ & $36.61 \pm  25.47$  & $42.35 \pm 25.07$ & $0.12 \pm 0.81$ & ---             & 1.73\\
14:19:55.62 & +53:40:07.2 & g &  $2.59 \pm 2.54 $ & $8.34  \pm 10.76$ & $42.57 \pm  26.21$  & $43.91 \pm 26.02$ & $2.70 \pm 0.24$ & ---             & 1.43\\
14:20:10.25 & +52:40:29.6* & i &  $1.39 \pm 1.66 $& $5.28  \pm 5.96$  & $61.91 \pm  23.02$  & $39.83 \pm 25.76$ & $2.95 \pm 0.04$  & $0.84 \pm 0.39$ & 7.09\\
14:20:23.88 & +53:16:05.1 & g &  $2.62 \pm 2.42 $ & $11.47 \pm 15.60$ & $53.48 \pm  26.47$  & $45.38 \pm 25.72$ & $0.59 \pm 0.63$  & ---             & 1.47\\
14:20:23.88 & +53:16:05.1 & i &  ---              & $14.01 \pm 14.43$ & $52.70 \pm  25.91$  & $44.96 \pm 26.07$ & ---                &  ---            & 1.37\\
14:20:38.52 & +53:24:16.5 & --- & $1.00 \pm 1.14$ & $25.92 \pm 9.33$ & $50.10 \pm 23.85$ & $51.24 \pm 25.20$ & $0.70 \pm 0.22$ & --- & 1.05\\
14:20:39.80 & +52:03:59.7 & --- & --- & $29.14 \pm 11.07$ & $22.77 \pm 22.81$ & $27.69 \pm 23.85$ & --- & --- & 1.56\\
14:20:39.80 & +52:03:59.7* & --- & $0.81 \pm 1.00$ & $27.17 \pm 9.51$ & $27.89 \pm 24.09$ & $33.61 \pm 24.82$ & $1.10 \pm 0.36$ & $1.90 \pm 1.92$ & 1.02\\

14:20:49.28 & +52:10:53.3 & g &  ---              & $47.93 \pm 7.19$  & $14.34 \pm  14.98$  & $19.19 \pm 17.13$ & ---                & ---             & 2.01\\
14:20:49.28 & +52:10:53.3 & i &  $2.30 \pm 2.60 $ & $46.52 \pm 7.44$  & $17.42 \pm  17.38$  & $25.05 \pm 19.50$ & $-0.34 \pm 0.12$ & $3.72 \pm 4.33$ & 1.85\\
14:20:49.28 & +52:10:53.3* & i &  ---             & $42.94 \pm 8.59$  & $26.97 \pm  21.62$  & $34.91 \pm 23.34$ & ---                & ---             & 1.21\\
14:20:52.44 & +52:56:22.4 & g  & $0.67 \pm 0.94 $ & $16.84 \pm 7.07$  & $49.01 \pm  24.60$  & $48.93 \pm 25.68$ & $0.82 \pm 0.24$  & ---             & 0.71\\
14:21:03.53 & +51:58:19.5 & g  & $1.88 \pm 2.20 $ & $20.63 \pm 12.99$ & $37.81 \pm  26.62$  & $38.66 \pm 25.71$ & $2.61 \pm 0.32$ & $1.31 \pm 1.07$ & 2.83\\
14:21:12.29 & +52:41:47.3 & g  & ---              & $18.33 \pm 9.35$  & $41.37 \pm  25.47$  & $44.44 \pm 26.07$ & ---                &   ---              & 1.13\\
14:21:12.29 & +52:41:47.3 & i  & $1.88 \pm 2.34 $ & $15.16 \pm 9.25$  & $42.83 \pm  25.63$  & $45.21 \pm 26.13$ & $2.09 \pm 0.59$ & ---             & 1.15\\
14:21:35.90 & +52:31:38.9 & g  & ---              & $7.6 \pm 4.79$    & $15.08 \pm  16.91$  & $20.51 \pm 18.85$ & ---                & ---             & 8.06\\
14:21:35.90 & +52:31:38.9*$\bullet$ & g  & $2.88 \pm 2.42$ & $6.54 \pm 3.14$   & $15.67 \pm  18.49$  & $21.10 \pm 20.72$ & $2.96 \pm 0.04$ & $4.94 \pm 7.20$ & 8.14\\
14:24:17.22 & +53:02:08.9 & g  & $2.70 \pm 2.80 $ & $19.36 \pm 22.46$ & $47.18 \pm  26.15$  & $46.45 \pm 26.20$ & $2.01 \pm 0.45$ & ---             & 2.38\\

\hline
\end{tabular}
}
\end{table*}

\begin{table*}
\contcaption{}

\resizebox{18cm}{!}{
\begin{tabular}{|cccccccccc|}\hline
$\alpha_{2000}$  & $\delta_{2000}$ & continuum & $\log(\tau_d)$ (days) & $\tau_{\mathrm{model}}$ (days) & $\theta_{\mathrm{inc}}$ (degree) & $\theta_{\mathrm{opn}}$ (degree) & $\gamma$ & $f_{\mathrm{BLR}}$ & $\chi^2 / dof$\\ \hline
15:36:38.40 & +54:33:33.2 & --- & $1.16 \pm 1.22$ & $26.89 \pm 5.92$ & $63.46 \pm 21.38$ & $47.87 \pm 26.73$ & $0.33 \pm 0.15$ & --- & 1.29 \\
15:59:09.62 & +35:01:47.6  & --- & $1.12 \pm 1.16 $ & $13.32 \pm 3.46$ & $57.50 \pm 22.92$ & $49.20 \pm 25.99$ & $0.12 \pm 0.16$ & --- & 1.42 \\
17:19:14.49 & +48:58:49.4 & --- & $1.53 \pm 1.54$ & $65.12 \pm 16.93$ & $64.18 \pm 20.44$ & $45.31 \pm 26.54$ & $-0.63 \pm 0.03$ & --- & 1.12\\
17:19:14.49 & +48:58:49.4* & --- & --- & $65.78 \pm 17.10$ & $58.85 \pm 22.23$ & $49.24 \pm 26.21$ & --- & --- & 1.21\\
23:03:15.67 & +08:52:25.3  & --- & --- & $52.67 \pm 16.86$ & $54.83 \pm 10.46$ & $10.76 \pm 14.0$ & --- & --- & 1.07 \\
23:03:15.67 & +08:52:25.3*  & --- & $1.17 \pm 1.22 $ & $7.06 \pm 1.41$ & $42.84 \pm 25.72$ & $44.15 \pm 25.45$ & $0.07 \pm 0.08$ & --- & 2.16 \\	
\hline
\end{tabular}
}
\end{table*}

\begin{table*}
\caption{Results for H${\alpha}$ and  Mg II lines. For objects with *, detrending was done, while for others detrending was not done. For the objects with $\bullet$ the BLR model fits are shown in Fig. \ref{fig:fig-2}  and Fig. \ref{fig:fig-rep}}.
\begin{tabular}{|ccccccccc|}\hline
\label{tab:table-4}
$\alpha_{2000}$ & $\delta_{2000}$  & continuum & $\log(\tau_d)$ (days) & $\tau_{\mathrm{model}}$ (days) & $\theta_{\mathrm{inc}}$ (degree) & $\theta_{\mathrm{opn}}$ (degree) & $\gamma$ &  $\chi^2 / dof$\\ \hline

13:42:08.39 & +35:39:15.3* & --- & $0.02 \pm 0.02$ & $2.29 \pm 2.24$ & $42.20 \pm 26.05$ & $42.36 \pm 25.29$ & $1.13 \pm 0.28$ & 1.56\\
14:05:18.02 & +53:15:30.0 & g & $1.09 \pm 1.09$ & $20.43 \pm 30.03$ & $43.42 \pm 25.63$ & $45.66 \pm 25.51$ & $0.58 \pm 1.06$ & 1.63\\
14:05:51.99 & +53:38:52.1 & g &  $1.46 \pm 1.60$ & $41.06 \pm 27.92$ & $40.04 \pm 25.22$ & $45.38 \pm 25.43$ & $0.82 \pm 0.75$ &  1.36\\
14:08:12.09 & +53:53:03.3 & g & $2.21 \pm 2.21$ & $11.48 \pm  6.54$ & $40.70 \pm 25.48$ & $44.39 \pm 26.00$ & $1.98 \pm 0.37$ & 1.08\\
14:09:15.70 & +53:27:21.8 & g & $1.44 \pm 1.52$  & $46.08 \pm 12.90$ & $32.25 \pm 22.35$ & $38.86 \pm 24.43$ & $2.24 \pm 0.46$ & 0.92\\
14:10:04.27 & +52:31:41.0 & g & $1.04 \pm 1.14$  & $ 7.15 \pm 13.72$ & $47.96 \pm 25.67$ & $45.76 \pm 26.02$ & $0.68 \pm 1.04$ & 1.60\\
14:10:18.04 & +53:29:37.5 & g & $1.37 \pm 1.51$ & $23.92 \pm 10.05$ & $39.83 \pm 24.69$ & $45.95 \pm 25.30$ & $1.01 \pm 0.72$ & 1.03\\
14:10:41.25 & +53:18:49.0 & g &  --- & $26.13 \pm 10.19$ & $32.70 \pm 22.69$ & $39.85 \pm 24.91$ & --- & 1.14\\
14:10:41.25 & +53:18:49.0 & i & $2.53 \pm 2.41$ & $21.61 \pm  8.21$ & $37.80 \pm 24.96$ & $43.35 \pm 25.42$ & $-0.12 \pm 0.08$ & 0.94\\
14:11:15.19 & +51:52:09.0 & g & $0.98 \pm 1.10$ & $1.07 \pm   2.05$ & $45.87 \pm 25.93$ & $45.06 \pm 25.99$ & $0.64 \pm 0.93$ & 1.39\\
14:11:23.42 & +52:13:31.7 & g & $2.73 \pm 2.35$ & $39.38 \pm 12.60$ & $46.32 \pm 24.38$ & $48.79 \pm 25.25$ & $0.16 \pm 0.10$ & 0.94\\
14:13:24.28 & +53:05:27.0 & g & $2.74 \pm 2.26$ & $47.20 \pm 10.86$ & $35.84 \pm 22.00$ & $43.54 \pm 24.66$ & $-0.64 \pm 0.32$ & 0.58\\
14:16:25.71 & +53:54:38.5 & g & $2.04 \pm 2.24$ & $38.73 \pm 18.20$ & $38.66 \pm 23.54$ & $45.94 \pm 25.59$ & $-0.52 \pm 0.19$ & 0.95\\
14:16:45.15 & +54:25:40.8 & g & $2.85 \pm 2.51$ & $17.49 \pm 8.40$ & $51.18 \pm   24.10$ & $49.99 \pm 25.71$ & $2.42 \pm 0.32$ & 1.16\\
14:16:45.58 & +53:44:46.8 & g & $1.90 \pm 2.14$ & $18.62 \pm 12.47$ & $44.58 \pm  25.35$ & $45.83 \pm 26.07$ & $1.36 \pm 0.84$ & 1.32\\
14:17:12.30 & +51:56:45.5$\bullet$ & g & $2.71 \pm 2.35$ & $3.25 \pm   7.12$ & $43.8^ \pm 26.21$  & $44.56 \pm 26.02$ & $-0.63 \pm 0.48$ & 1.51\\
14:17:24.59 & +52:30:24.9 & g & $2.05 \pm 2.21$ & $6.94 \pm 12.55$ & $43.75 \pm 25.72$   & $44.91 \pm 25.77$ & $-0.69 \pm 0.27$ & 1.34\\
14:17:29.27 & +53:18:26.5 & g & $1.17 \pm 1.27$ & $3.81 \pm 7.92$ & $45.95 \pm 25.83$   & $45.58 \pm 25.93$ & $0.23 \pm 0.77$ & 1.67\\
14:17:51.14 & +52:23:11.1 & g &  --- & $17.16 \pm 9.44$ & $48.20 \pm 25.32$   & $46.14 \pm 26.07$ & --- & 0.94\\
14:17:51.14 & +52:23:11.1 & i &  $1.87 \pm 2.17$ & $13.91 \pm 6.68$  & $47.23 \pm 25.29$  & $45.59 \pm 25.99$ & $0.75 \pm 0.26$ & 0.86\\
14:19:23.37 & +54:22:01.7 & g & $2.88 \pm 2.50$ & $62.08 \pm 14.74$ & $40.01 \pm 26.54$  & $42.12 \pm 26.31$ & $0.53 \pm 0.24$ & 1.21\\
14:19:55.62 & +53:40:07.2 & g & $2.48 \pm 2.39$ & $3.64 \pm 7.67$   & $46.31 \pm 25.89$  & $45.82 \pm 25.41$ & $0.61 \pm 0.66$ & 1.17\\
14:20:10.25 & +52:40:29.6 & g & $2.63 \pm 2.37$ & $46.67 \pm 9.8$ & $39.52 \pm 24.53$ & $44.69 \pm 25.25$ & $0.70 \pm 0.39$ & 1.73\\
14:20:38.52 & +53:24:16.5 & g & $1.68 \pm 2.14$ & $24.17 \pm 13.53$ & $49.87 \pm 25.44$ & $46.47 \pm 25.83$ & $0.39 \pm 0.34$ & 0.94\\
14:20:39.80 & +52:03:59.7 & g & $2.81 \pm 2.57$ & $35.95 \pm 14.38$ & $37.17 \pm 24.81$ & $42.06 \pm 25.36$ & $-0.26 \pm 0.09$ & 1.26\\
14:20:43.53 & +52:36:11.4 & i & $1.61 \pm 1.73$ & $3.78 \pm 4.65$ & $44.33 \pm 26.20$  & $43.91 \pm 26.30$ & $2.54 \pm 0.35$ & 1.52\\
14:21:03.53 & +51:58:19.5 & g & $1.76 \pm 2.12$ & $5.57 \pm 6.29$   & $45.6 \pm 25.76$ & $44.97 \pm 25.91$ & $0.97 \pm 0.58$ & 1.23\\
14:21:35.90 & +52:31:38.9 & g & --- & $14.84 \pm 8.46$  & $29.49 \pm 20.67$ & $37.35 \pm 24.13$ & --- & 2.30\\
14:21:35.90 & +52:31:38.9* & g & $2.80 \pm 2.24$ & $13.98 \pm 7.27$ & $30.83 \pm 21.3$ & $39.16 \pm 24.54$ & $2.94 \pm 0.06$ & 2.14\\
\hline
14:10:18.04 & +53:29:37.5 & --- & $0.61 \pm 0.61$ & $32.62 \pm 11.74$ & $44.73 \pm 27.21$ & $43.98 \pm 26.54$ & $2.65 \pm 0.26$ & 1.30\\
14:12:14.20 & +53:25:46.7*$\bullet$ & --- & $0.86 \pm 0.86$ & $38.59 \pm  6.95$ & $ 4.00 \pm  2.79$ & $ 6.02 \pm  3.97$ & $2.96 \pm 0.03$ & 4.91\\
14:14:17.13 & +51:57:22.6 & --- & $0.37 \pm 0.49$ & $28.15 \pm 10.42$ & $46.02 \pm 25.77$ & $46.68 \pm 25.84$ & $0.52 \pm 0.24$ &  0.92\\
14:16:44.17 & +53:25:56.1 & --- & $-0.13 \pm 0.10$ & $19.80 \pm  9.90$ & $32.44 \pm 25.46$ & $36.60 \pm 25.58$ & $1.38 \pm 0.36$ & 1.13\\
14:16:50.93 & +53:51:57.0 & --- & $0.08 \pm 0.04$ & $17.41 \pm 15.67$ & $45.00 \pm 27.06$ & $44.00 \pm 26.23$ & $2.07 \pm 0.39$ & 1.07\\
14:20:49.28 & +52:10:53.3 & --- & $1.18 \pm 1.20$ & $34.12 \pm 10.24$ & $62.23 \pm 22.69$ & $47.69 \pm 27.15$ & $1.10 \pm 0.26$ & 1.23\\ 

\hline

\end{tabular}
\end{table*}

\begin{figure}
\resizebox{8cm}{9cm}{\includegraphics{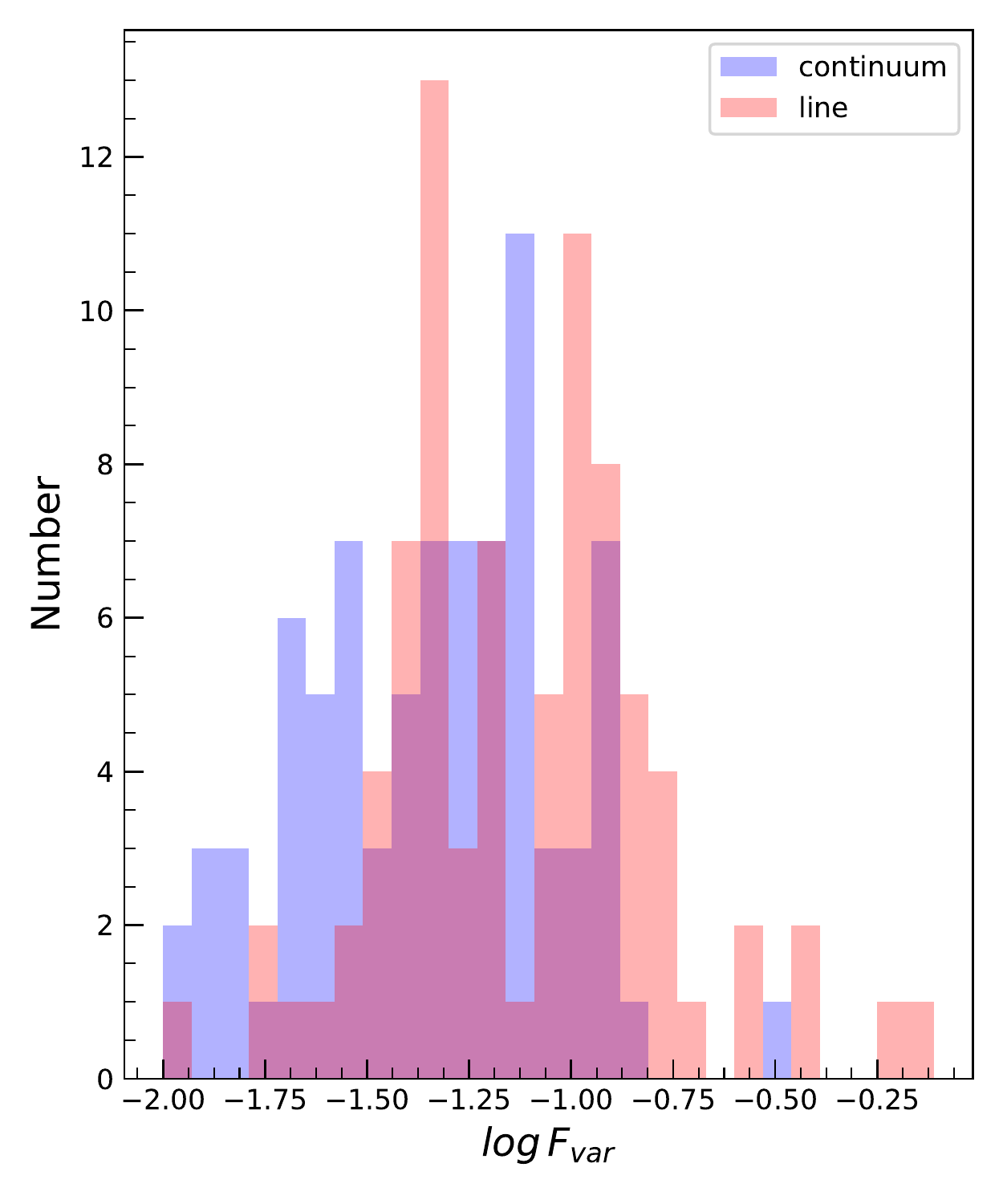}}
\caption{Distribution of excess variance, $\mathrm{F_{var}}$, of the objects studied here.}
\label{fig:fig-1}
\end{figure}

\section{Results and discussions}\label{sec:results}
\subsection{Flux variability}
AGN have been extensively studied for their variability and it has been found that 
optical variability of AGN correlates with many of their physical properties.
Most of these studies concentrate on photometric monitoring, however, in 
such studies, broad emission lines too can fall in the photometric pass band.
The best way to study line and continuum variability separately is through
spectroscopic monitoring observations, however, it is time consuming. A data set
suitable for such a study is the one accumulated for RM studies.
Though, the line and continuum light curves accumulated from RM
observations and used in this work  are primarily used  to understand the BLR, 
 they are also a good data set to investigate the variability of AGN. 
As we have both continuum and line light curves, we characterized the 
line and continuum variability of our sample using F$_{var}$  \citep{2002ApJ...568..610E,2003MNRAS.345.1271V}. The light curves for some objects analyzed here, were corrected for the constant host galaxy contamination to the continuum and the narrow-line contamination to the line fluxes. For few sources the host galaxy and the narrow line contribution to the continuum and line fluxes were not removed. This will have some effect on the derived F$_{var}$ values reported here as the flux contribution from host galaxy could lead to low values of F$_{var}$ \citep{2001sac..conf....3P}. However, this will not lead to biases on the comparative analysis of the F$_{var}$ values between different emission lines and the continuum. The distribution of 
the amplitude of variability, F$_{var}$ for both the continuum and line 
(that includes MgII, H$\beta$ and H$\alpha$) are 
given in Fig. \ref{fig:fig-1}. A two sample Kolmogorov-Smirnov (KS) test indicates
that the two distributions are indeed different with a statistic of
0.317 and a $p$ value of 4.0 $\times$ 10$^{-4}$. 
 We found mean F$_{var}$  values of $0.057 \pm 0.001$ and  $0.106 \pm 0.012$ 
for continuum and line, respectively. We have a total of 50 measurements for $\mathrm{H\beta}$, 26 for $\mathrm{H\alpha}$ and 6 for MgII line. Separating the sample based on 
different lines, for $\mathrm{H\beta}$ sample, we found mean F$_{var}$ values
of $0.065 \pm 0.002$ and $0.119 \pm 0.012$ for continuum and line, respectively. For H$\alpha$ sample, the mean F$_{var}$
values for continuum and line are $0.041 \pm 0.001$ and $0.076 \pm 0.018$, while for MgII sample, we found
mean F$_{var}$ of $0.057 \pm  0.001$ and $0.120 \pm 0.002$ for continuum and line, respectively. The F$_{var}$ in line is thus found to exceed than that of the continuum. Such increased variations in
emission lines relative to  the continuum support the deviation of BLR from 
the linear response to the ionizing continuum 
\citep{2015MNRAS.454.2918R, 2013ApJ...779..110L}.  Though photoionization 
models predict that MgII line should be less responsive to the continuum than 
Balmer lines \citep{2000ApJ...536..284K, 2004ApJ...606..749K},  
\cite{2008AJ....135.1849W} found $\mathrm{F_{var}}$ in Mg II line higher than the continuum too, similar to what is found in this work. For intermediate-redshift quasars the MgII line may originate almost in the same region as $\mathrm{H\beta}$, as can be seen in the cases of NGC 3783 and NGC 4151, for which similar time lags were obtained using
$\mathrm{H\beta}$ and MgII lines \citep{1994ApJ...425..582R, 2004ApJ...613..682P, 2006ApJ...647..901M, 2008AJ....135.1849W}. Therefore, it is possible to detect a similar kind of line variability in 
both the $\mathrm{H\beta}$ and MgII lines in these objects.

\subsection{BLR characteristics}
\begin{figure}
\begin{center}
\resizebox{9cm}{4cm}{\includegraphics{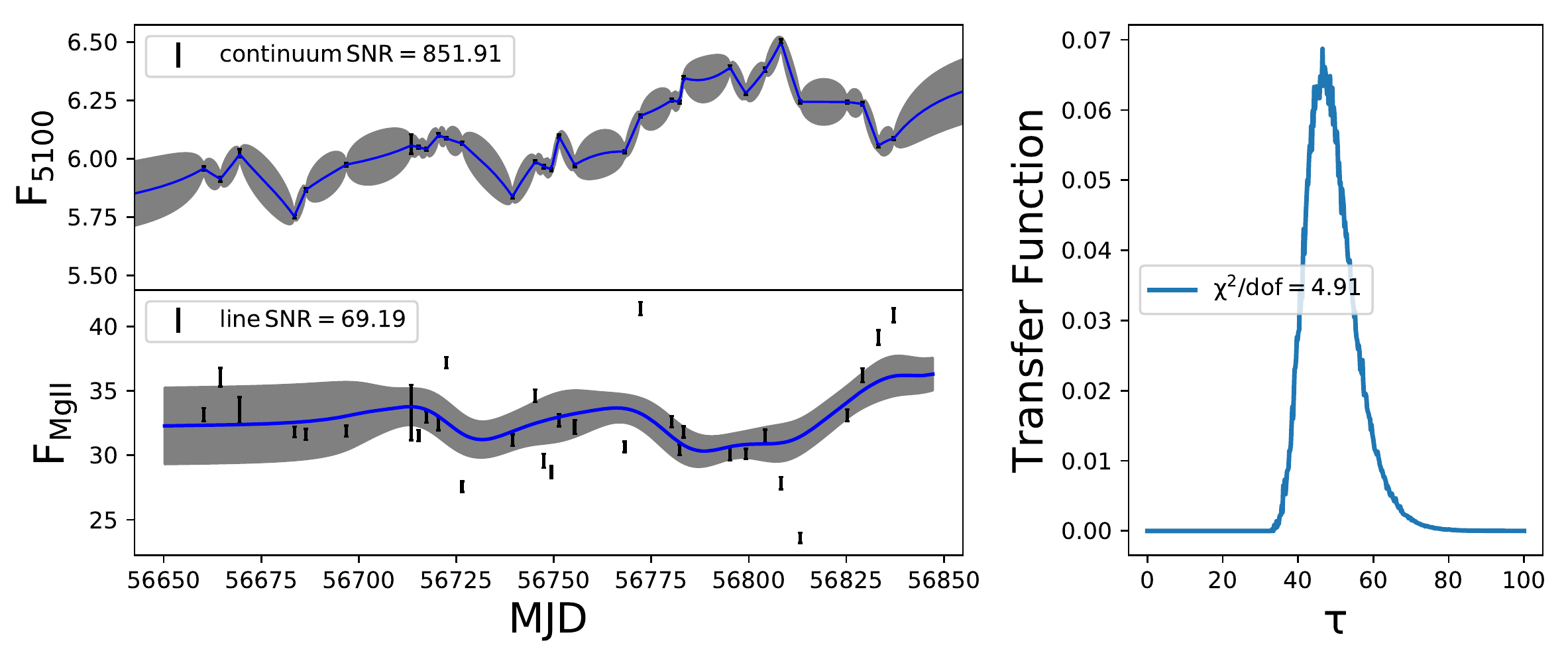}}
\resizebox{9cm}{4cm}{\includegraphics{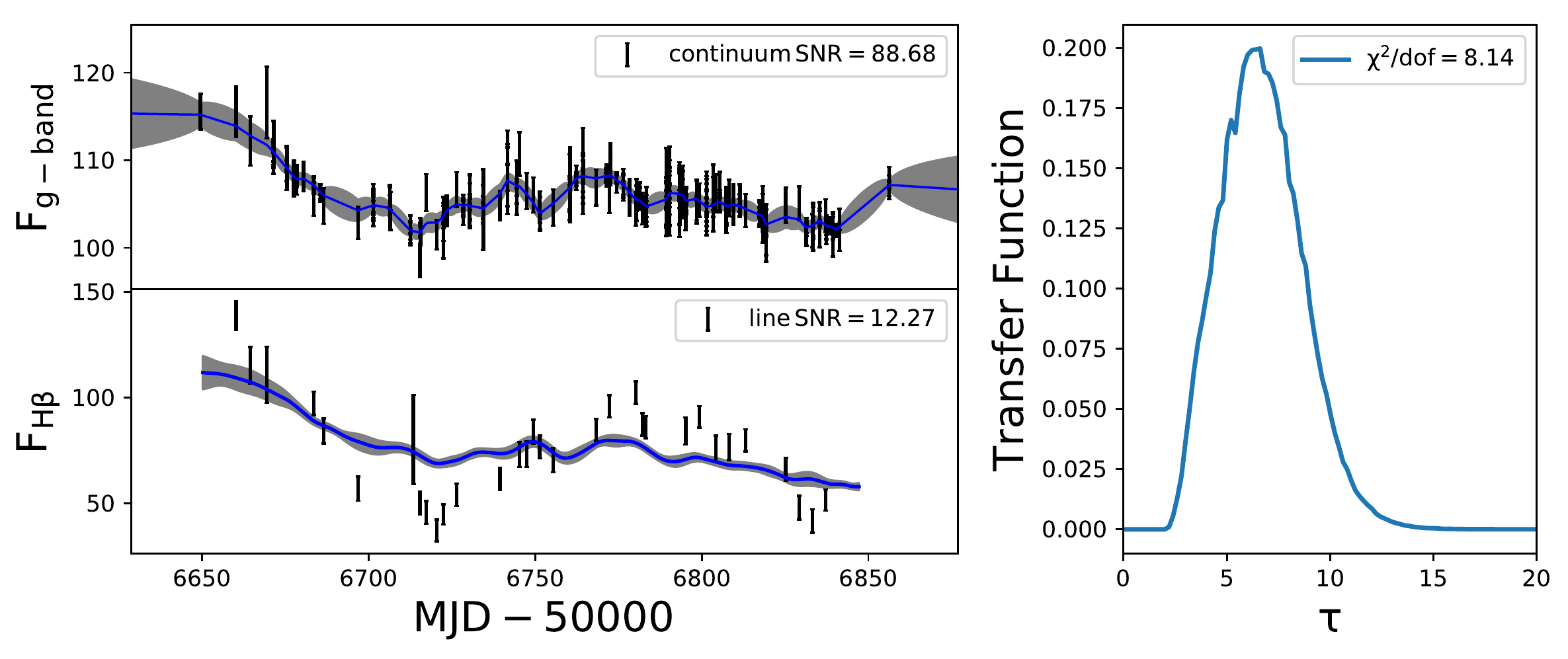}}
\resizebox{9cm}{4cm}{\includegraphics{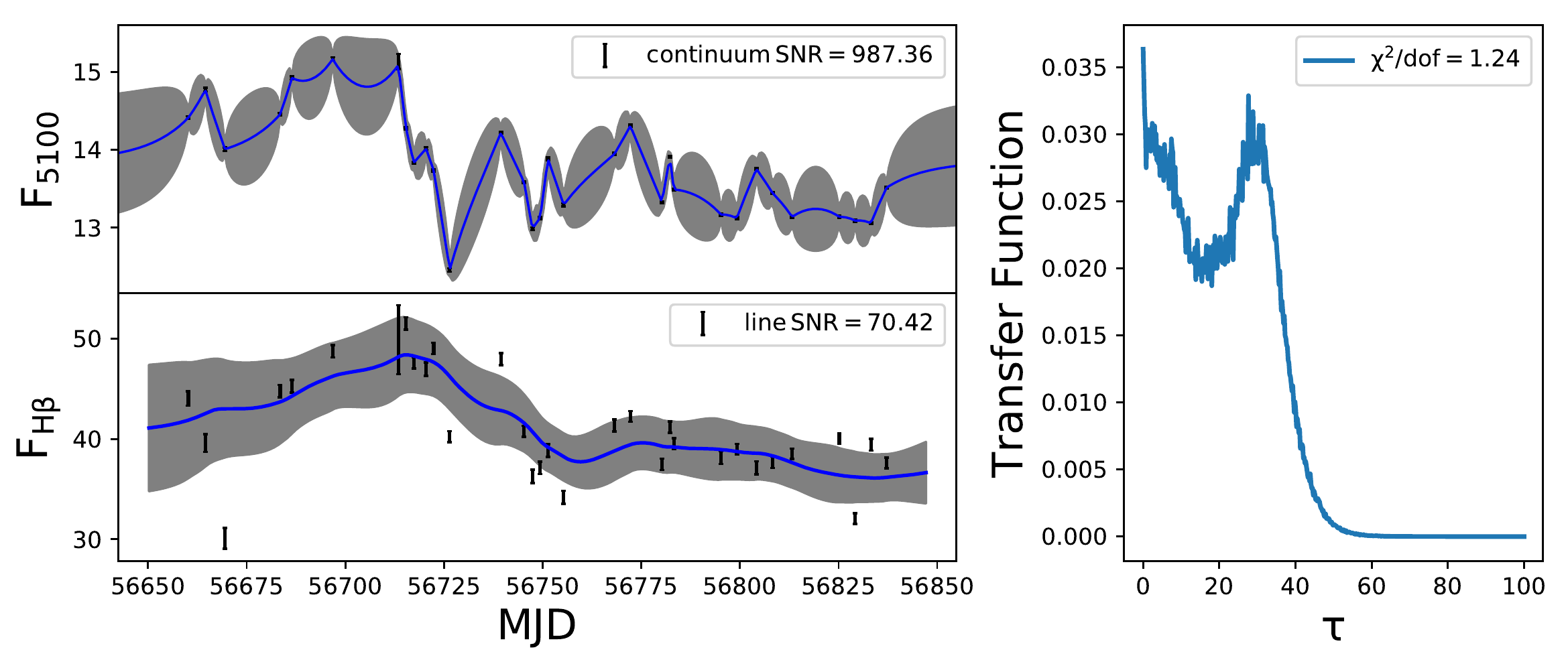}}
\resizebox{9cm}{4cm}{\includegraphics{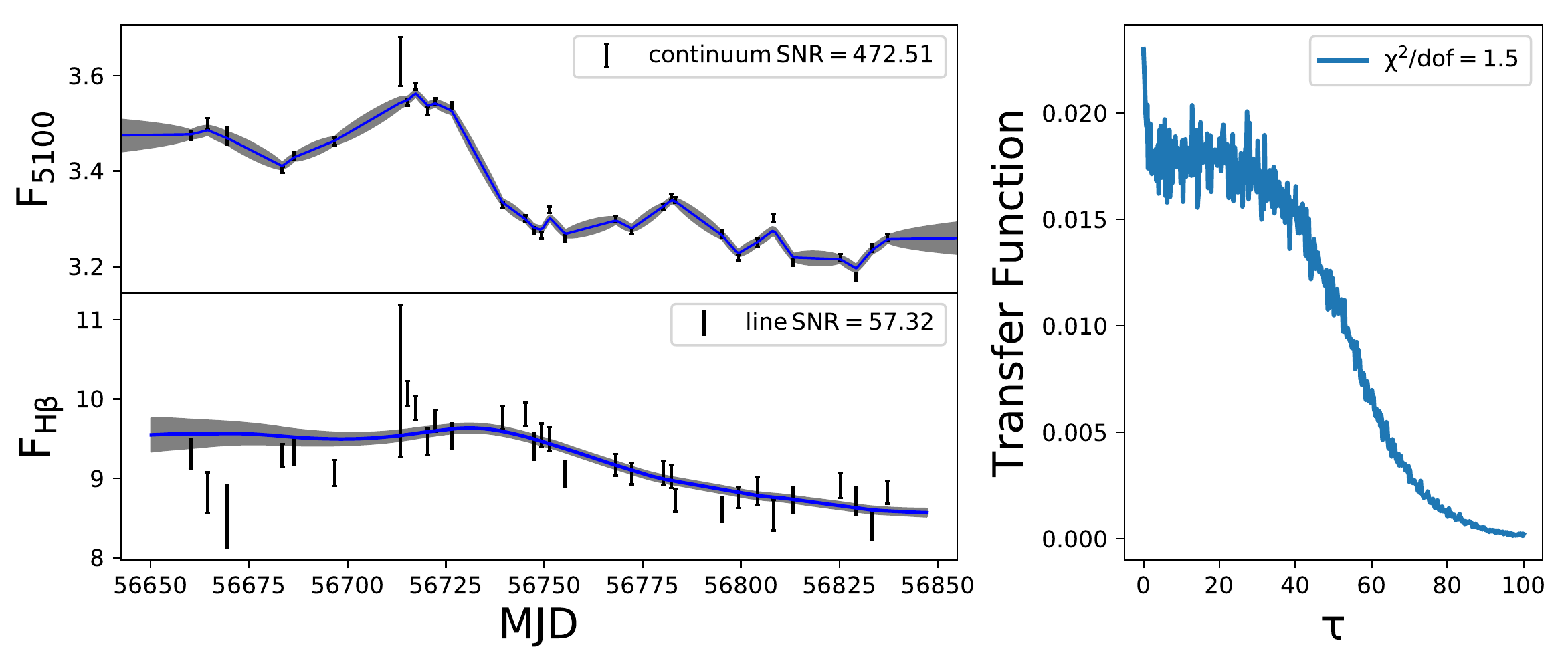}}
\end{center}
\caption{Examples of  BLR model fits to four objects J1412+534, J1421++525, 
J1407+537 and J1417+517 from top to bottom, respectively. In the left hand panels, the  data 
points with error bars are the observed light curves. The thick solid lines are 
the reconstructed light curve. The grey shaded areas represent the uncertainties 
in the reconstructed light curves. The corresponding transfer function for each 
objects are shown on the right hand panels. 
}
\label{fig:fig-2}
\end{figure}

Modeling of the BLR using a Bayesian approach was developed by
\cite{2011ApJ...730..139P} and subsequently applied to
Arp 151 \citep{2011ApJ...733L..33B} and Mrk 50 \citep{2012ApJ...754...49P}. 
In addition to Mrk 50 and Arp 151, more AGN were subjected to BLR modeling 
\citep{2014MNRAS.445.3073P,2017ApJ...849..146G,2018ApJ...866...75W,2013ApJ...779..110L}.
The BLR modeling used in this work is based on \cite{2013ApJ...779..110L}, which
is an independent implementation of the approach of \cite{2011ApJ...730..139P}, however,
with the additional inclusion of (a) non-linear response of emission lines to the continuum
variations and (b) option to carry out a detrending of the light curves.
Here, we analysed the data for 57 objects with 82 independent measurements for $\mathrm{H\beta}$, $\mathrm{H\alpha}$ and MgII lines, which is  twice the number of AGN studied
earlier \citep{2013ApJ...779..110L} for uncovering the characteristics of BLR.  We show in Fig. \ref{fig:fig-2} 
 few examples to illustrate 
our BLR model fits to their observed continuum and line reverberation data. The 
model reproduces the observed light curves
for all the objects. In these plots the data points with error bars are
the observed light curves and the thick solid lines are the reconstructed
light curves by the model. For the object J1412+534, few points of the observed 
line light curve deviate from the reconstructed light curve. These points are also found to deviate from the general trend of the observed line light curve that results of larger $\chi^2 / dof$ value of 4.91 pointing to poor fitting to the light 
curve.  For the object J1421+525, there is a 
discrepancy between the observed and modeled line light curve. The 
$\chi^2 / dof$ of 8.14 obtained here could be due to poor sampling and/or SNR of 
the emission line measurements. The corresponding transfer functions for those four objects
are given in the right hand panel of Fig. \ref{fig:fig-2}. We found the transfer 
functions to have different shapes. For example in J1412+534 
(top panel; $\alpha_{2000}$ = 14:12:14.20, $\delta_{2000}$ = 53:25:46.7), the 
transfer function is single peaked at $\tau$ = $\tau_{lag}$. For this object
the BLR modeling gives $\theta_{inc}$ = 4.0 $\pm$ 2.8 deg and $\theta_{opn}$ = 6.0 $\pm$ 3.9 deg. 
The transfer function for J1407+537 
($\alpha_{2000}$ = 14:07:59.07, $\delta_{2000}$ = 53:47:59.8) is double peaked,
for which we obtain $\theta_{inc}$ = 65.0 $\pm$ 22.1 deg and $\theta_{opn}$ = 37.0 $\pm$ 26.5 deg. 
For J1417+517 (bottom panel; $\alpha_{2000}$ = 14:17:06.68, $\delta_{2000}$ = 51:43:40.1), the transfer function  has a top hat structure with derived  $\theta_{inc}$ and  $\theta_{opn}$ of 57.9 $\pm$ 21.6 deg and 51.8 $\pm$ 25.8 deg, respectively. 
 We note that the peak of the transfer function is not always a reliable indicator for the size of the BLR. For example, peak can represent BLR size for a single peaked transfer function but it is not a reliable indicator for a double peaked transfer function or a transfer function with long tail. The first moment of transfer function, which represents the time lag in CCF analysis, gives a better estimate of the average size of the BLR \citep{1986ApJ...305..175G, 1987ApJS...65....1G, 2014AdSpR..54.1414K}. We found a) for a thick disk, larger $\theta_{\mathrm{inc}}$ 
tends to produce a double-peaked transfer function, as seen in  the case of 
object J1407+537 (third panel of Fig. \ref{fig:fig-2} from the top) with  
$\theta_{\mathrm{inc}}$ = 65.0 $\pm$ 22.1 deg. As $\theta_{\mathrm{inc}}$ increases, 
the object appears more edge on and the radiation coming both from the front 
and back surfaces makes a double peak transfer function where the stronger peak 
appears closer to the center. b) a larger opening angle $\theta_{\mathrm{opn}}$ 
tends to broaden the transfer functions toward top-hat as seen in case of 
J1417+517 (bottom panel of Fig. \ref{fig:fig-2})  which  has a $\theta_{\mathrm{opn}}$ of 
51.8 $\pm$ 25.8 degree.  As $\theta_{\mathrm{opn}}$ increases the BLR tends to 
a spherical geometry and the 
virial motion of the clouds contributes to the transfer function making the transfer 
function broaden towards a top-hat structure. 

\begin{figure}
\resizebox{7cm}{7cm}{\includegraphics{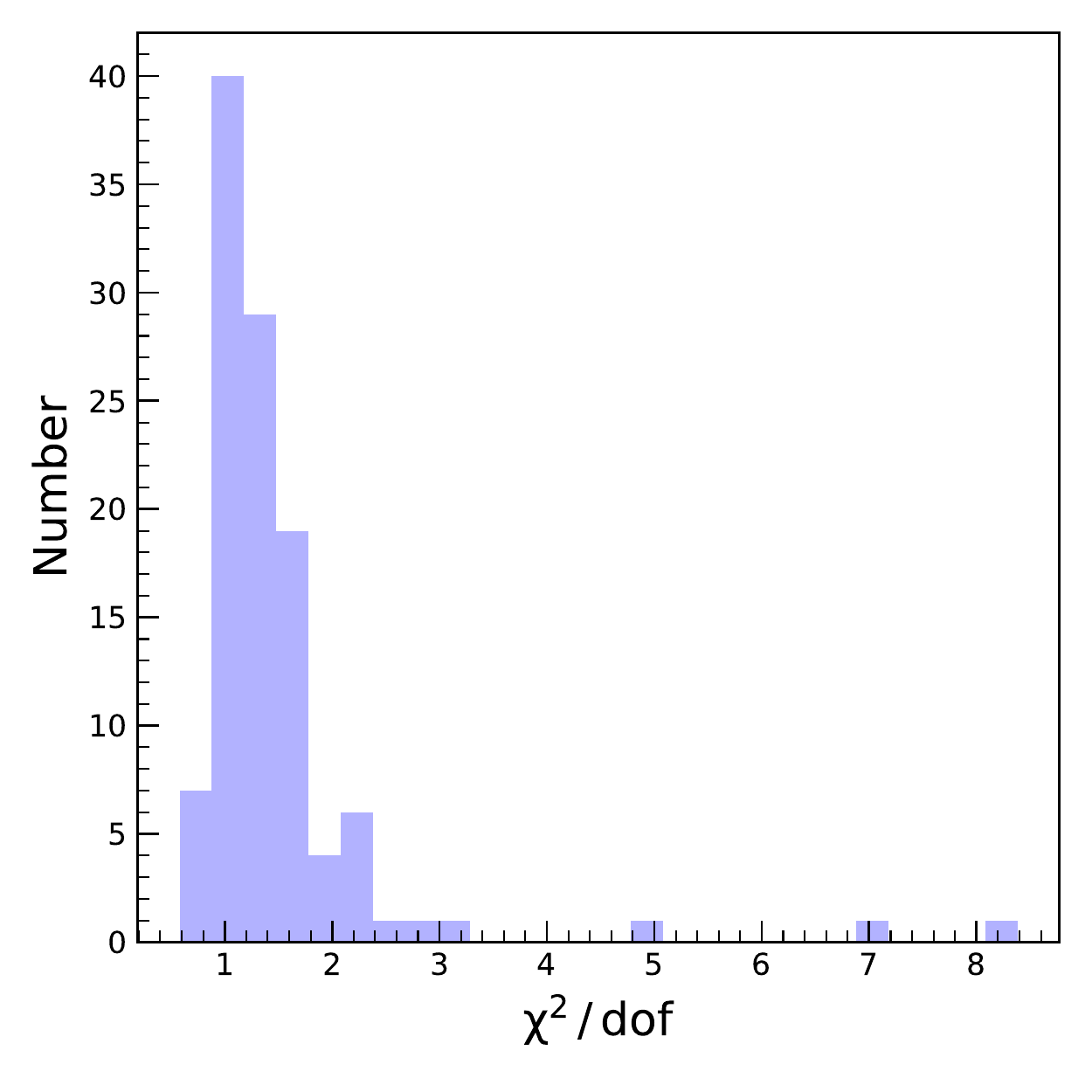}}
\caption{Distribution of $\chi^2$/dof returned by the models for the objects analysed in this work.}
\label{fig:chi2}
\end{figure}

We show in Fig. \ref{fig:chi2} the distribution of $\chi^2 / dof$ obtained for the best fit model returned by PBMAP. 
The fits are indeed bad (with $\chi^2 /dof$ $>$ 4) for a few sources, namely 
J141214.20+532546.7, J141941.11+533649.6 and J142135.90+523138.9. These sources have poor quality (less number of points and sparsely sampled) data and is the likely cause for large  $\chi^2$/dof. Though the overall distribution of $\chi^2 /dof$ seems skewed to values larger than 1.0, in majority of the sources, we obtained a low $\chi^2 / dof$ close to 1.0. For about 60\% of the light curves we obtained $\chi^2 / dof \leq$ 1.2. The poorer $\chi^2 / dof$ in some objects is due to them having 
continuum and line light curves with SNR $<$ 50. Also, systematic errors due to 
calibration which are usually not included in the reported uncertainties of the 
original data could affect the $\chi^2$ value.    
Overall, the continuum and emission line light curves generated by the model 
are in good agreement with the observed data.

\begin{figure}
\resizebox{8cm}{7cm}{\includegraphics{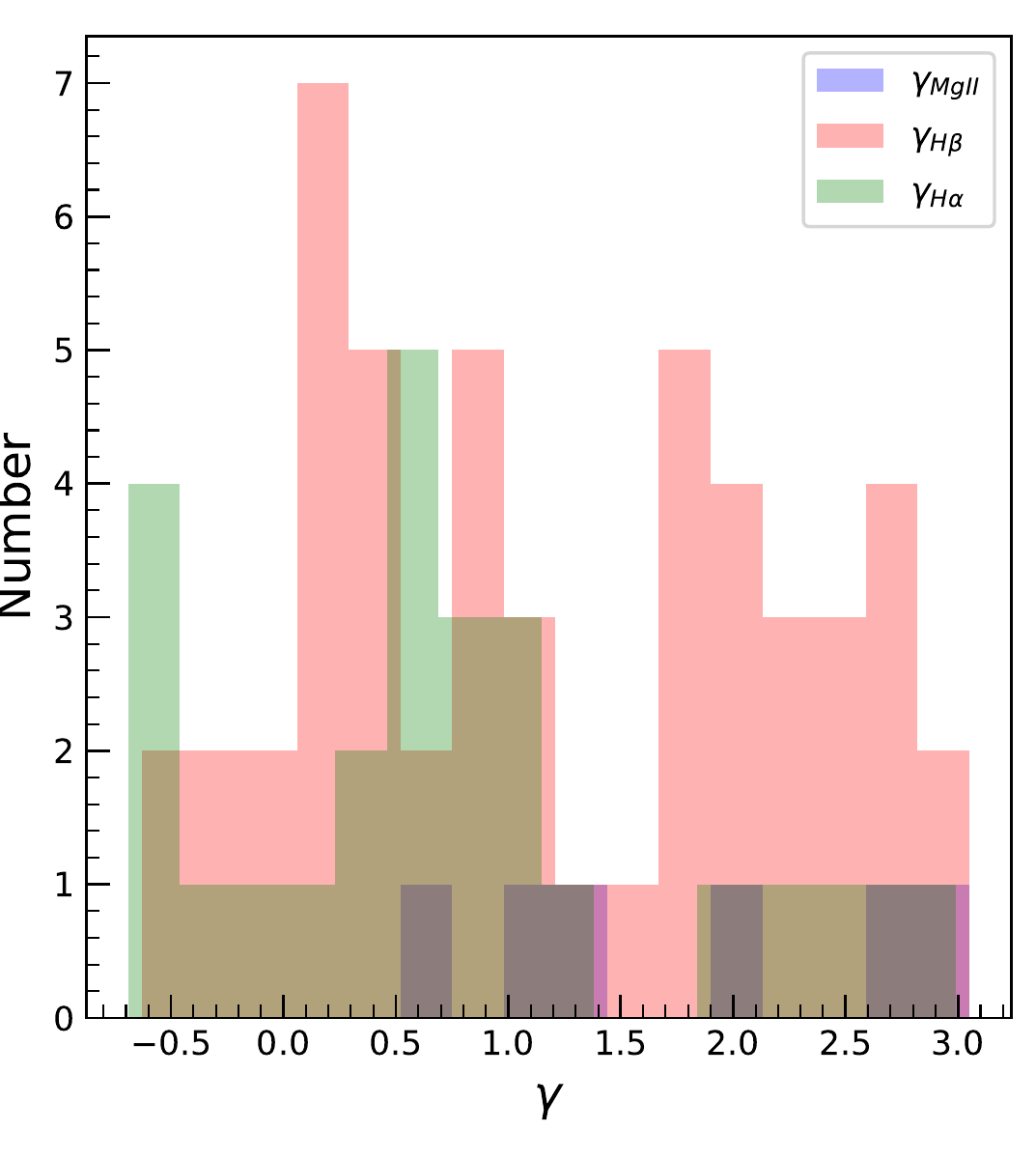}}
\caption{ Distribution of non-linearity parameter $\gamma$ for different emission lines.}
\label{fig:fig-3}
\end{figure}

Fig. \ref{fig:fig-3} shows the distribution of the non-linearity parameter $\gamma$ obtained from modeling, we found $<\gamma_{MgII}> = 1.78 \pm  0.86$, $<\gamma_{H\beta}> = 1.18 \pm 1.03$ and $<\gamma_{H\alpha}> = 0.76 \pm 0.99$ for MgII, $\mathrm{H\beta}$ and $\mathrm{H\alpha}$, respectively. This clearly indicates a non-linear response of emission-lines from BLR to the ionizing optical continuum. Such non-linear response of line flux to the ionizing continuum can be due to the anisotropic and non axis-symmetric emission coming from different spectral regions in AGN \citep{2000ApJ...536..284K, 2004ApJ...606..749K, 2019arXiv190906275G}. Note that the shorter  wavelength UV continuum usually vary larger compared to the longer wavelength optical continuum, therefore, depending on the continuum, the response of a given emission line could be different \citep{1995MNRAS.275.1125O,Zhu_2017}.




\subsection{Dependency of damping time scale on Luminosity at $5100 \, $\AA$ $ for H${\beta}$ line fitting}

\citet{2009ApJ...698..895K} modeled  the light curves of 100 quasars using DRW and found the time scale of variability to  correlate with luminosity. Recently, \citet{2019ApJ...877...23L} performed DRW modeling of 73 AGN including high-accreting sources, which are also studied here. They found that the damping time scale is strongly correlated with luminosity with a slope of $0.46\pm0.09$. However, \citet{2010ApJ...721.1014M} using SDSS stripe 82 data, did not find any strong correlation with luminosity.  In our fitting, emission line and continuum model parameters are fitted simultaneously allowing us to study this relation.

\begin{figure}
\begin{center}
\resizebox{9cm}{8cm}{\includegraphics{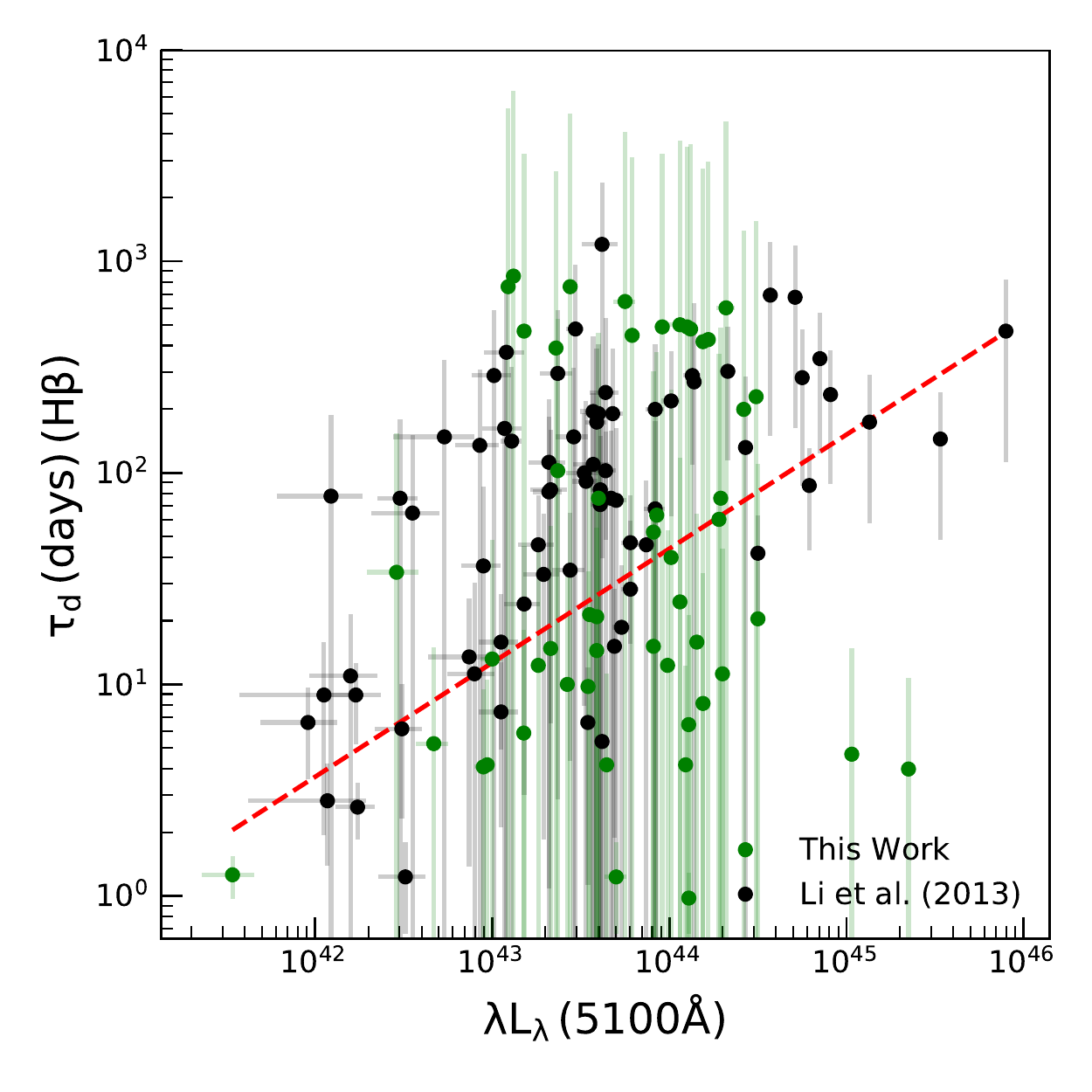}}

\end{center}
\caption{ Relation between the damping time scale ($\tau_d$) and the monochromatic continuum 
luminosity at 5100 \AA. Here, the filled green circles are the objects studied in this work, while 
filled black circles represent the objects taken from \citet{2013ApJ...779..110L}. The dashed red line is the best fit to the data points including measurements from this work and \citet{2013ApJ...779..110L}}.
\label{fig:fig-4}
\end{figure}

We show in Fig. \ref{fig:fig-4} the dependence of the derived rest frame damping time scale on the observed host-galaxy corrected continuum luminosity at 5100 \AA. We found that
the damping time scale $\tau_d$ is positively correlated with the luminosity at 5100 \AA. From linear least squares fitting to the data we found

\begin{equation}\label{eq:damp}
\log\left(\frac{\tau_{d(H\beta)}}{\mathrm{1day}}\right) = \beta + \alpha \log(\lambda L_{\lambda})(5100 $\r{A}$)
\end{equation}

with $\alpha = 0.54 \pm 0.06$ and $\beta = -22.09 \pm 2.67$. The slope of the 
correlation is similar to the value of $\alpha = 0.60 \pm 0.06$  found by  \citet{2013ApJ...779..110L}, who performed BLR modeling, in the same fashion like we did here, from an analysis of 50 AGN with H$\beta$ lags. We note that the scatter in the relation is much higher than \citet{2019ApJ...877...23L}, mainly because they modeled continuum light curves with only two main parameters while we fitted both continuum and BLR model parameters simultaneously. Moreover, their sample does not include SDSS RM sample, which has relatively less time sampling and variability. A carefully analysis suggests that the more deviant points have lower variability and hence the model parameters are not well constrained. 

 To check the correlation between the damping time scale and luminosity at 
5100 \AA, we estimated the Spearman rank correlation coefficient ($\mathrm{r_s}$) using Monte 
Carlo simulation where each point in the $\mathrm{\tau_d - \lambda L_{\lambda}}$ 
plane is modified by a random Gaussian deviate consistent with the measured 
uncertainty. From the distribution obtained for 10000 Monte Carlo iterations, the median value of $\mathrm{r_s}$ is found to be $0.218^{+0.065}_{-0.066}$ with a probability ($p$) of 
no correlation of $0.016^{+0.079}_{-0.015}$. The upper and lower errors are the 
values at the 15.9 and 84.1 percentile of the distributions of those 10000 
iterations. \citet{2017A&A...597A.128K, 2017ApJ...835..250K} suggested that 
deriving damping time from short duration light curves leads to biased results and the 
time length of the light curve must be 10 times the true damping time scale. We 
note that reverberation light curves are usually shorter in length compared to 
the long-term survey light curves such as those from the Sloan
Digital Sky Survey and the Catalina Real Time Transient Survey. In fact, the median ratio 
of the total span ($\mathrm{\Delta t}$) of the light curves to the damping time 
scale $\tau_d$ is 4.84 and 4.35 for continuum and line light curves, respectively. 
Considering objects with light curve length $>10\times \tau_d$, which includes a total 
20 objects from our sample and 21 objects from \citet{2013ApJ...779..110L}, the Spearman 
rank correlation coefficient is found to be $0.304^{+0.104}_{-0.104}$ with $p$-value of 
$0.053^{+0.155}_{-0.045}$ (10000 iterations) for the 
$\mathrm{\tau_d - \lambda L_{\lambda}}$ relation. The least-square fit using 
Equation \ref{eq:damp} provides $\alpha=0.39 \pm 0.08$ and $\beta = -15.99 \pm 3.50$. 
The correlation thus obtained between $\mathrm{\tau_d}$ and 
$\mathrm{\lambda L_{\lambda}}$ is significant at greater than 90\% level. This 
result is consistent with \cite{2019ApJ...877...23L}, who studied the optical 
variability characteristic of reverberation mapped AGN and found 
$\alpha=0.46\pm0.09$ and $\beta=-18.52\pm4.06$ in the 
$\mathrm{\tau_d - \lambda L_{\lambda}}$ relation, for sources with light curve 
lengths greater than 10 times $\tau_d$.

\begin{figure}
\begin{center}
\resizebox{9cm}{8cm}{\includegraphics{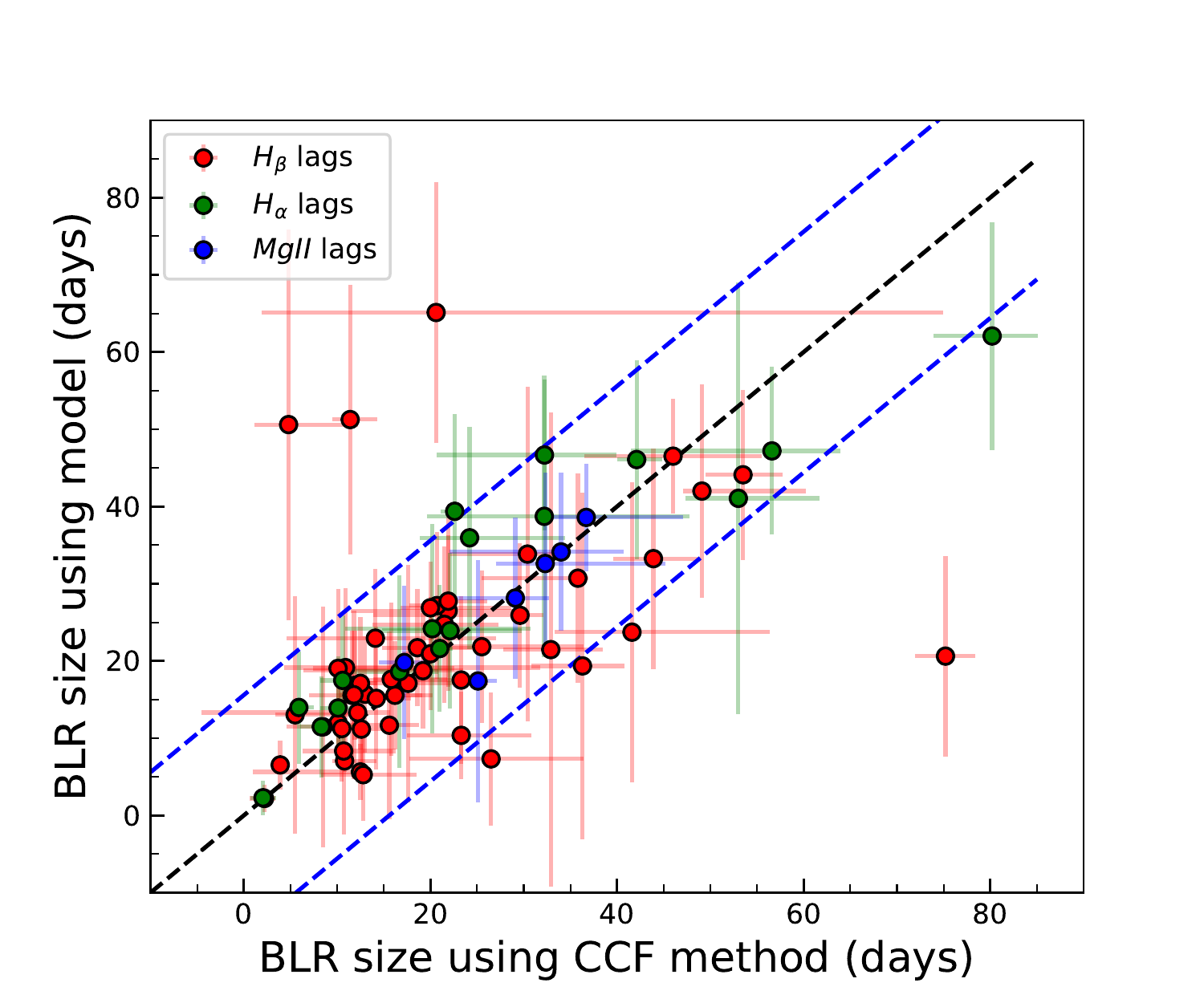}}
\resizebox{9cm}{8cm}{\includegraphics{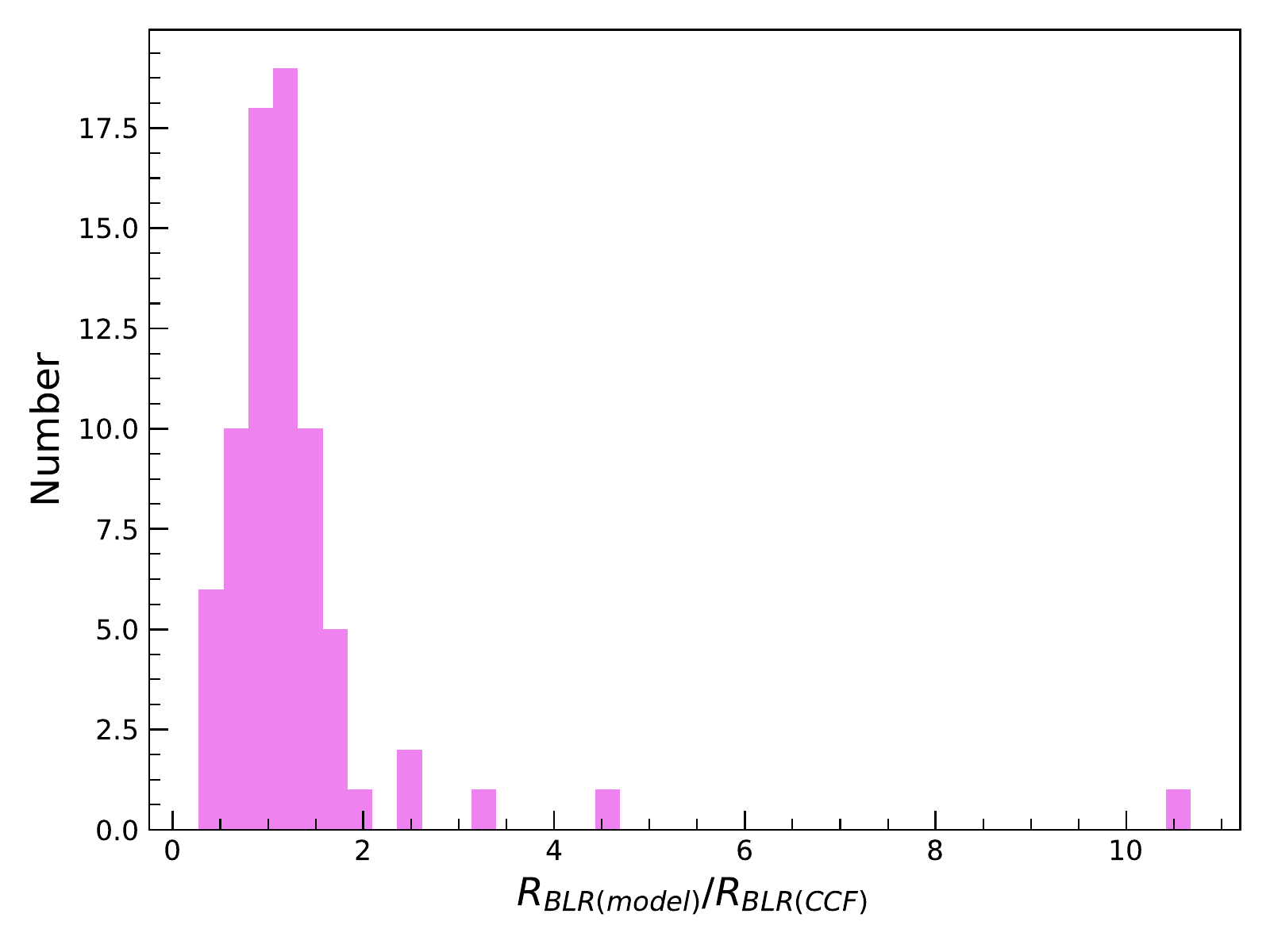}}
\end{center}
\caption{(Top) Comparison of the BLR size obtained in this work from the model and that from CCF analysis taken from literature. Red colored circles correspond to H$\beta$ 
lags, whereas green and blue circles correspond to H$\alpha$ and Mg II lags, 
respectively. The black dashed line shows y=x, while the blue dashed lines are 
y = x $\pm$ $\sigma$, where $\sigma = 15.59$ days is the standard deviation of the BLR sizes 
obtained from CCF analysis. (bottom) Distribution of the ratio of the BLR size using model ($R_{BLR(model)}$) to that obtained from CCF  ($R_{BLR(CCF)}$).}
\label{fig:fig-5}
\end{figure}

\subsection{Relation between $R_{\mathrm{BLR}}^{\mathrm{mod}}$ and $R_{\mathrm{BLR}}^{\mathrm{CCF}}$}
In Fig. \ref{fig:fig-5} we show a comparison of the size of the BLR derived by 
the modeling approach ($R_{\mathrm{BLR}}^{\mathrm{mod}}$) with that obtained using the conventional 
cross-correlation function (CCF) analysis ($R_{\mathrm{BLR}}^{\mathrm{CCF}}$). The model BLR size is in general consistent with that obtained from CCF, however, with a large scatter. The median of the ratio between BLR size estimated by modeling and CCF is $1.09$ with standard deviation of $1.24$. 

\cite{2013ApJ...779..110L} found that the BLR sizes obtained by CCF analysis are underestimated by 20$\%$. Although we found a few objects to have model $R_{\mathrm{BLR}}$ larger than that obtained by CCF analysis, a few others have model $R_{\mathrm{BLR}}$ smaller than that of CCF. The median of the ratio of $R_{\mathrm{BLR}}$ model to $R_{\mathrm{BLR}}$ CCF (see lower panel of Fig. \ref{fig:fig-5}) is found to be $1.09 \pm 1.24$, where 3 objects are found to deviate from the unit ratio by a factor larger than 3. The objects that show larger deviation from the $R_{\mathrm{BLR}}^{mod}$ = $R_{\mathrm{BLR}}^{CCF}$ line also have large $\chi^2 / dof$ ($>$ 1.5) values giving ample indication that the model fitting is improper which could be due to the following reasons (a) poor sampling of the light curves 
(b) low SNR in the light curves and (c) multiple peaks in the CCF leading to ambiguity in 
the determination of the peak of the CCF and subsequently $R_{\mathrm{BLR}}$. 

\subsection{BLR size-luminosity relation}
\begin{figure}
\begin{center}
\resizebox{9cm}{8cm}{\includegraphics{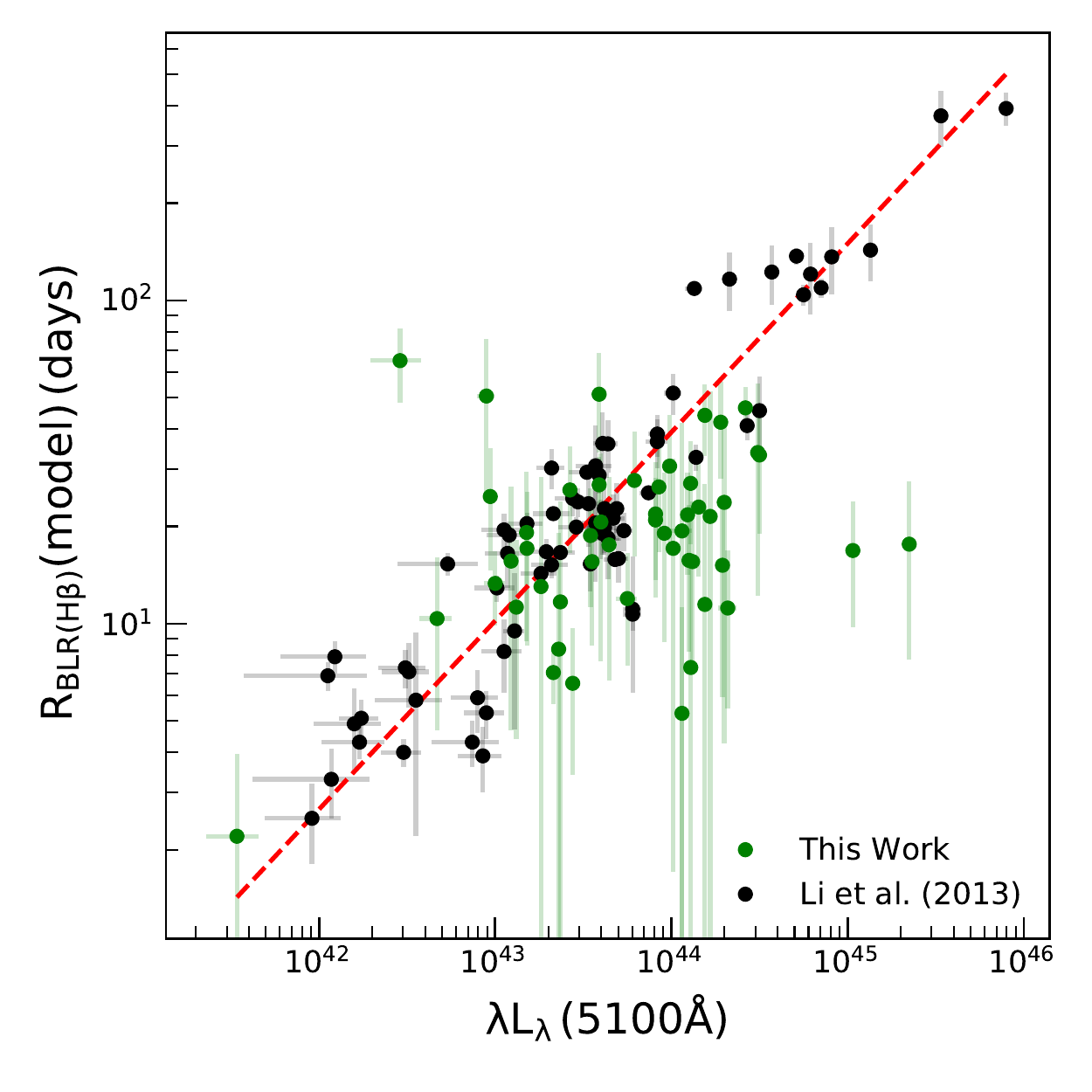}}
\end{center}
\caption{Relation between the radius of the BLR obtained from the model for sources with H$\beta$ light curves and
their continuum luminosity at 5100 \AA. Here, filled green circles are the objects studied in this work, while the filled black circles are the objects 
from \citet{2013ApJ...779..110L}. The dashed red line is the best fit to the data points including measurements from this work and \citet{2013ApJ...779..110L}}.
\label{fig:fig-6}
\end{figure}

Reverberation mapping observation over the years have led to a power law relation ($R_{\mathrm{BLR}} \propto L^{\alpha}$) 
between the size of the BLR and the optical luminosity of the AGN. The $R_{\mathrm{BLR}} - L$ relation is very 
important as it enables the determination of M$_{\mathrm{BH}}$ from single epoch spectroscopic observations. Also
the $R_{\mathrm{BLR}} - L$ relation can provide a means to consider AGN as standard candles \citep{2019arXiv190309687L}. Therefore, it
is important to check if the derived $R_{\mathrm{BLR}}$ from fitting shows the power law dependence with luminosity
that we know from observation. As we have sources over a varied range of redshift, $R_{\mathrm{BLR}}$ from
model fitting has been found using lines of H${\beta}$, H${\alpha}$ and Mg II. The relation between
$R_{\mathrm{BLR}}$ and luminosity for H${\beta}$ is shown in 
Fig. \ref{fig:fig-6}. Note that we adopted  host-corrected luminosities from the original 
literature \footnote{ In Fig. \ref{fig:fig-6}, we excluded one measurement for which the host-corrected luminosity is not available in the literature}. Details on host-subtraction can be found in the original literature.

Using weighted linear least squares fit we 
obtained the following relation
\begin{equation}
\log\left(\frac{R_{\mathrm{BLR}(H\beta)}}{\mathrm{1day}}\right) = \beta + \alpha \log(\lambda L_{\lambda})(5100 $\r{A}$)
\end{equation}
with $\alpha$ = 0.58 $\pm$ 0.03 and  $\beta$ = $-$24.08 $\pm$ 1.13. This is similar to the value
of $\alpha$ = $0.519^{+0.063}_{-0.066}$ and $\beta$ = $-21.3^{+2.9}_{-2.8}$ obtained
by \cite{2009ApJ...697..160B} with $R_{\mathrm{BLR}}$ obtained by CCF analysis of the observed continuum and
line light curves. \citet{2013ApJ...767..149B} found a slope of $\mathrm{\alpha = 0.533^{+0.035}_{-0.033}}$ and $\mathrm{\beta = 1.527^{+0.031}_{-0.031}}$ considering lag-luminosity relation of $\mathrm{\log\left(\frac{R_{\mathrm{BLR}(H\beta)}}{\mathrm{1day}}\right) = \beta + \alpha \log(\lambda L_{\lambda}/10^{44} \, erg \, s^{-1})}$ (5100\AA). Our values closely match with those obtained by \citet{2013ApJ...767..149B} considering the uncertainties. \cite{2013ApJ...779..110L} using the approach adopted in this work for
40 quasars with $H\beta$ measurements found a value of $\alpha$ = 0.55 $\pm$ 0.03, which 
 again is in agreement with the one found by us using a different sample of 
51 AGN for the $\mathrm{H{\beta}}$ line.

\begin{figure}
\begin{center}
\resizebox{9cm}{8cm}{\includegraphics{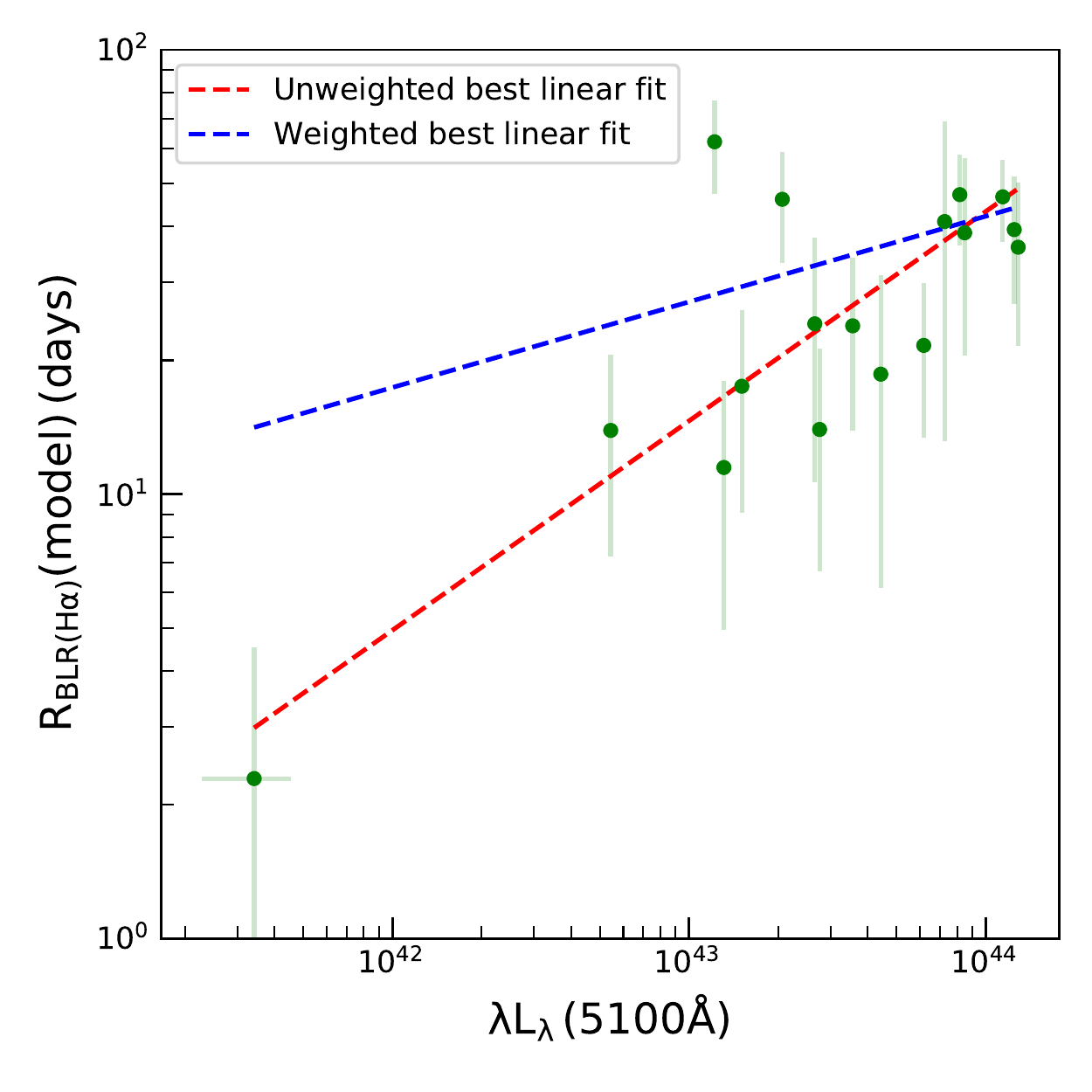}}
\end{center}
\caption{R$_{\mathrm{BLR}}$ v/s luminosity relation for sources with H$\alpha$ light curves. The dashed blue and red lines are the weighted and unweighted linear
least squares fit, respectively, to the data points.}
\label{fig:fig-7}
\end{figure}

Similarly, the relation between  R$_{BLR}$ and $L_{5100\AA}$ for objects with $\mathrm{H\alpha}$ measurements
is shown in Fig. \ref{fig:fig-7}. We used only measurements with fractional error less than 1. Using weighted 
linear least squares fit to the data we found
\begin{equation}
\log\left(\frac{R_{\mathrm{BLR}(H\alpha)}}{\mathrm{1day}}\right) = \beta + \alpha \log(\lambda L_{\lambda})(5100 $\r{A}$)
\end{equation}
with $\alpha = 0.19 \pm 0.12$ and $\beta = -6.86 \pm 5.16$ as shown by the dashed blue line. Using unweighted linear least squares fit to the data as shown by dashed red line, we found
\begin{equation}
\label{Eq:eqn-alp}
\log\left(\frac{R_{\mathrm{BLR}(H\alpha)}}{\mathrm{1day}}\right) = \beta + \alpha \log(\lambda L_{\lambda})(5100 $\r{A}$)
\end{equation}
with $\alpha = 0.47 \pm 0.08$ and $\beta = -19.10 \pm 3.67$ which closely matches 
with $\alpha$=0.5 based on simple photoionization arguments. We note that the unweighted fit is driven by a single data point at low luminosity. This point corresponds to the object J1342+356 (NGC 5273), which has a luminosity of $\mathrm{log \, L_{AGN} = 41.534 \pm 0.144}$ erg s$^{-1}$ and a BLR size of $2.06^{+1.42}_{-1.31}$ days based on $\mathrm{H\alpha}$ line obtained from traditional CCF analysis by \citet{2014ApJ...796....8B}. The BLR size obtained from our modeling approach is $2.29 \pm 2.24$ days which is consistent with that obtained by \citet{2014ApJ...796....8B}.

\begin{figure}
\begin{center}
\resizebox{9cm}{8cm}{\includegraphics{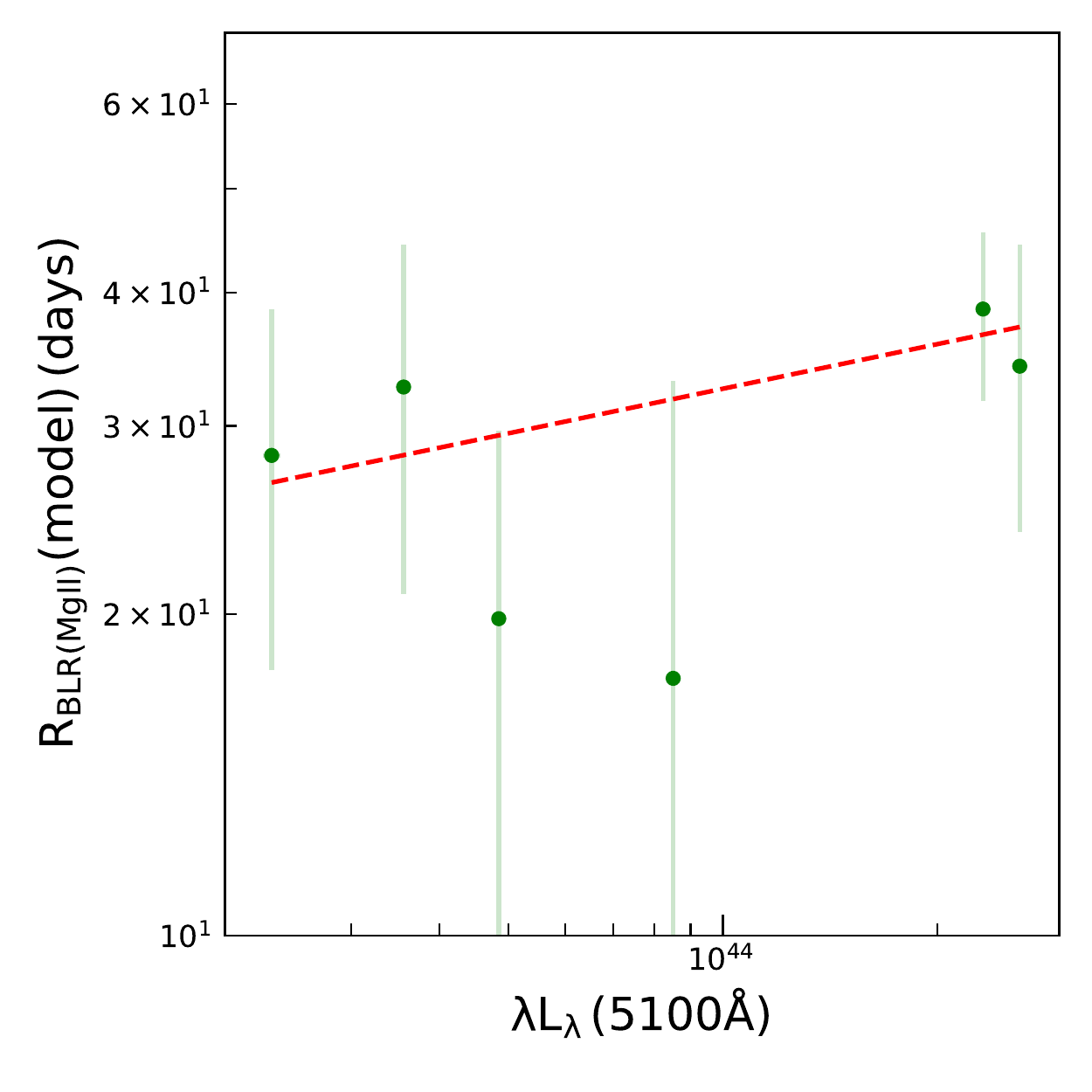}}
\end{center}
\caption{Relation between the radius of the BLR and the continuum luminosity for objects with MgII line light curves. 
The dashed red line is the weighted linear least squares fit to the data points.}
\label{fig:fig-8}
\end{figure}

We have six objects with Mg II line light curves. For those objects the relation between R$_{\mathrm{BLR}}$ and $L_{5100\AA}$ is
given in Fig. \ref{fig:fig-8}. We plotted the data with the form 
\begin{equation}
\label{Eq:eqn-mg}
\log\left(\frac{R_{\mathrm{BLR}(Mg II)}}{\mathrm{1day}}\right) = \beta + \alpha \log(\lambda L_{\lambda})(5100 $\r{A}$)
\end{equation}
and we found $\alpha$ = $0.14 \pm 0.08$ and $\beta$ = $-4.59 \pm 3.58$.  The 
relation between $R_{\mathrm{BLR}}$ and luminosity of Mg II deviates from the 
value expected from photoionization argument. This is only due to the poor 
quality of measurement available on small number of sources. We note that the $R_{\mathrm{BLR}} - L$ relation of $\mathrm{H\alpha}$ line has a luminosity range of $10^{41.5}$ to $10^{44.1}$ $\mathrm{erg \, s^{-1}}$ and majority of them are above $10^{43}$ $\mathrm{erg \, s^{-1}}$, whereas for Mg II line the luminosity ranges only between $10^{43.4}$ to $10^{44.4}$ $\mathrm{erg \, s^{-1}}$.  $\mathrm{R_{BLR}}$ measurements on large number of objects spanning over a wide range of luminosities are needed to firmly establish the  relationship between $\mathrm{R_{BLR}}$ and luminosity based on $\mathrm{H\alpha}$ and Mg II emission lines and therefore, the coefficients of Equations \ref{Eq:eqn-alp} and \ref{Eq:eqn-mg}, should be taken  with caution.


\begin{figure}
\begin{center}
\resizebox{9cm}{7cm}{\includegraphics{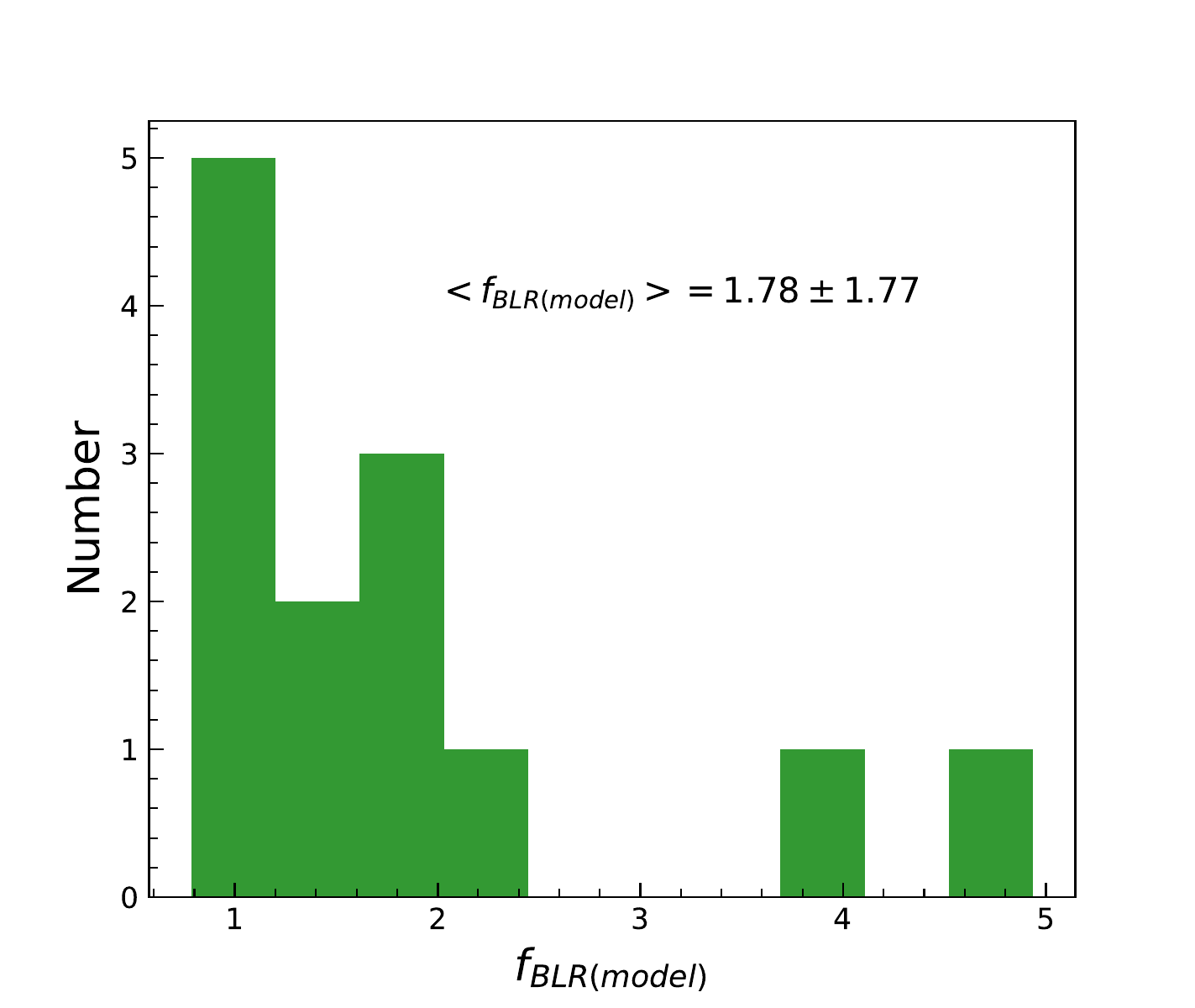}}
\end{center}
\caption{Distribution of the virial factor $f_{\mathrm{BLR}}$ obtained for the objects analysed in this work}
\label{fig:fig-9}
\end{figure}

\subsection{The Virial factor $f_{\mathrm{BLR}}$}
The virial factor $f_{\mathrm{BLR}}$ given in Equation \ref{eq:blc_v} depends on factors such as the kinematics, geometry and inclination of the BLR. One of the many factors
provided by the Bayesian based modeling approach carried out here is the capability to estimate $f_{\mathrm{BLR}}$. For a  disk like BLR, \citep[see][]{2006A&A...456...75C,2013ApJ...779..110L,2015MNRAS.447.2420R} $f_{\mathrm{BLR}}$ can be written as

\begin{equation}
f_{\mathrm{BLR}} \approx (\sin^2\theta_{\mathrm{opn}} + \sin^2\theta_{\mathrm{inc}})^{-1} 
\end{equation}

where $\theta_{\mathrm{inc}}$ is the inclination angle and $\theta_{\mathrm{opn}}$ is the opening angle of the disk. Following
\cite{2013ApJ...779..110L} we calculated $f_{\mathrm{BLR}}$ for only those objects with $\theta_{opn}$ $<$ 40$^{\circ}$. Our
calculated values of $f_{\mathrm{BLR}}$ range from 0.79 to 4.94, with a mean value of $1.78 \pm 1.77$. A distribution of $f_{\mathrm{BLR}}$ is 
shown in Fig. \ref{fig:fig-9}. \cite{2006A&A...456...75C} found a value of $<\log(f_{\mathrm{BLR}})>$ = 0.18. Our average value of 
$<\log(f_{\mathrm{BLR}})>$ = 0.17 closely matches with that found by \cite{2006A&A...456...75C}. The large error bars
in our $f_{\mathrm{BLR}}$ values are due to large uncertainties in both $\theta_{\mathrm{inc}}$ and $\theta_{\mathrm{opn}}$.

It is also possible to get an estimate of $f_{\mathrm{BLR}}$ for sources that 
have stellar velocity dispersion measurements.
For local inactive galaxies a tight correlation is known to exist 
between M$_{BH}$ and bulge or spheroid stellar velocity dispersion ($\sigma_*$).
This correlation  
\citep{2000ApJ...539L...9F,2000ApJ...539L..13G} is  given as
\begin{equation}\label{eq:m-singma}
\log \left(\frac{M_{BH}}{M_{\odot}}\right) = \alpha + \beta \log \left(\frac{\sigma_*}{200 \, \mathrm{km s^{-1}}} \right)
\end{equation}
with $\alpha = 8.13\pm0.06$ and $\beta = 4.02\pm0.32$ \citep{2002ApJ...574..740T}.
Assuming AGN too follow the above equation, one can
estimate M$_{BH}$. Comparing this M$_{BH}$ with the virial product VP = $\left(\frac{\Delta V^{2}R_{\mathrm{BLR}}}{G}\right)$ obtained by reverberation 
mapping, we can get an estimate of $f_{\mathrm{BLR}}$ as 
\begin{equation}\label{eq:fblr}
f_{\mathrm{BLR}} = \frac{M_{BH}^{\sigma_*}} {\mathrm{VP}}.
\end{equation}

\begin{figure}
\begin{center}
\resizebox{9cm}{7cm}{\includegraphics{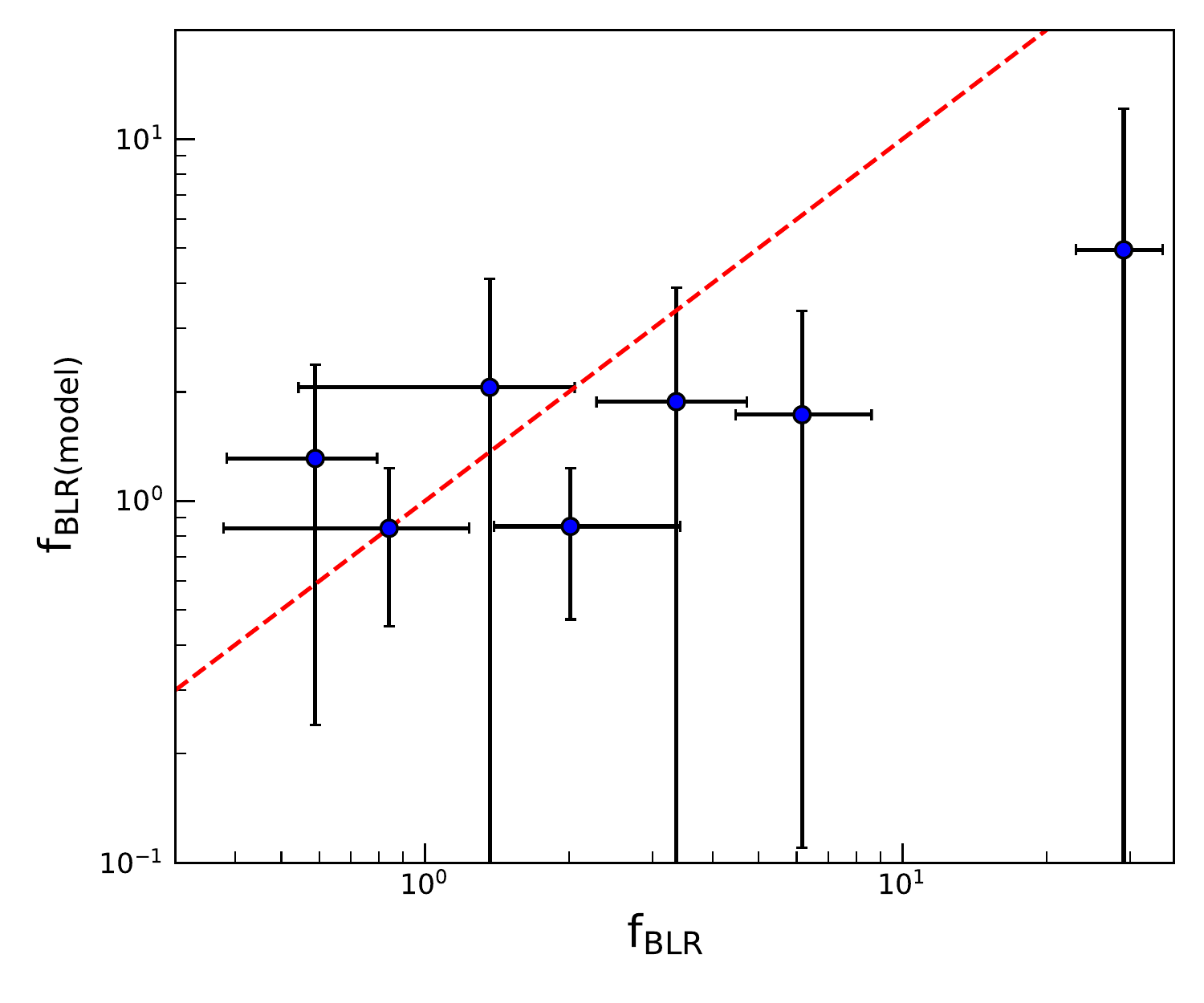}}
\end{center}
\caption{Relation between $f_{\mathrm{BLR}(model)}$ obtained from model fits and $f_{\mathrm{BLR}}$ calculated from the ratio of M$_{BH}$ from stellar velocity dispersion ($\sigma_{*}$) to the virial product (VP). The dashed red line represents the y=x line.}
\label{fig:fig-10}
\end{figure}

For a total of seven sources in our sample, we could obtain both the $f_{\mathrm{BLR}}$ measurements, one based on the 
Bayesian based BLR modeling approach and the other obtained from the ratio of  $M_{\mathrm{BH}}$ based on Equation \ref{eq:m-singma} to the virial product obtained from RM. We found a good correlation between the two virial factors (see Fig. \ref{fig:fig-10}).  Considering the dispersion of $\sim0.4$ dex in the $M_{\mathrm{BH}}-\sigma_{*}$ relation our 1D modeling approach is able to provide $f_{\mathrm{BLR}}$ consistent with that obtained from RM method and
M$_{\mathrm{BH}}$ and $\sigma_*$  relation. From linear least squares fit 
to the data points in Fig. \ref{fig:fig-10}, we found a Spearman rank correlation 
coefficient of $0.29^{+0.39}_{-0.54}$ and a p-value of $0.38^{+0.44}_{-0.31}$. Removing the 
data point with $\mathrm{f_{BLR}}$ $>$ 10, also the one with very large uncertainty, linear least squares
fit gave a linear correlation coefficient of  0.14$^{+0.46}_{-0.51}$ and a $p$ value of $0.47^{+0.40}_{-0.31}$. Though, the points are scattered around the dotted line in Fig. \ref{fig:fig-10}, the derived $\mathrm{f_{BLR(model)}}$ have large error bars and this could be the reason for no tight correlation between the scale factors obtained by both the methods. Most of the values of $\mathrm{f_{BLR(model)}}$ are found to be lesser than 3, which points to a BLR with a thick geometry and viewed at an inclination angle. Given the fact that $f_{\mathrm{BLR}}$ has a large range, the M$_{\mathrm{BH}}$ values 
obtained from single epoch measurements adopting a single $f_{\mathrm{BLR}}$ are bound to have large uncertainties.
\cite{2018NatAs...2...63M} by comparing $M_{\mathrm{BH}}$ obtained by accretion disk model fitting and virial methods found that $f_{\mathrm{BLR}}$ is correlated with the width of the broad emission lines as $f_{\mathrm{BLR}} \propto \mathrm{FWHM}^{-1}$. 

\begin{figure}
\begin{center}
\resizebox{8cm}{8cm}{\includegraphics{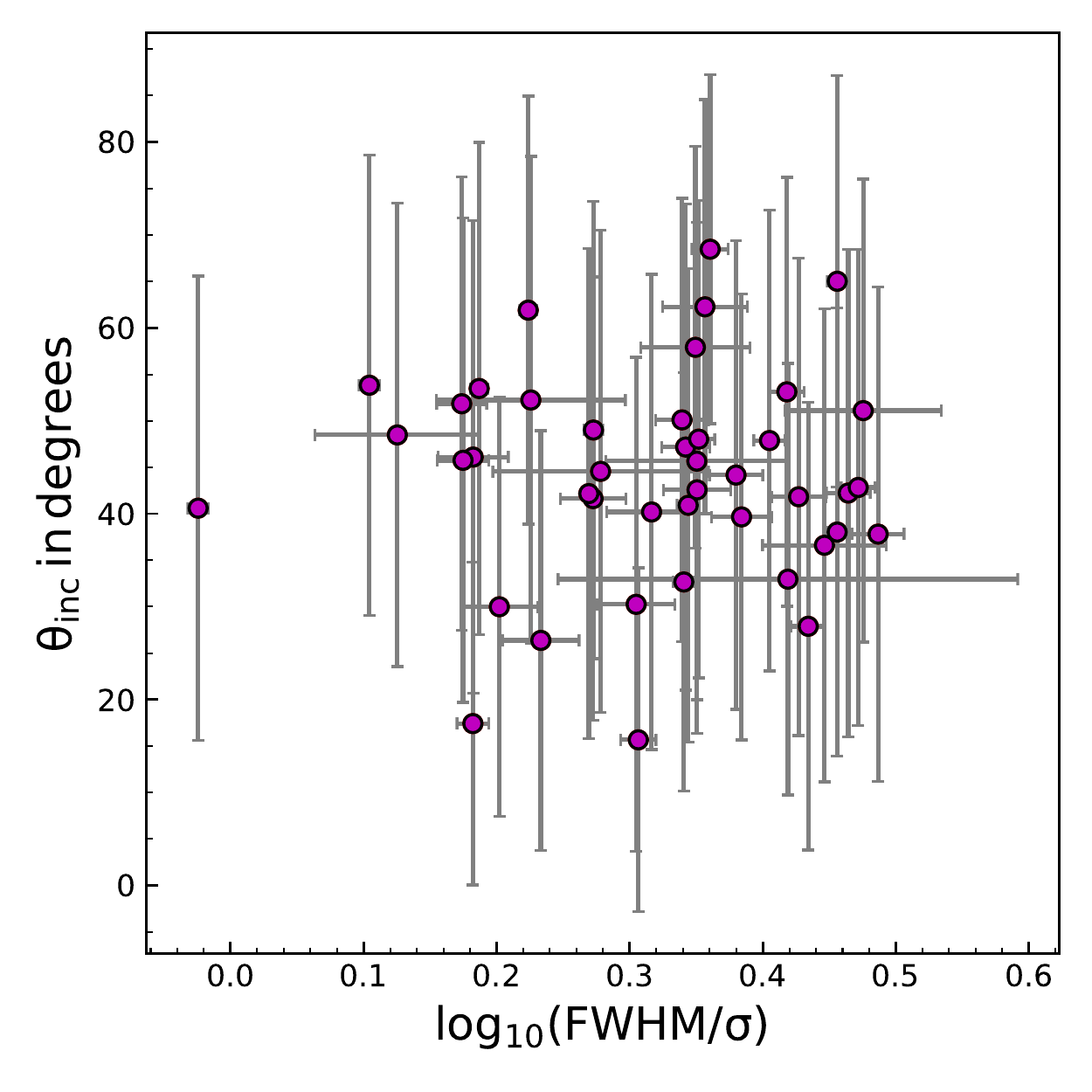}}
\end{center}
\caption{Inclination angle from model 
as a function of the ratio of the FWHM to the line dispersion $\sigma$ of the 
$\mathrm{H\beta}$ line.}
\label{fig:fig-11}
\end{figure}
 
Also, the ratio of FWHM to the line dispersion of broad H${\beta}$ line is 
suggested to be correlated with the inclination angle 
\citep{2006A&A...456...75C,2012MNRAS.426.3086G}. However, \citet{2014MNRAS.445.3073P} 
could not find any correlation using a small sample of 5 objects having good quality 
measurements. 
We used the available FWHM and line dispersion measurements from the RMS 
spectra of broad $\mathrm{H\beta}$ line collected from the literature. We plot 
the inclination angle from the model as a function of the 
ratio of the FWHM to the line dispersion in Fig. \ref{fig:fig-11}. 
We do not find any strong correlation. 
Though our measurements 
have large error bars, the results agree with the finding of 
\citet{2014MNRAS.445.3073P}. 

\subsection{Measurement of $M_{BH}$ and accretion rates} 

\begin{figure}
\begin{center}
\resizebox{9cm}{8cm}{\includegraphics{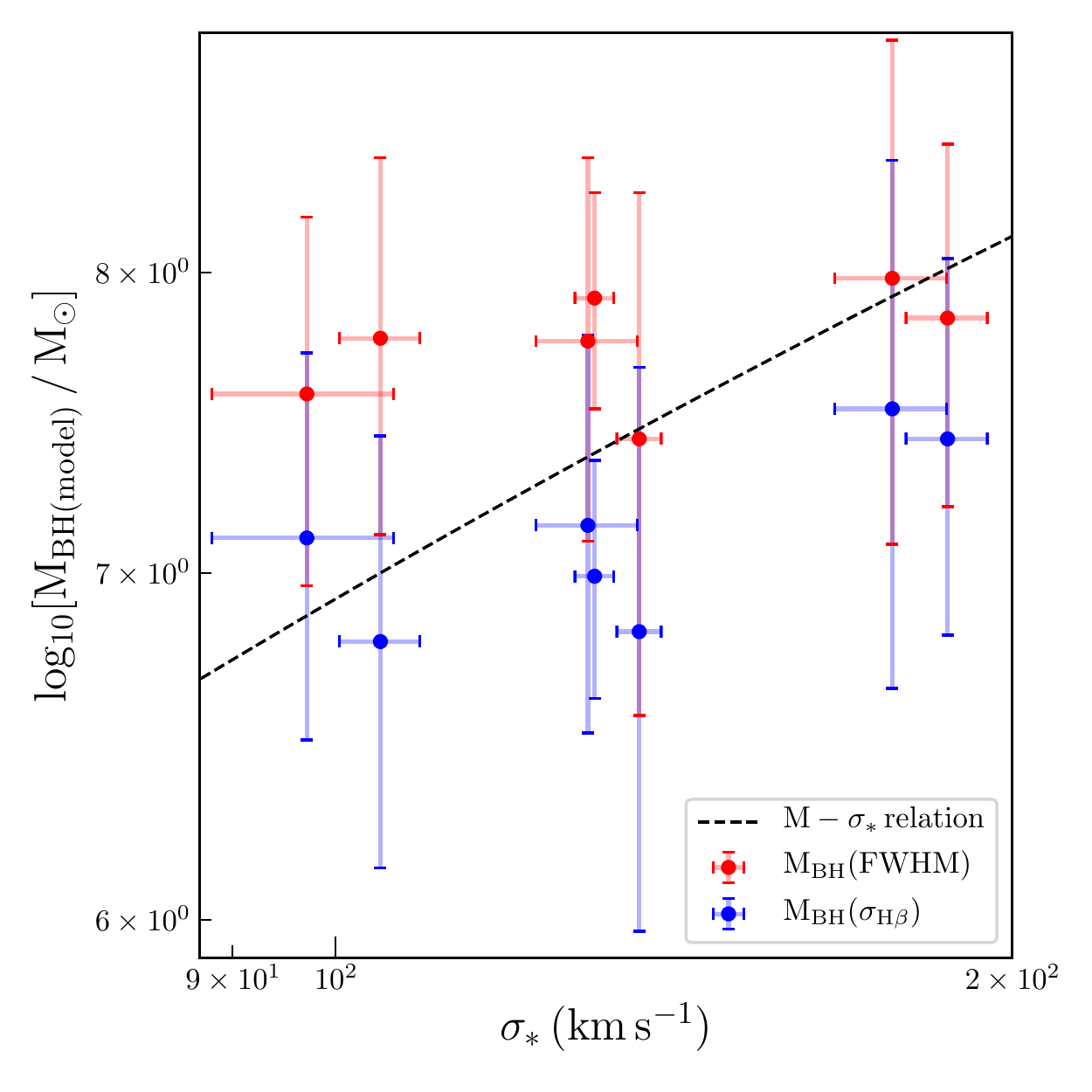}}
\end{center}
\caption{Comparison of black hole masses obtained from model with the $\mathrm{M-\sigma_*}$ relation. The dashed black line represents the $\mathrm{M-\sigma_*}$ relation. Only the objects having stellar velocity dispersion ($\sigma_{*}$) are included in the plot.}
\label{fig:fig-mbh}
\end{figure}

Black hole masses are calculated using Equation \ref{eq:blc_v}, where we adopted $f_{\mathrm{BLR}}$ and $R_{\mathrm{BLR}}$ of the $\mathrm{H\beta}$ line obtained from the model. The velocity width $\Delta V$ can be measured either from the full width at half maximum (FWHM) or from the line dispersion $\sigma_{\mathrm{H\beta}}$. We estimated the black hole masses for those 11 objects which have $f_{\mathrm{BLR}}(\mathrm{model})$ measurements using both FWHM and line dispersion $\sigma_{\mathrm{H\beta}}$ separately as $\sigma_{\mathrm{H\beta}}$ gives less biased $\mathrm{M_{BH}}$ measurement than using the FWHM \citep{2011arXiv1109.4181P,2012ApJ...755...60G}.  In Fig. \ref{fig:fig-mbh}, we compared the $\mathrm{M_{BH (model)}}$ values with the $\mathrm{M-\sigma_*}$ relation as given in Equation \ref{eq:m-singma}. We found that most of the $\mathrm{M_{BH (model)}}$ measurements using FWHM lie above $\mathrm{M-\sigma_*}$ relation, whereas, most of the $\mathrm{M_{BH (model)}}$ values obtained using $\sigma_{\mathrm{H\beta}}$ lie below the $\mathrm{M-\sigma_*}$ line. But considering the uncertainties all $\mathrm{M_{BH (model)}}$ measurements are found to be consistent with the $\mathrm{M-\sigma_*}$ relation.  


We also calculated the dimensionless accretion rate as given by \citet{2018ApJ...856....6D}    

\begin{equation}
\dot{M} = 20.1 \left(\frac{L_{44}}{\cos i}\right)^{3/2}m_7^{-2}
\end{equation}
where $m_7 = M_{\mathrm{BH}}/10^7 M_{\odot}$, $L_{44} = L_{5100}/10^{44} erg s^{-1}$ and $i$ is the inclination angle. Our obtained  values for those 11 objects as mentioned in Table \ref{tab:table-acc} indicate low to moderately accreting black holes with $\dot{M}$ ranging from 0.002 to 2.266. 

\begin{table*}
\caption{$M_{\mathrm{BH}}$ and accretion rate $\dot{M}$ measurements.}
\label{tab:table-acc}
\begin{tabular}{ccccc} \hline
$\alpha_{2000}$  &  $\delta_{2000}$ & $\log(M_{\mathrm{BH}})(\mathrm{FWHM})$ & $\log(M_{\mathrm{BH}})(\sigma_{\mathrm{H\beta}})$ & $\dot{M}$ \\ \hline

02:30:05.52 & -08:59:53.2 & $7.98 \pm 0.70$ &$7.53 \pm 0.70$  & 0.018 \\
06:52:12.32 & +74:25:37.2 & $8.51 \pm 0.85$ &$7.67 \pm 0.54$  & 0.011 \\ 
14:07:59.07 & +53:47:59.8 & $7.91 \pm 0.38$ &$6.99 \pm 0.37$  & 0.232 \\
14:10:31.33 & +52:15:33.8 & $7.84 \pm 0.63$ &$7.43 \pm 0.62$  & 0.510 \\
14:11:12.72 & +53:45:07.1 & $7.58 \pm 0.62$ &$7.11 \pm 0.61$  & 2.266 \\
14:13:18.96 & +54:32:02.4 & $7.76 \pm 0.66$ &$7.15 \pm 0.63$  & 0.563 \\
14:16:25.71 & +53:54:38.5 & $7.75 \pm 0.35$ &$7.03 \pm 0.33$  & 2.188 \\
14:20:39.80 & +52:03:59.7 & $8.14 \pm 0.60$ &$7.27 \pm 0.60$  & 0.187 \\
14:20:49.28 & +52:10:53.3 & $9.29 \pm 0.59$ &$8.93 \pm 0.58$  & 0.002 \\
14:21:03.53 & +51:58:19.5 & $7.77 \pm 0.65$ &$6.79 \pm 0.65$  & 0.211 \\
14:21:35.90 & +52:31:38.9 & $7.43 \pm 0.86$ &$6.82 \pm 0.85$   & 0.420 \\

\hline

\end{tabular}
\end{table*}

\begin{figure*}
\begin{center}
\resizebox{18cm}{11cm}{\includegraphics{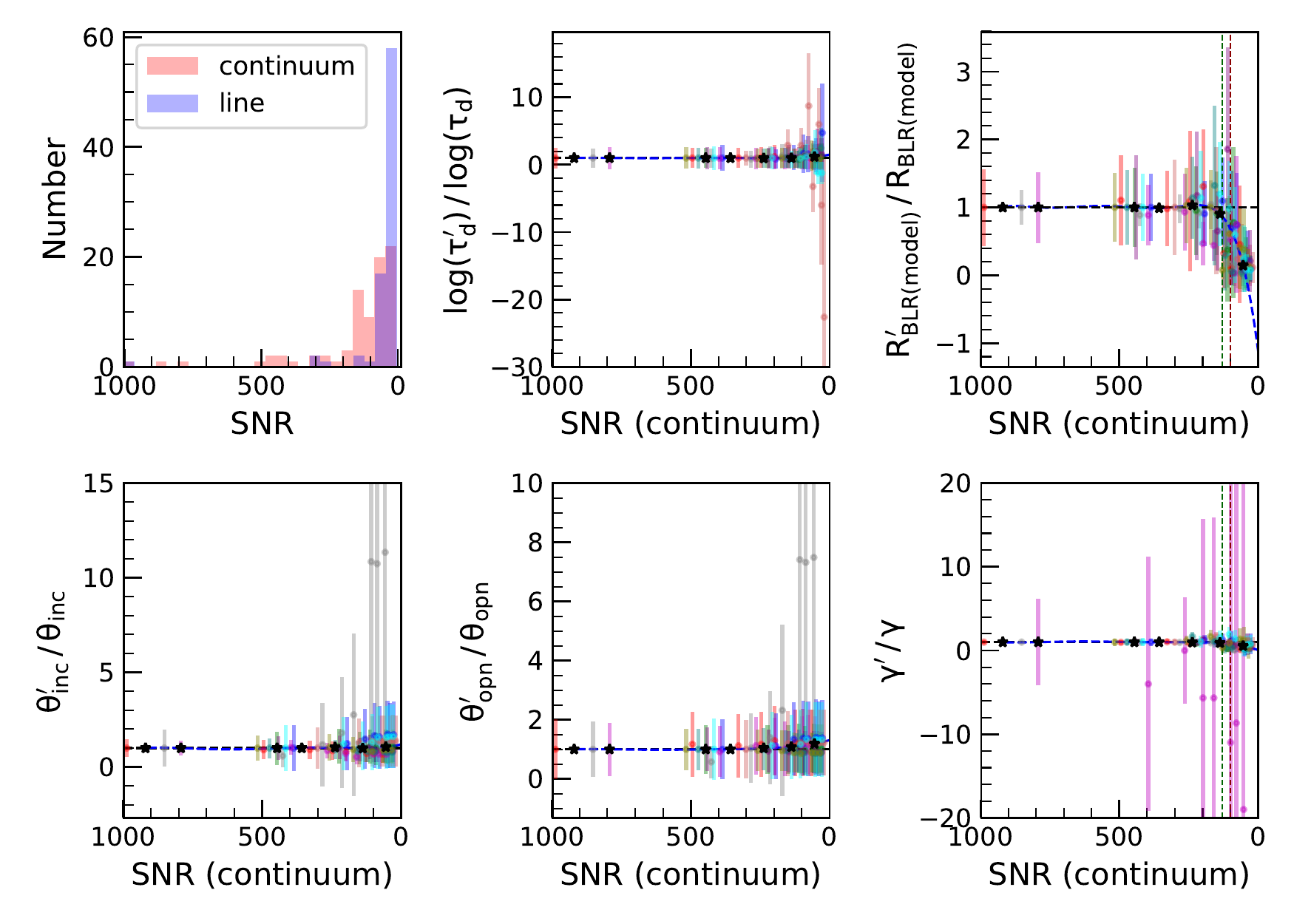}}
\end{center}
\caption{ From top left to bottom right: SNR distribution of all the objects 
	in continuum and line and comparison of the recovered model parameters 
	$\mathrm{\tau_d^{\prime}}$, $\mathrm{R_{BLR(model)}^{\prime}}$, 
	$\mathrm{\theta_{inc}^{\prime}}$, $\mathrm{\theta_{opn}^{\prime}}$ 
	and $\mathrm{\gamma^{\prime}}$ from the SNR degraded simulated light curves to 
	those obtained from the original light curves. Measurements from each object 
	are shown by an unique colour. The sample median is also shown by a star marker. 
	The dashed black lines represent the y=1 lines, whereas the dashed blue lines 
	represent the best polynomial fit to the sample median values in each panel. 
	The vertical green and red lines correspond to the SNR values where the comparisons 
	deviate from the unit ratio by 10\% and 30\%, respectively.}
\label{fig:fig-snr}
\end{figure*}

\subsection{Reliability of the recovered model parameters: effect of signal to noise ratio}

Modeling of the continuum and line light curves to estimate various BLR parameters 
depends on the SNR of the light curves. \citet{2001ASPC..224..457C} and 
\citet{2004PASP..116..465H} suggested that BLR parameters can be well recovered 
(a) with continuum light curves of SNR $\sim$ 100 and (b) line light curves of SNR $\sim$ 30. But it is often difficult to find RM data satisfying the above mentioned qualities.

The distribution of the SNR of the continuum and line light curves used in this 
work is shown in Fig. \ref{fig:fig-snr} (top left). They span a wide range from 
as low as 5 to as high as 1000. The median SNR of the continuum and line light 
curves used in this work is about 84 and 30, respectively. To access the effects 
of SNR on the derived BLR parameters, we carried out simulations. We firstly 
selected objects with continuum light curves with SNR greater than 300. We 
arrived at a total of 10 objects. We then degraded the SNR of the continuum and 
line light curves of those 10 objects by multiplying a factor of 2, 3, 4, 5, 8, 
10 and 15 to the original flux errors and adding a Gaussian random deviate of 
zero mean and standard deviation given by the new flux errors. We then applied 
the PBMAP code on the simulated light curves and extracted the BLR parameters. 
Here, the ratio of the recovered to the original BLR parameters are plotted 
against the continuum SNR. It is evident from Fig. \ref{fig:fig-snr}, that this 
ratio is close to unity for most of the BLR parameters, except for the radius of 
the BLR, where it is found to deviate by 10\% and 30\%, when the SNR of the 
continuum are 130 and 100, as shown by the vertical green and red lines, 
respectively. Similarly, for the line light curves too, we found that the ratio 
of the recovered to the original BLR parameters are close to unity except for 
the BLR size, which deviates by 10\% and 30\%, when the SNR of the line are 25 
and 15, respectively. This is also in agreement with the continuum SNR cut-off 
of 100 suggested by \citet{2001ASPC..224..457C} and \citet{2004PASP..116..465H} 
to extract BLR parameters from RM data and line SNR cut-off of 15, considering 
an accuracy of 70\% to the original recovered parameters.

For the 57 objects studied in this work, we have a total of 82 different 
measurements of $\mathrm{H\beta}$,  $\mathrm{H\alpha}$ and MgII lines, out of 
which $\sim 42$\% of objects have continuum and line SNR greater than 100 and 
15, respectively. Also, for all the objects the BLR sizes obtained from model 
fits are consistent with those obtained from conventional CCF analysis within 
the errors.  However, as the SNR is found to have a major effect on the derived 
sizes of the BLR from the simulations, we note that the values of the size of 
the BLR  obtained for sources with continuum and line light curves with SNR 
lesser than 100 and 15, respectively, needs to be used with caution. 
We also carried out an analysis of the correlation of  $\tau_d$ and  the BLR sizes
obtained from model to the luminosity to only those sources that have 
the continuum and line light curves SNR greater than 100 and 15, respectively. Though the 
trend of the correlation is similar to that of the full sample, the significance
of the correlation is not strong due to the low number of sources.

We show in Appendix B (see Fig. \ref{fig:fig-rep}), sample light curves 
and the recovered transfer functions for two sources. One belongs to J1411+537 that 
has good quality light curves with continuum and line SNR of about 386 and 60, 
respectively. The other light curves belong to J1417+519, that has 
SNR of about 9 and 3 for the continuum and line light curves, respectively. From these light 
curves it is clear that BLR parameters are well constrained only for sources with 
good SNR data.

\section{Summary}\label{sec:summary}
We analysed RM data collected from the literature for a total of 57 AGN that includes
51 AGN with $\mathrm{H\beta}$ data, 26 AGN with $\mathrm{H\alpha}$ data and 6 AGN with MgII line data. The main motivation is to constrain the structure and
dynamics of the BLR that emits MgII, $\mathrm{H\beta}$ and $\mathrm{H\alpha}$. We summarize our results below
\begin{enumerate}
\item The estimated BLR sizes using our approach are in general consistent with that 
calculated from conventional CCF analysis. 
\item The best-fitted model $\mathrm{H\beta}$ BLR size is correlated with $L_{5100{\AA}}$ 
having a slope of $0.58\pm0.03$. This is similar to what is known in literature from
CCF analysis. We also examined the correlation of R$_{\mathrm{BLR}}$ (H$\alpha$) with the continuum luminosity at
5100 \AA \ and found a slope of $0.47\pm0.08$ similar to what is expected from photo-ionization calculations. However more $\mathrm{H\alpha}$ measurements are needed to better constrain this correlation.
\item We estimated virial factor using geometrical parameters and obtained a mean of 
1.78 $\pm$ 1.77. The model virial factor is consistent with the virial factor obtained by
the ratio of M$_{\mathrm{BH}}$ from $M-\sigma_*$ relation to the virial product (VP) obtained from RM. Using line light curves only it is not possible to constrain the virial factor $f_{BLR}$ \citep{2013ApJ...779..110L}. For that reason our measured $f_{BLR}$ have large uncertainties because of large errors present in both $\theta_{inc}$ and $\theta_{opn}$ obtained from the model fitting. 
\item We found a close correspondence between the BLR size found from model and that estimated from
CCF analysis, however, some objects do show large deviation. The objects that show large deviation
from the R$_{\mathrm{BLR}} (\mathrm{model})$ = R$_{\mathrm{BLR}} (\mathrm{CCF})$ line have poor quality light curves.
\item The mean value of the non linearity parameter $\gamma$ is found to be non zero for different lines indicating deviation from linear response of the line emission to the optical ionizing continuum. This may be due to a) the anomalous behaviour of the BLR region because of the poor correlation between optical continuum variability and the ionizing continuum variability \citep{1996ApJ...470..364E,2002AJ....124.1988M,2019arXiv190906275G} and b) anisotropic line emission from the partially optically thick BLR. This anisotropic effect is not considered in the model used here. 

\item Variability analysis of the sample indicates that line varies more than the continuum. The damping time scale obtained from modeling is found to be positively correlated with the continuum luminosity at 5100 \AA.

\item  From the analysis of the simulated light curves, we conclude that reliable 
	estimation of BLR size as well as other parameters via the modeling approach
	requires continuum and line light curves with SNR greater than 100 and 15, respectively.


\end{enumerate}

This work has considerably increased the number of objects investigated through geometrical modeling
of the BLR. Despite that, we were able to estimate $f_{\mathrm{BLR}}$ for only about a dozen objects. 
Analysis of high quality data sets for more number of AGN are needed to find precise estimates of $f_{\mathrm{BLR}}$ which can then be used with the conventional RM techniques to estimate more accurate M$_{\mathrm{BH}}$ values.


\section*{Acknowledgements}
 We thank the referee for valuable comments and suggestions that helped to 
improve the quality of the manuscript. We are thankful to 
Yan-Rong Li (IHEP, CAS) for making the code PBMAP available and providing 
instruction to run the code. AKM and RS acknowledge support from the 
National Academy of Sciences, India.

\section*{Data availability}

The data used in this article are taken from the literature, the references of which are given in Table \ref{tab:table-1}.



\clearpage

\appendix
\section{BLR lag measurement of J1420+526 for $\mathrm{H\alpha}$ line using ICCF }\label{sec:ID301}

\begin{figure}
\begin{center}
\resizebox{7cm}{6cm}{\includegraphics{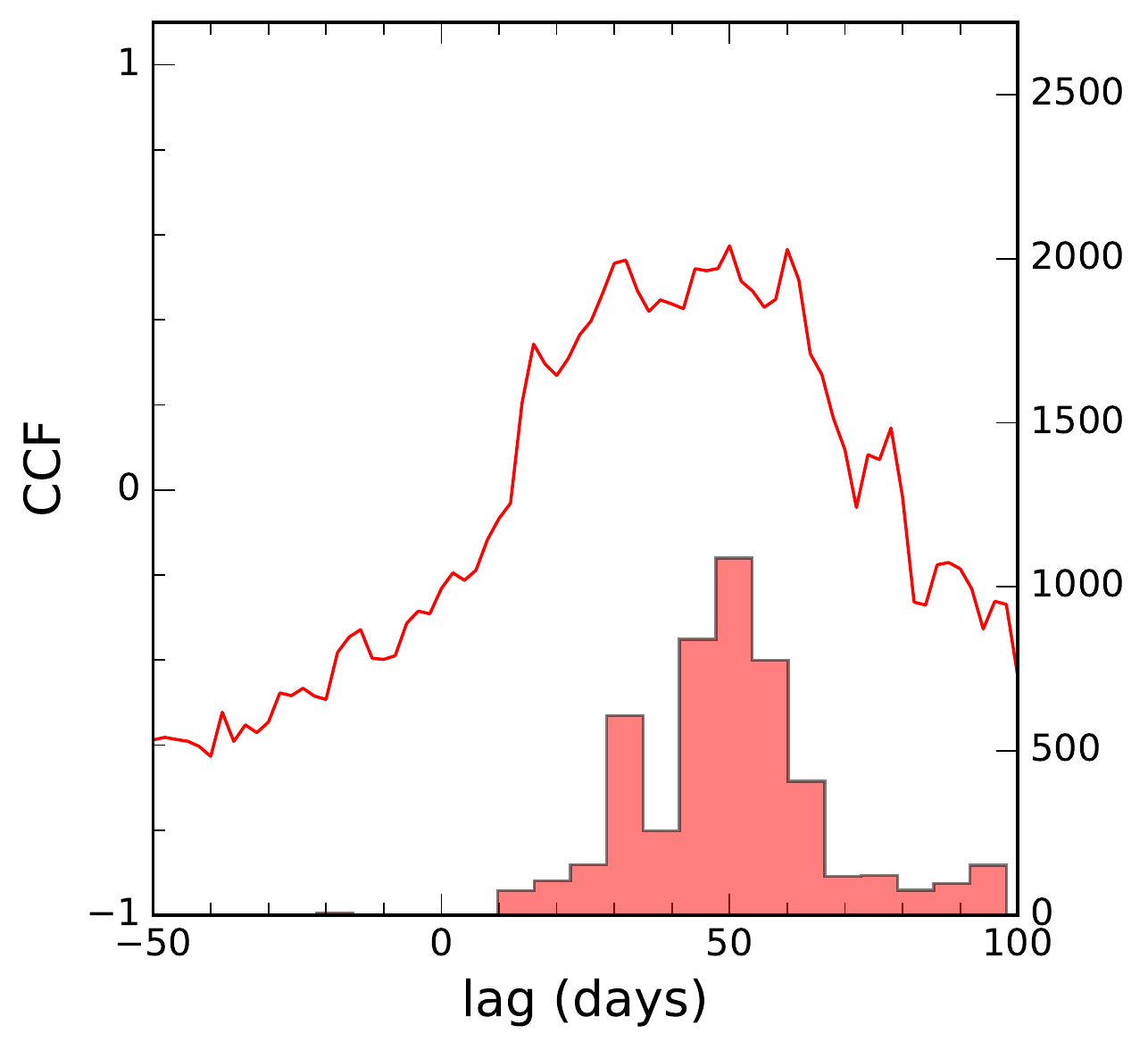}}
\end{center}
\caption{The red solid line represents the average ICCF between g-band and H$\alpha$ line while the histogram shows the centroid lag distribution for object J1420+526 having a $\tau_{\mathrm{cent}} = 49.88^{+11.99}_{-17.88}$ days in observed frame of the object.
}
\label{fig:id301}
\end{figure}

\citet{2017ApJ...851...21G} did not perform CCF analysis to measure the H$\alpha$ 
lag for the object J1420+526. We estimated the H$\alpha$ lag for this object using Interpolated Cross-correlation Function (ICCF) analysis  method as shown in Fig. \ref{fig:id301}. The lag and its uncertainty  are estimated using a Monte Carlo simulation based on the flux randomization (FR) and random subset selection (RSS) described in \citet{1998PASP..110..660P}, \citet{1999ApJ...526..579W} and \citet{2004ApJ...613..682P}. The median of the centroid distribution is considered as final lag while uncertainties were estimated within a 68$\%$ confidence interval around the median value. We obtained the rest frame H$\alpha$ lag of $32.22^{+7.75}_{-11.55}$ days from ICCF method. The lag estimated based on modeling is $46.67 \pm 9.8$ days which is consistent with CCF lag within error bars.

\section{Examples of model fit light curves and transfer functions}
\begin{figure*}
\centering
\resizebox{8cm}{4.5cm}{\includegraphics{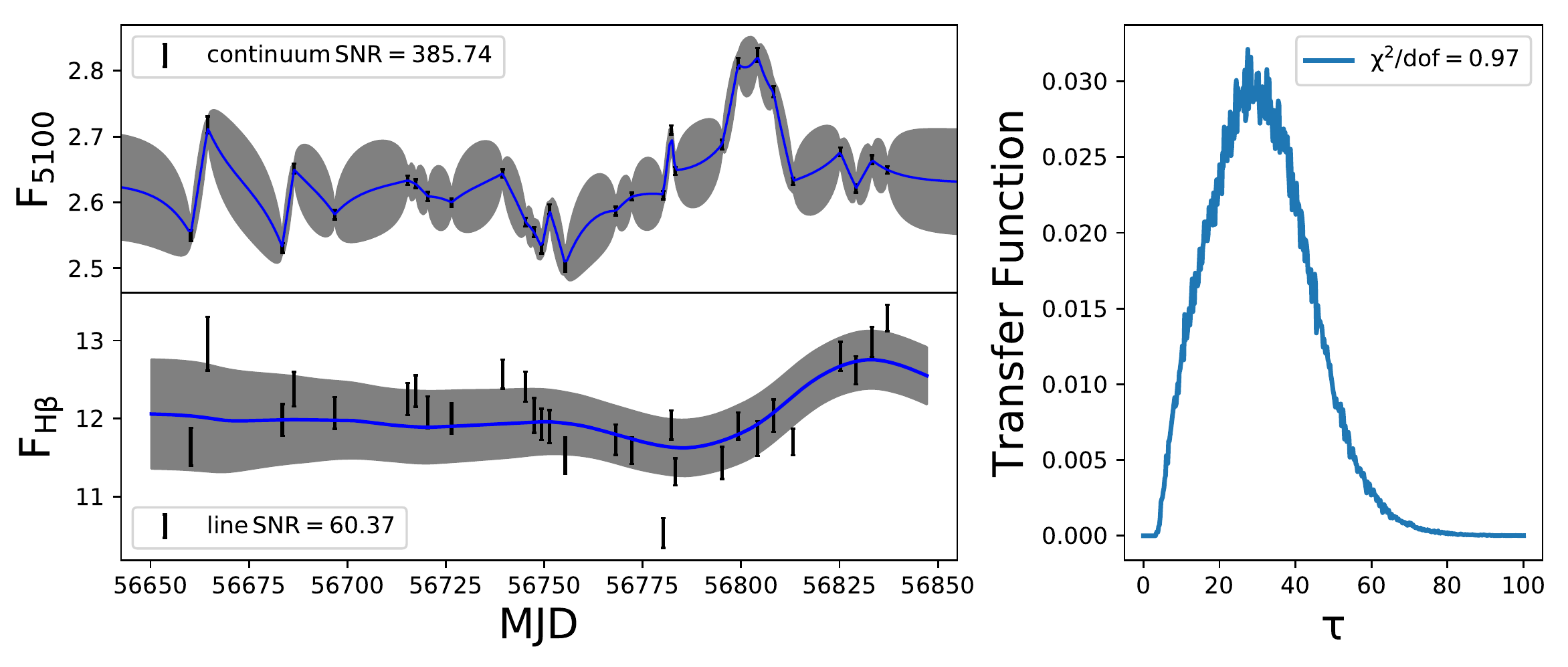}}
\resizebox{8cm}{4.5cm}{\includegraphics{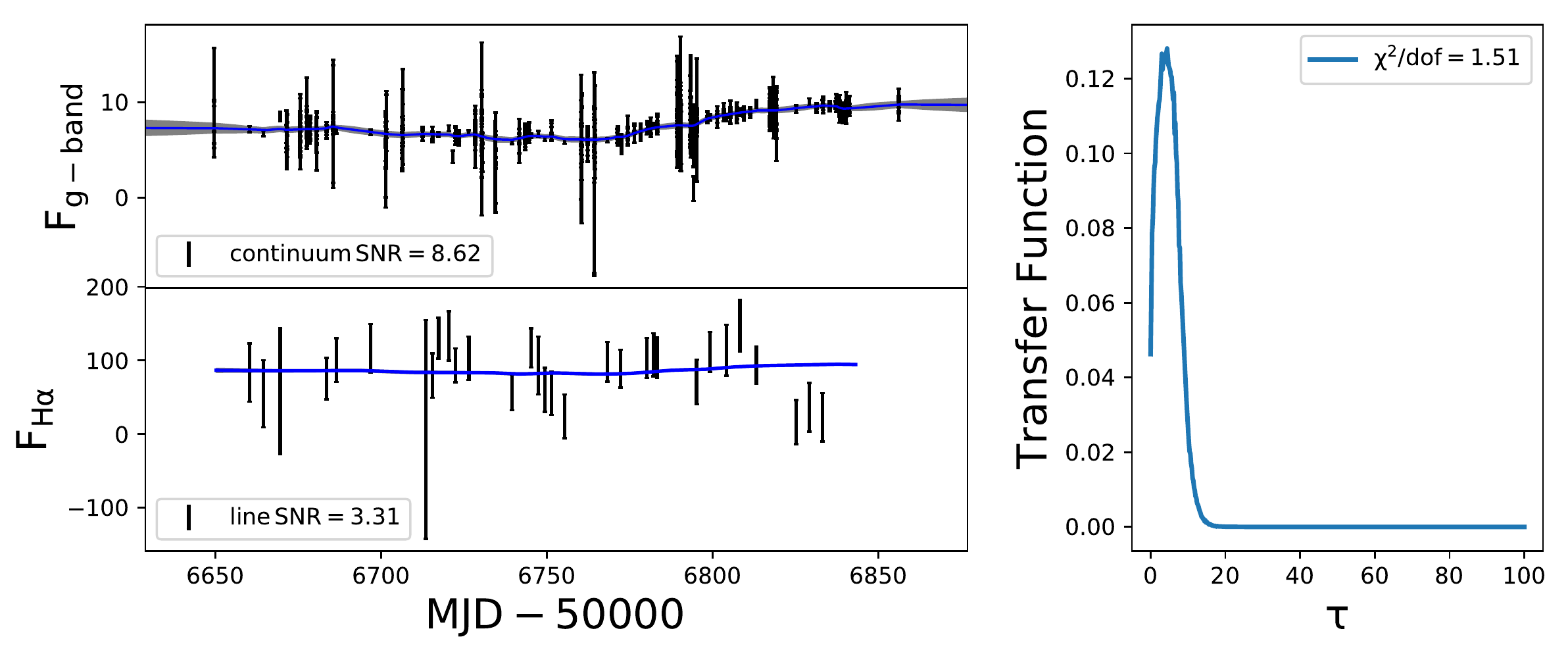}}
\resizebox{15cm}{8cm}{\includegraphics{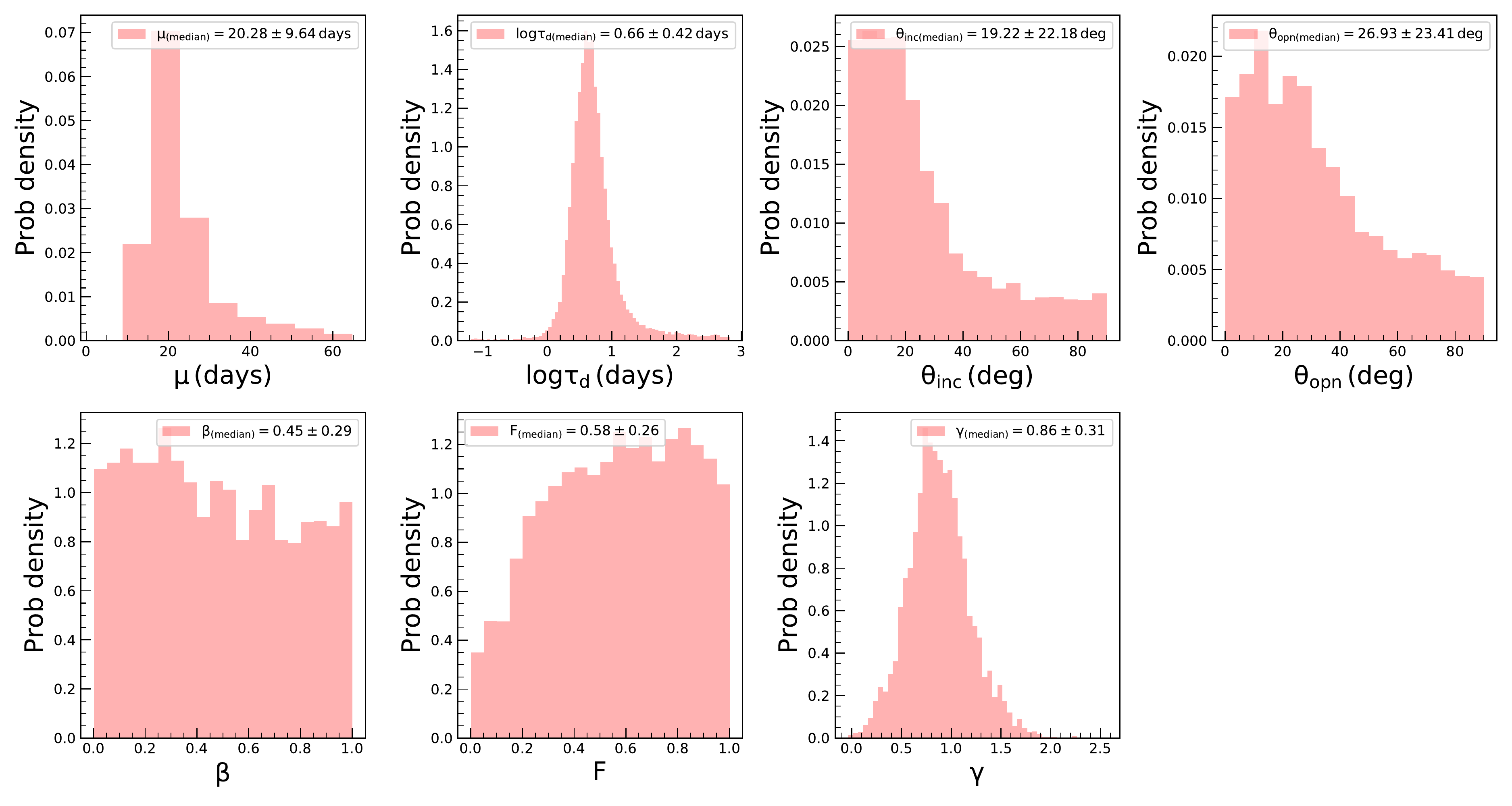}}
\resizebox{15cm}{8cm}{\includegraphics{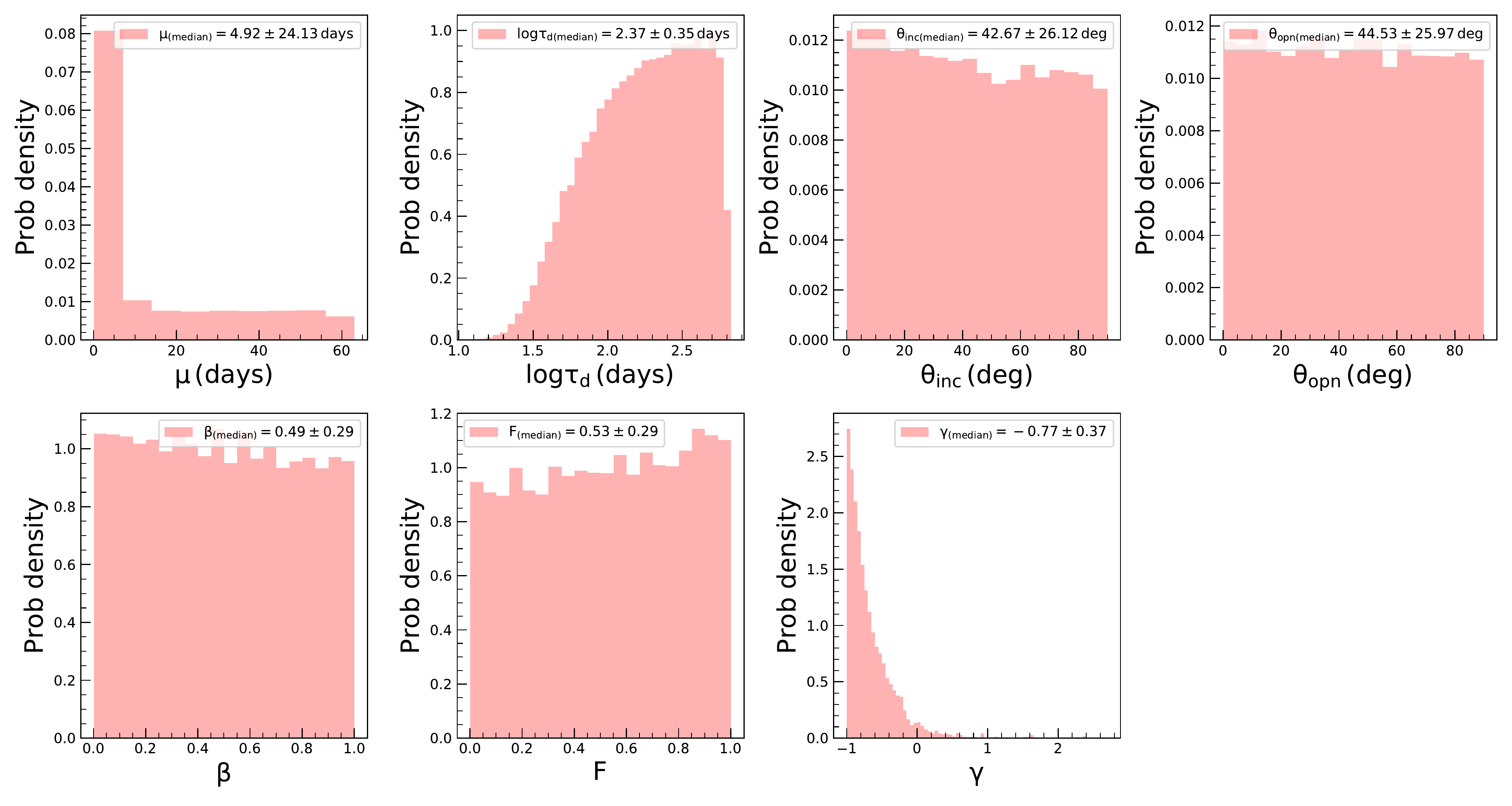}}

        \caption{(Top) Model fits for the 
	source (J1411+537) with high SNR light curves (left) and the 
	source (J1417+519) with low SNR light curve (right). 
	In the left-hand panels the data points with error bars are the observed light 
	curves and the thick solid lines are the reconstructed light curves. The gray 
	shaded areas represent the uncertainties in the reconstructed light curves. The 
	corresponding transfer function for each object is shown on the right hand 
	panels. The SNR of the light curve is mentioned at each panel. Posterior 
	probability distributions of different model parameters are also shown for J1411+537 (middle) and J1417+519 (bottom).}
\label{fig:fig-rep}
\end{figure*}

We show in Figure \ref{fig:fig-rep} light curves for two objects, namely 
J1411+537 having high SNR in the continuum and line, and J1411+537 having poor 
SNR in both the light curves. From the figure it is evident that the geometrical model 
parameters ($\theta_{\mathrm{inc}}$ and $\theta_{\mathrm{opn}}$) are 
well-constrained for J1411+537 but not constrained for J1411+537.
This is due to the low SNR of the continuum and line light curves in J1411+537.

\bsp	
\label{lastpage}
\end{document}